\documentclass[12pt]{book}

\usepackage{a4}
\usepackage{fancyheadings_cb}
\usepackage{kapitel}
\usepackage{thesis}
\usepackage{psfigpc}
\usepackage{subeqn}
\usepackage{epsfig}

\def\gsim{ \,\, \vcenter{\hbox{$\buildrel{\displaystyle >}\over\sim$}}
 \,\,}
\def\lton{ \,\, \vcenter{\hbox{$\buildrel{\displaystyle <}\over\sim$}}
 \,\,}

\newcommand{\bse}{\begin{subequations}}
\newcommand{\ese}{\end{subequations}}
\newcommand{\be}{\begin{equation}}
\newcommand{\ee}{\end{equation}}
\newcommand{\bea}{\begin{eqnarray}}
\newcommand{\eea}{\end{eqnarray}}
\newcommand{\ban}{\begin{eqnarray*}}
\newcommand{\ean}{\end{eqnarray*}}
\newcommand{\nn}{\nonumber}
\newcommand{\half}{\frac{1}{2}}

\renewcommand{\d}{\mbox{d}}
\renewcommand{\Im}{\rm{Im}}
\renewcommand{\Re}{\rm{Re}}
\begin{document}
%----------------------------------------------------------------------
\newcommand{\Kapitel}{\Roman{chapter}}
\pagestyle{fancy}

\addtolength{\headwidth}{\headheight}

\lhead[\fancyplain{}{\thepage}]%
{\fancyplain{}{\scshape\rightmark}}

\rhead[\fancyplain{}{\scshape\leftmark}]%
{\fancyplain{}{\thepage}}

\cfoot{}

\renewcommand{\chaptermark}[1]%
                {\markboth{#1}{}}
\renewcommand{\sectionmark}[1]%
                {\markright{\thesection\ #1}}

\newcommand{\clearemptydoublepage}{\newpage{\pagestyle{empty}\cleardoublepage}}

\begin{titlepage}
\begin{center}
\vspace*{\fill}
{\Large\bf Selfconsistent calculations of mesonic properties
at nonzero temperature}
\vspace{2cm}

{\large Dissertation \\
zur Erlangung des Doktorgrades\\
der Naturwissenschaften\\

%\vspace{1cm}
\vfill
vorgelegt beim Fachbereich Physik\\
der Johann Wolfgang Goethe--Universit\"at\\
in Frankfurt am Main\\

\vfill
von\\
Dirk R\"oder\\
aus Frankfurt am Main\\

\vfill
Frankfurt 2005\\
(D 30)\\}
\end{center}

\pagebreak
\thispagestyle{empty}
\vspace*{\fill}
{\large
{vom Fachbereich Physik der \\
Johann Wolfgang Goethe--Universit\"at als Dissertation angenommen}\\

\vspace{3cm}
Dekan: Prof. Dr. W. A\ss mus\\

Gutachter: Prof. Dr. Dirk H. Rischke\\
\hspace*{2.5cm}JProf. Dr. Adrian Dumitru\\

Datum der Disputation: 13.12.2005}

\end{titlepage}

\thispagestyle{empty}
\vspace*{2cm}
\begin{center}
\pagestyle{empty}
{\large {\it F\"ur meine Frau Julia \\
und meine S\"ohne Justus und Philipp.}}
\end{center}
\thispagestyle{empty}

\tableofcontents
\listoffigures 

%**********************************************************************
%*********Introduction*****************************************************

\chapter{Introduction}

\section{Four fundamental forces}
Modern physics believes that matter is held together through 
four fundamental forces, gravity, the weak interaction,
the electromagnetic force, and the strong interaction 
(cf. Tab.~\ref{4forces}). In this section, I briefly discuss some
aspects and differences of these forces.

The force with the smallest relative strength is gravity,
which couples two particles together via 
their mass, and which acts on all known particles.
It is the most ``common'' force in our
everyday life, and describes how the apple falls down from the tree.
Newton's theory of gravity (17th century) describes this force as a
long-range interaction between two bodies. In modern physics
interactions are described as an exchange of virtual 
particles, which ``carry'' the force from one particle to
another. Such a modern theory of gravity is Einstein's
theory of general relativity (20th century). The exchange particle
for the gravity is the graviton with
vanishing mass. The range $R$ of the force can be estimated
by the mass $m$ of the exchange particles \cite{KlapdorStaudt}
\be\label{range}
R\approx \frac{h}{2\pi mc}\;\;,
\ee
where $h=6.626\,0693\times 10^{-34}\,Js$ is Planck's constant, 
and $c=299\,792\,458\,ms^{-1}$ the velocity of light.
The mass of the graviton is zero and therefore the range of gravity is
infinity, as one expects. As mentioned above, gravity is very
weak compared to the other
three forces, thus it does not play an important role in microscopic
processes. Indeed, gravity is the only one of the four forces
where the interaction between two particles with equal charge
is attractive, and not repulsive.
This fact, together with the infinite range of gravity, leads to
its extreme importance in all astrophysical phenomena.
\begin{table}
\begin{center}
\begin{tabular}{|l|c|c|c|c|}
\hline
force & relative & range $[m]$& exchange
               & interaction between\\
&strength & &   particle         & two particles \\
&&&            & with equal charge\\\hline
gravity          & $10^{-38}$ & $\infty$   & graviton & attractive\\
weak interaction & $10^{-5} $ & $10^{-18}$ & $W^\pm,\,Z^0$    & repulsive\\
EM interaction & $10^{-2} $ & $\infty$ & photon    & repulsive\\
strong interaction & $1$ & $10^{-15}$ & gluon    & repulsive\\\hline
\end{tabular}
\end{center}
\caption[The four fundamental forces of nature by comparison.]
{The four fundamental forces of nature by comparison.}
\label{4forces}
\end{table}

Another fundamental force is the so called
weak interaction, which acts
on all fermions (particles with spin $1/2$). 
The exchange particles of the weak interaction are the 
very massive $Z^0$ and $W^\pm$ bosons, their masses are $\sim$ 80 times
larger than the mass of the proton \cite{PDBook}. Therefore
this interaction is of short range $R\approx 10^{-18}$ m,
which is $\sim 1000$ times smaller than an atomic
nucleus. It is very important for all decay processes (involving
fermions), e.g., the beta decay of the neutron
\be
n\rightarrow p+e^-+\bar{\nu_e},
\ee
where $n$ is a neutron, $p$ a proton, $e^-$ an electron, and $\bar{\nu_e}$
an anti-neutrino.

The electromagnetic (EM) force acts on all particles with electric
charge. Maxwell was the first (19th century) who could unify
the electric and the magnetic interactions in one theory. A modern
description of this interaction is quantum electrodynamics (QED)
(20th century). The mass of the exchange particles, the
photons, is zero,
and therefore the range of the electromagnetic interaction
is infinity. It is, in addition to gravity, very
common in our everyday life, and is used to light our rooms and
to power our hi-fi systems. In the 20th century, Glashow, Weinberg, 
and Salam worked out that the electromagnetic and the weak interaction
can be unified in one theory, the electroweak theory.

Last but not least, the strongest of the fundamental forces is the
strong interaction, all particles which experience this force
are called hadrons. The modern theory of strong interaction
is quantum chromodynamics (QCD) (this name can be traced back
to the name of the QCD charge: colour). 
The exchange particles are 8 different massless gluons, thus 
one would expect, from Eq.\ (\ref{range}), that the range of this force
is infinity. However, this is not the case, the range of the strong interaction
is $R\approx 10^{-15}$ m (of the order of an atomic nucleus).
The reason for this is that the gluons can interact with {\it themselfs},
which leads to the finite range of the force
and many other complications. 
QCD is the theory that this thesis is based on, and I will focus
on it in the next sections.

\section{Quantum Chromodynamics}\label{secQCD}
In this section I discuss some aspects of QCD which are important
for my thesis, especially the chiral and the
$Z(N_c)$ symmetry. Certainly, this introduction is far from
complete, for more details see e.g. 
\cite{SWeinberg1,PeskinSchroeder,SWeinberg2,
Kugo,TaLing,GreinerQCD,HalzenMartin}.

As mentioned above, modern physics
believes that QCD is the best theory we have
to describe the strong interaction between the bosonic gluons
and the fermionic quarks. Like all fundamental forces
(apart from gravity) it is based on a local
gauge symmetry, the $SU(3)_c$ colour symmetry.
As shown in \cite{'tHooft:1972fi}
it is a renormalizable theory, which means that divergences
can be regularised via the introduction of a finite number
of counter terms to
the Lagrangian of the theory. Another important fact is
that QCD is an asymptotically free theory 
\cite{Gross:1973ju,Politzer:1973fx},
i.e., the interaction between gluons and quarks becomes weaker with
increasing energy. One expects that this leads to a so-called quark-gluon
plasma (QGP) at large temperatures and/or chemical potentials
 \cite{Collins:1974ky}, which means
that the quarks and the gluons are no longer confined together but
can behave as free particles in a plasma. The
phase transition between hadronic matter, which is the state of
matter under ``normal'' conditions in our universe today,
and the QGP is extensively discussed in this thesis. 

The QCD Lagrangian is given by
\bse\be\label{qcd_lag}
{\cal L}=\bar\psi(i\gamma^\mu D_\mu-m)\psi
-\frac{1}{4}F_a^{\mu\nu}F^a_{\mu\nu},
\ee
where
\be
D_\mu\equiv\partial_\mu-igA_\mu^aT_a,
\ee
is the covariant derivative, and
\be
F_a^{\mu\nu}=\partial^\mu A_a^\nu-\partial^\nu A_a^\mu
+gf_{abc}A_b^\mu A_c^\nu,
\ee\ese
the gluonic field strength tensor.
For $N_c$ colours and $N_f$ quark flavours, $\psi$ is a 
$4N_cN_f$-dimensional spinor of quark fields, $\bar\psi$ the
Dirac conjugate spinor, $\gamma^\mu$ the $4$-vector of Dirac matrices,
$m$ the quark mass matrix, $g$ the strong coupling constant,
$A_a^\mu$ the gluonic fields, and $f_{abc}$ the structure constants
of the local $SU(N_c)_c$ colour
symmetry. As mentioned above, the QCD Lagrangian
foots on this symmetry, which means that it is
invariant under transformations of the fields
${\cal L}(\psi,A_a^\nu)
={\cal L}(\Omega^\dagger\psi,\Omega^\dagger A_a^\nu\Omega)$,
where
\be
\Omega(X)\equiv\exp\left[i\sum_{a=1}^{8}\alpha^a(X)T_a\right]
\ee
describes local $SU(3)_c$ transformations. The $T_a$'s are the
generators of the symmetry group, and $\alpha^a$ are the
corresponding
parameters. This transformation of the fields is called
a {\em gauge transformation}.
Note that, in contrast to global transformations, 
cf.~Eq.(\ref{trans_rl}), a local transformation
depends on the space-time point $X$. In most cases I suppress
the argument of $\Omega$ for simplicity. The matrices $\Omega$
satisfies
\be
\Omega^\dagger \Omega=1\!\!1,\quad \det\Omega=1.
\ee

\subsection*{The chiral $U(N_f)_r\times U(N_f)_\ell$ symmetry}
For massless quarks ($m=0$ in Eq.\ref{qcd_lag}) the Lagrangian 
is invariant under the global chiral symmetry. To see this, 
the spinor and the conjugate spinor of the quark fields are decomposed
into right- and left-handed spinors
\be
\psi\to \psi_r+\psi_\ell\equiv
\frac{1+\gamma_5}{2}\psi+
\frac{1-\gamma_5}{2}\psi.
\ee
Using this, the Lagrangian becomes
\be
{\cal L}=i(\bar\psi_r\gamma^\mu D_\mu\psi_r
+\bar\psi_\ell\gamma^\mu D_\mu\psi_\ell)
-\frac{1}{4}F_a^{\mu\nu}F^a_{\mu\nu}.
\ee
Invariance under chiral symmetry means that the 
Lagrangian does not change, if the spinors are transformed via
a global $U(N_f)_r\times U(N_f)_\ell$ symmetry transformation,
${\cal L}(\psi_r,\psi_\ell)={\cal L}(\Omega_r\psi_r,\Omega_\ell\psi_\ell)$,
where the right- and left-handed transformations are given by
\be\label{trans_rl}
\Omega_r\equiv\exp\left(i\sum_{a=0}^{N_f^2-1}\alpha_r^aT_a\right),\quad
\Omega_\ell\equiv\exp\left(i\sum_{a=0}^{N_f^2-1}\alpha_\ell^aT_a\right),
\ee
$\alpha_{r,\ell}$ are the parameters, and $T_a$ the generators of
the $U(N_f)_{r,\ell}$ symmetry group. The chiral symmetry group is
isomorphic to the vector and axial vector group,
$U(N_f)_r\times U(N_f)_\ell\cong U(N_f)_V\times U(N_f)_A$, with
$V\equiv r+\ell$, $A\equiv r-\ell$. Every unitary group can be decomposed
into a direct product of a special unitary group and a complex
phase, $U(N)\cong SU(N)\times U(1)$, hence the chiral group can
be written as
$U(N_f)_r\times U(N_f)_\ell
\cong SU(N_f)_r\times SU(N_f)_\ell\times U(1)_r\times U(1)_\ell
\cong SU(N_f)_V\times SU(N_f)_A\times U(1)_V\times U(1)_A$. 

The $U(1)_V$ symmetry corresponds to baryon number conservation, and
is always respected. In the vacuum, 
a non-vanishing expectation value of the quark condensate,
$\langle \bar{q}_{\ell} \, q_{r} \rangle \neq 0$, spontaneously
breaks the above symmetry to $SU(N_f)_{V}$.
This gives rise to $N_f^2$ Goldstone bosons which dominate
the low-energy dynamics of the theory.
As shown by 't Hooft \cite{'tHooft:1976fv,'tHooft:1976up},
instantons break the $U(1)_{A}$ symmetry explicitly to
$Z(N_{f})_{A}$ \cite{Pisarski:1984ms}. (For the low-energy dynamics
of QCD, however, this discrete symmetry is irrelevant.)
Consequently, one of the $N_f^2$ Goldstone bosons becomes massive, leaving
$N_f^2-1$ Goldstone bosons.

The $SU(N_f)_r \times SU(N_f)_{\ell} \times U(1)_A$ symmetry of the 
QCD Lagrangian is also explicitly broken
by nonzero quark masses. The $N_f^2-1$ low-energy degrees of freedom 
then become pseudo-Goldstone bosons. For $M \leq N_{f}$ degenerate quark
flavours, a $SU(M)_{V}$ symmetry is preserved. 
In nature the quark masses are not equal to zero, i.e., the chiral
symmetry is only approximatively conserved. Normally, one uses models
which foot on the chiral symmetry only to describe
the light quarks. The two lightest quarks are the up and down quark,
$1.5$ MeV $\le m_u\le 4.5$ MeV and $4$ MeV $\le m_d\le 8$ MeV.
Treating these quarks as massless, i.e., 
using a $SU(2)_r\times SU(2)_\ell\cong O(4)$ symmetry, yields six
Noether currents,
\bse\be
J_{r,a}^\mu=\bar{\psi}_r\gamma^\mu\frac{\tau_a}{2}\psi_r,\quad
J_{\ell,a}^\mu=\bar{\psi}_\ell\gamma^\mu\frac{\tau_a}{2}\psi_\ell,
\ee
with
\be
\partial_\mu J^\mu_r=\partial_\mu J^\mu_\ell=0
\ee\ese
where $\tau_a$ are the three Pauli matrices \cite{ItzyksonZuber}.
It is common to introduce
the vector and the axial-vector currents
\be
V_a^\mu\equiv J_{r,a}^\mu+J_{\ell,a}^\mu=\bar\psi\gamma^\mu\frac{\tau_a}{2}
\psi,\quad
A_a^\mu\equiv J_{r,a}^\mu-J_{\ell,a}^\mu
=\bar\psi\gamma^\mu\gamma_5\frac{\tau_a}{2}.
\ee
The corresponding conserved charges are
\be
Q_a^V=\int d^3x \psi^\dagger(x)\frac{\tau_a}{2}\psi(x),\quad
Q_a^A=\int d^3x \psi^\dagger(x)\gamma_5\frac{\tau_a}{2}\psi(x).
\ee
To extend this, one can consider also the mass of the strange quark as
``small'', $80\mbox{ MeV}\le m_s\le 130$ MeV. Then one has to replace the
three Pauli matrices by the eight Gell-Mann matrices 
$\lambda_a$ \cite{ItzyksonZuber}.
The masses of the charm, bottom, and top quarks are 
very large, $1.15\,{\rm GeV}\le m_c\le 1.35$ GeV, 
$4.1\,{\rm GeV}\le m_b\le 4.4$ GeV, and 
$ m_t=174.3\pm 5.1$ GeV, hence introduction of these degrees of freedom 
into a chirally symmetric model is not reasonable.

In section 1.4, I present the linear $\sigma$-model with 
$O(N)$ symmetry, which foots on the chiral symmetry for two 
quark flavours.

\subsection*{The $Z(N_c)$ symmetry}

As mentioned above, QCD is based on a local $SU(3)_c$ colour symmetry. In the
following I show how to construct from this the global $Z(3)$ 
symmetry. A particularly simple element
of the $SU(N_c)_c$ symmetry is a constant phase times the unit matrix,
\be
\Omega_c\equiv \exp(i\phi)1\!\!1.
\ee
The constraint
\be
\det\Omega_c=\det[\exp(i\phi)1\!\!1]=\exp(iN_c\phi)=
\cos(N_c\phi)+i\sin(N_c\phi)\stackrel{!}{=}1
\ee
leads to a restriction of the parameter $\phi$
\be
\phi=\frac{2\pi}{N_c}j,\quad j=0,1,\dots,(N_c-1).
\ee
Hence $\Omega_c$ is just the $N_c$ th root of unity
times the unit matrix. Since $j$ is an integer, $\Omega_c$
describes a global $Z(N_c)$ rotation (symmetry) of the fields. 
Indeed, we will see that this is not a symmetry of the theory with quarks,
because this transformation violates the periodicities of the 
quark fields.

To introduce the temperature $T$, I work in Euclidean space-time and use 
the imaginary-time formalism \cite{das,kapusta,leBellac}. From
the laws of quantum statistic, 
cf.~\cite{Gross:1980br}, we know that the (bosonic) gluon fields $A_a^\nu$ 
have to be periodic in $(0,1/T)$,  
\be
A_a^\nu\left(\frac{1}{T},{\bf x}\right)=A_a^\nu(0,{\bf x}),
\ee
and the (fermionic) quark spinor and conjugated spinor,
$\psi$ and $\bar\psi$, anti-periodic, 
\be
\psi\left(\frac{1}{T},{\bf x}\right)=-\psi(0,{\bf x}),\quad
\bar\psi\left(\frac{1}{T},{\bf x}\right)=-\bar\psi(0,{\bf x}).
\ee
Obviously, any gauge transformation of 
these fields which are periodic in $(0,1/T)$, i.e.,
$\Omega({\bf x},1/T)=\Omega({\bf x},0)$, does not change these
periodicities. 't Hooft \cite{'tHooft:1977hy} was the first to
notic that one can also consider a gauge transformation which is only
periodic up to a constant equal to the identity matrix times the 
$N$ th root of unity:
\be
\Omega\left({\bf x},\frac{1}{T}\right)=\Omega_c,\quad
\Omega({\bf x},0)=1.
\ee
Using this transformation the periodicity of the gluon fields remain
\be
\Omega_c^\dagger A_a^\nu\left({\bf x},\frac{1}{T}\right)
\Omega_c
=A_a^\nu\left({\bf x},\frac{1}{T}\right)
=A_a^\nu({\bf x},0)
\ee
where one uses the fact that $\Omega_c$, as a constant phase times
a unitary matrix, commutes with the fields $A_a^\nu$, and the relation
$\Omega_c^\dagger\Omega_c=1$. But the periodicities of the quark fields are
violated
\bse\bea
\Omega_c^\dagger\psi\left({\bf x},\frac{1}{T}\right)
&=&\exp(-i\phi)\psi\left({\bf x},\frac{1}{T}\right)
\neq\psi({\bf x},0),\\
\Omega_c^\dagger\bar\psi\left({\bf x},\frac{1}{T}\right)
&=&\exp(-i\phi)\bar\psi\left({\bf x},\frac{1}{T}\right)
\neq\bar\psi({\bf x},0).
\eea\ese
Hence, the $Z(N_c)$ is an exact symmetry only for the pure gauge theory,
i.e., without quark fields included. In section 1.4, I present a
theory which is based on this symmetry, the Polyakov-loop model.

\section{Phase transitions in QCD}
In this section I discuss the possible phases of strongly interacting
matter and the transitions between them. Therefore, I will briefly
motivate some major concepts for the well-known example of 
water.

Thermodynamic systems can show different types
of macroscopic behaviour (or in other words can be in
different {\it phases}), depending on external degrees of freedom 
(e.g. temperature $T$, pressure $p$, chemical potential $\mu$).
Water can be in three different
phases, the liquid phase, the gas phase, and the solid phase.  
By changing an external degree of freedom
the system can change from one phase into another. Let's consider the 
following example of our everyday life. If
one puts a pot with (liquid) water on the stove and switches the latter on,
the temperature of the water 
increases up to a (critical) temperature 
$T_c=100^\circ$ C, where the water becomes vapour (changes the phase).
\begin{figure}
\begin{center}
\includegraphics[height=8cm]{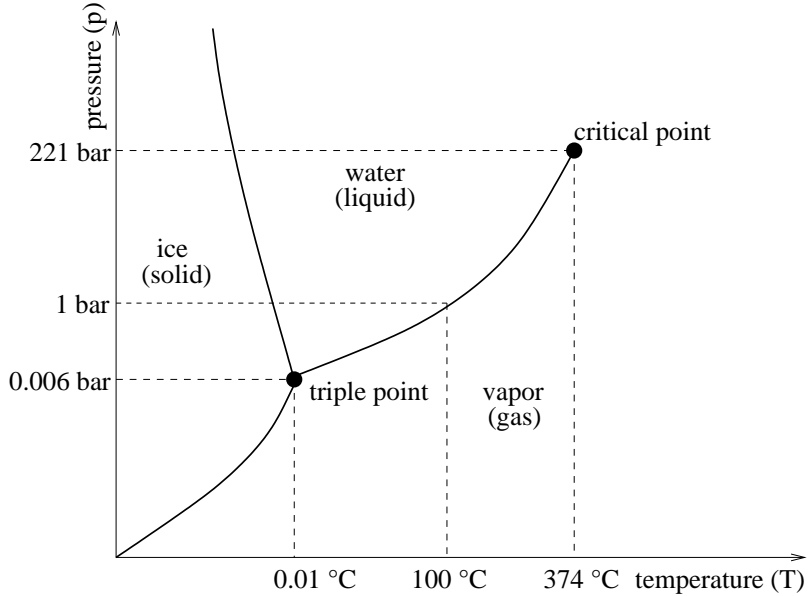}
\caption[A sketch of the water phase diagram 
in the pressure vs. temperature plane.]
{\it A sketch of the phase diagram of water
in the pressure vs. temperature plane.}
\label{phase_water}
\end{center}
\end{figure}
The {\it order} of a
phase transition is defined via the properties of the pressure, and
its nth derivatives with respect to the
temperature and/or chemical potential,
\be
\left.\frac{\partial^n p}{\partial T^n}\right|_{\mu={\rm const.}},\quad
\mbox{and}\quad
\left.\frac{\partial^n p}{\partial \mu^n}\right|_{T={\rm const.}}.
\ee
If the pressure is continuous, but there is a discontinuity in the 
first derivatives, we call this a first-order phase transition.
If the pressure and its first derivatives are continuous but there is
a discontinuity in its second derivatives, we call this a second-order
phase transition. In principle, one can define even higher-order
phase transitions, but in nature only the first- and second-
(maybe third- \cite{Gross:1980he,Li:2005yv,Li:2005kf,Lang:1980ws})
order phase transitions occur.
If there is no discontinuity, but a rapid
change in the thermodynamic quantities, we call this a crossover
transition. 

An appropriative way to visualise the possible phases and phase transitions 
of a system is a {\it phase diagram}. 
In Fig.~\ref{phase_water} a sketch of the phase diagram for water 
is shown in the pressure vs. temperature plane.
The lines represent first-order phase transitions (directly
{\it on} the line both phases coincide). The first-order phase transition
between the liquid and the gas phase becomes a second-order phase
transition at a {\it critical point} $(p,T)=(221\mbox{ bar},374\,^\circ\mbox{C})$,
and for even higher pressures $p>221\mbox{ bar}$ a crossover. Another
remarkable feature of this diagram is the {\it triple point} at
$(p,T)\approx(0.006\mbox{ bar},0.01\,^\circ\mbox{C})$ where {\it all}
three phases coincide. The transition of the example with the
pot of water is a first-order transition at 
$(p,T)=(1\mbox{ bar},100\,^\circ\mbox{C})$ in this diagram. In the next
sections, we will see that most of the discussed features
can also be found in the QCD phase diagram.

\subsection*{The quark mass dependence of the phase transition}

An important parameter for the order of the phase transition 
in QCD is given by the mass of the quarks [$m$ in Eq.~(\ref{qcd_lag})].
As mentioned in Sec.~\ref{secQCD} nonvanishing quark masses break
the symmetry of the QCD Lagrangian explicitly. The principle of
universality states that a phase
is characterised by its macroscopic behaviour, which is mostly driven
by the global symmetries. Therefore one way to find the order 
of the transition under this broken symmetry
is to construct a model which respects the same (broken) global symmetry as 
QCD, and calculates the order in this model instead of in QCD.
In this section I consider the case of vanishing chemical potential
$\mu=0$ and three (possible) quark flavours $N_f=3$. Therefore, the mass
parameters are
the strange quark mass $m_s$ and the degenerate masses of the up and
down quark $m_{ud}\equiv m_u=m_d$.

\begin{figure}
\begin{center}
\includegraphics[height=8cm]{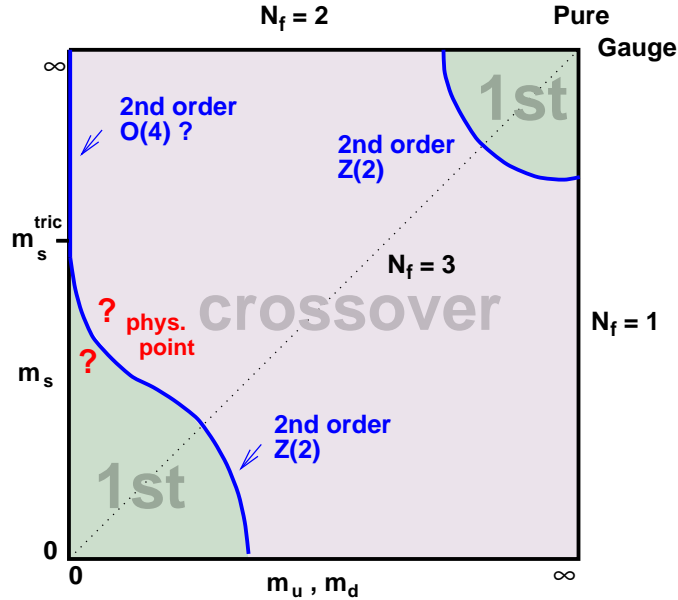}
\caption[The expected phase diagram  
in the plane of strange vs. degenerate up and down quark masses.]
{The expected phase diagram  
in the plane of strange vs. degenerate up and down quark masses
\cite{Laermann:2003cv}.}
\label{s_ud_plane}
\end{center}
\end{figure}
The expected phase diagram in the plane of these
two mass parameters is shown in Fig.~\ref{s_ud_plane}. This diagram
exhibits two areas of first-order phase transitions, near $m_s=m_{ud}=0$ and
$m_s=m_{ud}=\infty$, and a crossover regime between them, separated by
lines of second-order phase transitions.  The question marks
indicate the expected position for the physical quark masses in this
diagram. If one of the mass parameters is infinity (i.e., the 
quark is infinitely heavy) the corresponding quark flavours are thus 
removed from the spectrum of physical excitations. Therefore
one can distinguish between four different flavour cases.
First the pure gauge theory (upper right corner 
of Fig.~\ref{s_ud_plane}) where
$m_s=m_{ud}=\infty$, in this point all quarks are 
removed from the theory and only the gluons remains. Second, the one-flavour
case $N_f=1$ (right boundary of Fig.~\ref{s_ud_plane}) 
where $m_s<\infty$ but $m_{ud}=\infty$, on this line only the 
strange degree of freedom remains in the theory. Third, the two-flavour case
$N_f=2$ (upper boundary of Fig.~\ref{s_ud_plane}) where
$m_s=\infty$ but $m_{ud}<\infty$, here the strange quark
is removed from the theory and only the up- and down quarks remain.
And finally the full three-flavour case $N_f=3$ (the
rest of the plane in Fig.~\ref{s_ud_plane}) where 
$m_s<\infty$ and $m_{ud}<\infty$, 
in this area all quarks are included in the underlying theory.

One can use universality arguments to determine the order of the
phase transition. According to universality,
the order of the chiral transition for vanishing quark masses
in QCD is {\it identical} to that
in a theory with the same chiral symmetries as QCD.
This argument was employed by
Pisarski and Wilczek \cite{Pisarski:1984ms} who found that
for $N_f=2$ flavours of massless quarks (upper left corner of 
Fig.~\ref{s_ud_plane}), the transition
can be of second-order, if the $U(1)_A$ symmetry is explicitly
broken by instantons. It is driven first-order by fluctuations, if
the $U(1)_A$ symmetry is restored at the critical temperature $T_c$. 
For $N_f=3$ massless flavours (lower left corner of 
Fig.~\ref{s_ud_plane}), 
the transition is always first-order. In this case, the
term which breaks the $U(1)_A$ symmetry explicitly is a cubic
invariant, and consequently drives the transition first-order. In
the absence of explicit $U(1)_A$ symmetry breaking, the transition
is fluctuation-induced of first-order.

For nonzero quark masses, the chiral symmetry of QCD is explicitly 
broken. Nonzero quark masses act like a magnetic field in spin
systems, such that a second-order phase transition becomes a
crossover transition. When the quark masses increase,
a first-order phase transition may for a while remain of first-order, but it
will ultimately become a crossover transition, too. 
In order to decide whether this happens for a particular choice
of quark masses, universality arguments cannot be applied, and
one has to resort to numerical calculations. 

For the pure gauge theory (the theory without quarks, upper right corner
of Fig.~\ref{s_ud_plane}) 
the transition is of first-order 
\cite{Bacilieri:1988yq,Brown:1988qe,Ukawa:1989tn,Ukawa:1996ps,
Laermann:1998pf,Karsch:1999vy} as predicted by 
Svetitsky and Yaffe \cite{Svetitsky:1982gs}.
Nevertheless, recent results show that 
the transition is a weak first-order transition, thus it is more accurate 
to speak of a ``nearly second-order'' transition \cite{Kaczmarek:1999mm}.

\subsection*{The temperature vs. chemical potential plane}
\begin{figure}
\begin{center}
\includegraphics[height=8cm]{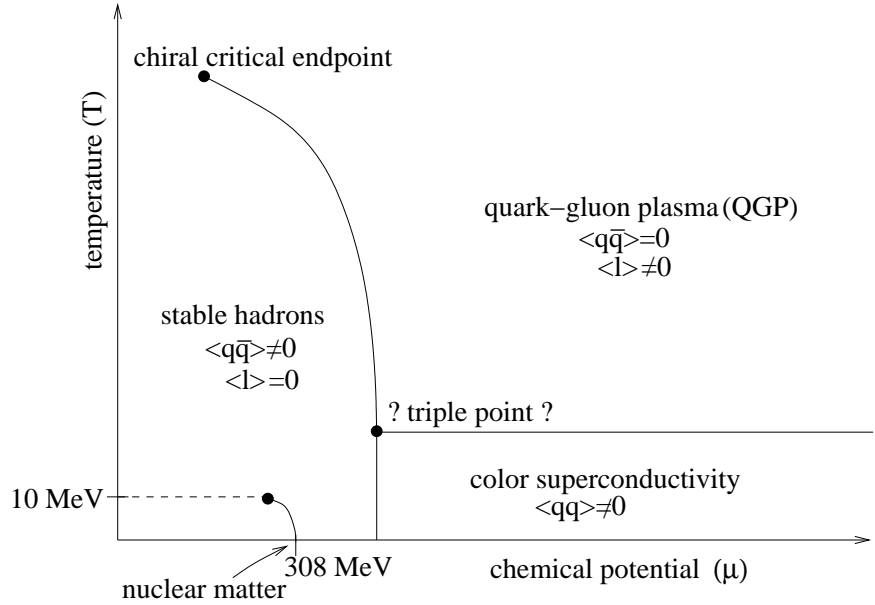}
\caption{A sketch of the QCD phase diagram in the
temperature vs. chemical potential plane.}
\label{T_mu_plane}
\end{center}
\end{figure}
In Fig.~\ref{T_mu_plane} a schematic plot of the QCD phase diagram in 
the temperature vs. chemical potential plane is shown. 
This diagram exhibits mainly three different
phases, the hadronic phase, the QGP, and the 
colour-superconductivity phase. In the hadronic phase, the energy is 
small enough for stable hadrons to exist, e.g., protons, neutrons, pions, etc.
As mentioned above, QCD is an asymptotically free theory, therefore for
large temperatures and/or chemical potential the quarks in the hadrons
become deconfined and behave more or less as free particles in
a plasma, the QGP. There is
probably a third phase at small temperatures and very large 
chemical potential, the phase of colour superconductivity, where
a quark-quark Cooper pair can be build, in analogy
to the well-known electron-electron Cooper pair of electromagnetic 
superconductivity. Note that this phase can be separated into many
other phases, e.g., the 2SC, CFL, CSL, or the polar phase, 
c.f. \cite{Bailin:1983bm,Alford:1997zt,Rapp:1997zu,Alford:2001dt,
Rischke:2003mt,Shovkovy:2004me,Rajagopal:2000wf}.

First, let's consider the transition between the hadronic phase and the QGP. 
The line between these two phases corresponds to 
a first-order phase transition which ends in the chiral critical endpoint of 
a second-order phase transition. The position of this chiral critical endpoint 
is still under investigation \cite{Fodor:2001pe,Fodor:2004nz,Stephanov:2004xs,
Gorenstein:2005rc,Barducci:1993bh,Asakawa:1989bq,Hatta:2002sj,
Antoniou:2002xq,Ejiri:2003dc}, the latest lattice results
from Fodor and Katz \cite{Fodor:2004nz}
found this point at $T=162\pm 2$ MeV and $\mu=120\pm 13$ MeV (indeed  
at nonzero chemical potential lattice QCD calculations are hampered by the 
fermionic sign problem and become increasingly unreliable with increasing
chemical potential \cite{deForcrand:2002ci}).
As indicated in this plot, there are two possibilities for an order 
parameter of this phase transition, the quark condensate 
$\langle q\bar q\rangle$ which links the transition to the restoration
of the chiral symmetry, and the expectation value of the Polyakov-loop
$\langle l\rangle$ which
links the transition to the restoration of the $Z(N_c)$ 
symmetry. In contrast to the quark condensate, the expectation value
of the Polyakov-loop is zero in the hadronic phase and nonzero in the
QGP, i.e., the $Z(N_c)$ symmetry is {\em broken} in the QGP and
restored in the hadronic phase. However, note that these order parameters
correspond to two very different Ans\"atze for the underlying symmetry.
The $Z(N_c)$ symmetry is exact for infinitely heavy quarks (in the
pure gauge theory), and the chiral $U(N_f)_r\times U(N_f)_\ell$
symmetry for quarks with vanishing mass. 
Although there are lattice QCD calculations which
indicate that these two Ans\"atze lead to the same phase
transition \cite{Karsch:2001cy}, the relationship of the 
symmetries is still an open question in modern high energy physics,
cf. e.g. \cite{Mocsy:2003qw,Sinclair:2003rm,Karsch:1998qj}.

The stable ground state of nuclear mater at vanishing temperature is at 
nonzero chemical potential $\mu_0=308$ MeV. For smaller
chemical potential, nuclear matter shows a similar behaviour to 
a gaseous phase and for larger the behaviour of a liquid. As the line
of first-order phase transitions in the phase diagram of water, this
line in the QCD phase diagram ends in an critical point of
second-order (at a temperature of $\sim 10$ MeV).

Note that the area of colour-superconductivity is still not well 
understood, especially for ``small'' chemical potential. Maybe there is
no direct phase transition between the hadronic phase and the colour
superconductivity phase without passing the QGP, i.e., maybe there is
no triple point between the three phases (indicated by the question
marks in Fig. 1.3). Or maybe there is no phase transition at all 
\cite{Alford:1999pa,Rajagopal:2000wf}.

\section{Effective models of QCD}
At temperatures of the order of $\sim \langle \bar{q} q \rangle^{1/3}$,
the thermal excitation energy is large enough to expect
the restoration of chiral symmetry.
At such energy scales, the QCD coupling constant is still large,
rendering perturbative calculations unreliable. Thus, one has to resort to
nonperturbative methods to study chiral symmetry restoration.
A first-principle approach is lattice QCD
\cite{Karsch:2001cy}. Lattice QCD calculations have determined the
temperature $T_c$ for chiral symmetry restoration to be of order 150 MeV
at zero quark chemical potential \cite{Laermann:2003cv}. These 
calculations, however, face several technical problems.

The first is that they become numerically very difficult for 
physically realistic, i.e., small, values of the up- and down-quark masses.
Although progress in this direction has been made \cite{Fodor:2004nz},
most studies use unphysically large values.
Another problem is that, at nonzero
chemical potential, lattice QCD calculations are hampered
by the fermion sign problem and
become increasingly unreliable for chemical potentials 
larger than (a factor $\pi$ times) the temperature \cite{deForcrand:2002ci}.

An alternative nonperturbative approach to study chiral symmetry restoration
is via chiral effective theories. These theories have
the same global $U(N_f)_r \times U(N_f)_\ell$ symmetry as QCD
but, since quark and gluons are integrated out, 
do not possess the local $SU(3)_c$ colour symmetry of QCD.
The effective low-energy degrees of freedom are 
the (pseudo-) Goldstone bosons of the QCD vacuum, i.e.,
the pseudoscalar mesons. However, in the chirally symmetric phase
these particles become degenerate
with their chiral partners, the scalar mesons. 
Therefore, an appropriate effective 
theory to study chiral symmetry restoration in QCD is the linear sigma model
\cite{Levy,Gell-Mann:1960np}
which treats both scalar and pseudoscalar degrees of freedom on the
same footing.
The advantage of chiral effective theories over lattice QCD
calculations is that their numerical treatment 
(within some many-body approximation scheme) is comparatively simple
and that there is no
problem to consider arbitrary quark chemical potential.

In the following, I introduce a reasonable effective model
which respects the chiral symmetry,
the linear $\sigma$-model with $O(N)$ symmetry,
and an effective model which is based on the $Z(N_c)$ symmetry,
the Polyakov-loop model.

\subsection*{The linear $\sigma$-model with $O(N)$ symmetry}\label{introon}

The Lagrangian of the $O(N)$ linear sigma model is given by 
\be
{\cal L}({\mbox{\boldmath$\phi$}})=
\frac 12\partial_\mu{\mbox{\boldmath$\phi$}}\cdot\partial^\mu
\mbox{\boldmath$\phi$}-U(\mbox{\boldmath$\phi$})
\ee
where $\mbox{\boldmath$\phi$}\equiv(\phi_1,\mbox{\boldmath$\pi$})$, 
with the first component $\phi_1$ corresponding to the scalar 
$\sigma$-meson and the other components 
$\mbox{\boldmath$\pi$}=(\phi_2,...,\phi_N)$
corresponding to the pseudoscalar pions. 
(However, note that the original $\sigma$-model introduced
by Gell-Mann and Levy \cite{Gell-Mann:1960np,ItzyksonZuber} 
incorporates 
the scalar and pseudoscalar mesons as well as a fermionic isodoublet
field $\psi$ of mass zero. In the linear $\sigma$ model discussed
in the following, the fermionic degrees of freedom are
neglected.) The function $U(\mbox{\boldmath$\phi$})$ is the
tree-level potential,
\be \label{U}
U(\mbox{\boldmath$\phi$})=\frac 12\mu^2\mbox{\boldmath$\phi$}\cdot
\mbox{\boldmath$\phi$}
+\frac \lambda N(\mbox{\boldmath$\phi$}\cdot\mbox{\boldmath$\phi$})^2
-H\phi_1\,\,,
\ee
where $\mu^2$ is the bare mass and
$\lambda>0$ the four-point coupling constant.
For $\mu^2 <0$ the $O(N)$ symmetry is spontaneously broken to $O(N-1)$,
leading to $N-1$ Goldstone bosons, the pions. The parameter $H$ breaks 
the symmetry explicitly, giving a mass to the pion.

The parameters can be expressed in terms of the vacuum
mass of the $\sigma$-meson, $m_\sigma$, the vacuum mass of the pion,
$m_\pi$, and the vacuum decay constant of the pion, $f_\pi$,
\be\label{def_para}
\mu^2=-\frac{m_\sigma^2-3\,m_\pi^2}{2}\; , \;\;\;\;
\lambda=\frac{N(m_\sigma^2-m_\pi^2)}{8f_\pi^2}\; , \;\;\;\;
H=m_\pi^2\,f_\pi\;.
\ee
With $m_\sigma=600$ MeV, $m_\pi=139.5$ MeV, and $f_\pi=92.4$ MeV 
\cite{PDBook} this
leads to $H=(121.60\,\mbox{MeV})^3,\lambda=19.943$, and
$\mu^2=-(388.34\,\mbox{MeV})^2$.

I assume translational invariance so that one may consider
the effective potential $V$ instead of the effective action
$\Gamma$, cf. Sec. 1.5.
For translationally invariant systems, these two
quantities are related via
\begin{equation}
\Gamma[\bar{\sigma},\bar{\mbox{\boldmath$\pi$}}, \bar{S},\bar{P}] 
= - \frac{\Omega_3}{T} \, 
V [\bar{\sigma},\bar{\mbox{\boldmath$\pi$}}, \bar{S},\bar{P}] \;,
\end{equation}
where $\Omega_3$ is the 3-volume of the system, and 
$\bar{\sigma}, \bar{\mbox{\boldmath$\pi$}}, \bar{S}, \bar{P}$ 
are the expectation values of the one- and two-point functions for 
the scalar and pseudoscalar fields in the presence of external sources
\cite{Cornwall:1974vz}.
I am interested in the case where these sources are zero, i.e.,
in the stationary points of $\Gamma$ or $V$.
Because the vacuum of QCD has even parity, the expectation values of the 
pseudoscalar fields are zero, $\bar{\mbox{\boldmath$\pi$}} = 0$,
and I shall simply omit the dependence of $V$ on $\bar{\mbox{\boldmath$\pi$}}$
in the following.

Then, the effective potential for the $O(N)$ model in the CJT formalism,
discussed in Sec. 1.5,  reads \cite{Roder:2003uz,Lenaghan:1999si}
\bea\label{CJT-potential}
V[\bar{\sigma},\bar{S},\bar{P}]
 & = & U(\bar{\sigma})  +  \half \int_Q \left[\, \ln \bar{S}^{-1}(Q) +
        S^{-1}(Q;\bar{\sigma})\, \bar{S}(Q)-1 \, \right] \nn \\
 & + &  \frac{N-1}{2} \int_Q \, \left[\,
        \ln \bar{P}^{-1}(Q) + P^{-1}(Q;\bar{\sigma})\, \bar{P}(Q)-
        1 \, \right]  \nn\\
 & + &  V_{2}[\bar{\sigma},\bar{S},\bar{P}]\,\, ,
\eea 
where $ U(\bar{\sigma})$
is the tree-level potential (\ref{U}), evaluated at
$\mbox{\boldmath$\phi$}=(\bar{\sigma},0,\ldots,0)$.
At tree-level
the one-point expectation value can be calculated analytically 
\bea\label{vev_tree}
\bar{\sigma}=f_\pi=\sqrt{-\frac{N\mu^2}{4\lambda}}\frac{2}{\sqrt{3}}
             \cos\frac{\theta}{3},\quad
\theta=\arccos\left[
\frac{HN}{8\lambda}\left(-\frac{12\lambda}{N\mu^2}\right)^{3/2}\right].
\eea
\begin{figure}
\begin{center}
\includegraphics[height=5cm]{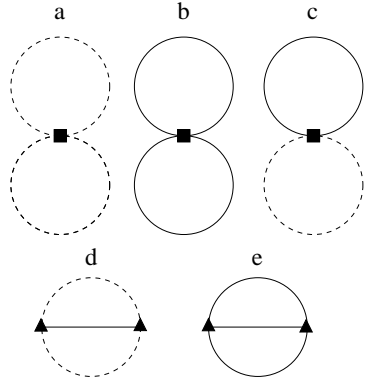}
\caption[The set of two-particle irreducible diagrams 
considered in the linear $\sigma$ model with $O(N)$ symmetry.]
{The set of two-particle irreducible diagrams 
considered in the linear $\sigma$ model with $O(N)$ symmetry.
The diagrams a,b, and c are the double-bubble
diagrams, and d and e are the sunset diagrams. A full line denotes the
full propagator for the $\sigma$-meson and a dashed line the full
propagator for the pion. The four-particle vertex 
$\sim \lambda$ is represented by a square and the three-particle
vertex $\sim \lambda\sigma$  by a triangle.}
\label{paper1}
\end{center}
\end{figure}
The quantities $S^{-1}$ and $P^{-1}$ are the inverse tree-level
propagators for scalar and pseudoscalar mesons,
\bse
\bea  \label{Dsigma} 
S^{-1}(K;\bar{\sigma}) & = & -K^{2} + m_\sigma^2(\bar{\sigma})\; , \\
P^{-1}(K;\bar{\sigma}) & = & -K^{2} + m_\pi^2 (\bar{\sigma})\; , 
\label{Dpi}
\eea
\ese
where the tree-level masses read
\bse
\bea\label{tree_level_masses}
m_\sigma^2(\bar{\sigma})  & = & \mu^{2} + 
        \frac{12\, \lambda}{N}\, \bar{\sigma}^{2}\; , \\
m_\pi^2(\bar{\sigma}) &  =  &   \mu^2 +
        \frac{4\, \lambda}{N}\, \bar{\sigma}^{2}\; . 
\eea
\ese
As explained in Sec. \ref{seccjt},
the functional $V_2$ in Eq.\ (\ref{CJT-potential}) is the
sum of all 2PI diagrams. The standard Hartree-Fock approximation
is defined by restricting this sum to 
the three double-bubble diagrams shown in Figs.~\ref{paper1} a, b, and c,
\bse\label{V2db}
\bea 
V_{2}^a[\bar{P}] &\equiv&
        (N+1)(N-1)\, \frac{\lambda}{N} \left[
        \int_Q \, \bar{P}(Q)\right]^{2}\,\,,\\
V_{2}^b[\bar{S}] &\equiv& 3\, \frac{\lambda}{N}
        \left[ \int_Q \, \bar{S}(Q) \right]^{2}\,\,, \\
V_{2}^c[\bar{S},\bar{P}]&\equiv& 2\,(N-1) \frac{\lambda}{N}
        \int_Q\, \bar{S}(Q) 
        \int_L\, \bar{P}(L) \,\,.
\eea
\ese 
Later, we will see that, in order to include the nonzero decay width
of the particles, one has to go beyond the Hartree-Fock approximation by
additionally including the sunset diagrams of
Figs.~\ref{paper1} d and e. These diagrams have an explicit dependence
on $\bar{\sigma}$, 
\bse\label{V2sunset}
\bea 
V_2^d[\bar\sigma,\bar{S},\bar{P}]&\equiv&\frac{1}{2} \, 2(N-1)
\left(\frac{4\lambda \bar\sigma}{N}\right)^2\int_L\int_Q
\bar{S}(L)\bar{P}(Q)\bar{P}(L+Q)\; ,\\
V_2^e[\bar\sigma,\bar{S}]&\equiv&\frac{1}{2}\, 3!
\left(\frac{4\lambda \bar\sigma}{N}\right)^2\int_L\int_Q
\bar{S}(L)\bar{S}(Q)\bar{S}(L+Q)\; . 
\eea 
\ese
The complete expression for $V_2$ is obtained by the sum of
the contributions in Eqs. (\ref{V2db}) and (\ref{V2sunset}),
\be
V_2=V_{2}^a+V_{2}^b+V_{2}^c+V_2^d+V_2^e\;.
\ee
The corresponding Dyson-Schwinger equations for the dressed
propagators and the condensate equations can be
derived by the minimisation of this effective potential, which
will be the topic of the following chapters. 

\subsection*{The Polyakov-loop model}

In this section, I discuss how to construct an effective model, based on
the $Z(N_c)$ symmetry, as discussed in Section 1.2. To find an
order parameter for the pure gauge theory with $N_c$ colours
let's first consider the Wilson line 
\cite{Polyakov:1978vu,'tHooft:1977hy,Yaffe:1982qf}
\be
{\bf L}\equiv{\cal P}\exp
\left[ig\int_0^{1/T}A_0({\bf x},\tau)d\tau\right],
\ee
where $A_0$ denotes the temporal component of the gauge field in the
fundamental representation, $g$ the gauge
coupling, and ${\cal P}$ denotes path ordering with respect
to imaginary time $\tau$.
The thermal Wilson line transforms under local $SU(N_c)$ transformations as:
\be
{\bf L(x)}\longrightarrow \Omega^\dagger\left({\bf x},\frac{1}{T}\right)
{\bf L(x)}\Omega^\dagger({\bf x},0).
\ee
An order parameter for the $Z(N_c)$ symmetry is
given by the trace over the Wilson line
\be \label{PloopDef}
\ell = \frac{1}{N_c} \; {\rm tr}\; {\cal P} 
\exp \left[ i g \int^{1/T}_0 A_0(\vec{x},\tau) \, 
d\tau \right]~.
\ee
Its expectation value, $\ell_0(T)$, vanishes when $T< T_c$, and
is nonzero above $T_c$.  Indeed, by asymptotic freedom,
$\ell_0 \rightarrow 1$ as $T\rightarrow \infty$.
The simplest guess for a potential for the Polyakov-loop is:
\begin{equation}
V(\ell) = - \frac{b_2}{2} |\ell|^2 
+ \frac{1}{4} \left( |\ell|^2 \right)^2, \; \quad\quad(N_c=2)~.
\label{e3}
\end{equation}
The  Polyakov-loop model~\cite{Pisarski:2000eq,Pisarski:2002ji,Dumitru:2000in,Dumitru:2001bf,Dumitru:2001xa,Scavenius:2001pa} is a mean-field
theory for $\ell$. In a mean-field analysis
all coupling constants are taken as constant with temperature, except
for the mass term, $\sim b_2 |\ell|^2$.  
About the transition, condensation
of $\ell$ is driven by changing the sign of the two-point coupling:
$b_2 > 0$ above $T_c$ [$b_2(T)\to 1$ for $T\to\infty$],
and $<0$ below $T_c$.

For two colours, (\ref{e3}) is a mean-field theory for a second-order
deconfining transition \cite{Engels:1998nv,Engels:1996kp,Engels:1995em,Engels:1994xj,Engels:1988ph}.
The $\ell$ field is real, and so the potential defines a mass:
$(m_\ell/T)^2 = (1/Z_s) \partial^2 V/\partial\ell^2$,
where $Z_s$ is the wave function renormalisation constant for
$\ell$~\cite{Wirstam:2001ka,Diakonov:2003yy}. The mass is measured
from the two-point function of Polyakov-loops in coordinate space, 
$\propto (1/r)\exp(- m_\ell\; r)$ as $r\rightarrow \infty$.

For three colours, $\ell$ is a complex valued field,
and a term cubic in $\ell$ appears in $V(\ell)$,
\begin{equation}
V(\ell) = - \frac{b_2}{2} |\ell|^2 
- \frac{b_3}{3}\frac{\ell^3 + \ell^*\!^3}{2}
+ \frac{1}{4} \left( |\ell|^2 \right)^2, \; \quad\quad(N_c=3)~.
\label{e3_N=3}
\end{equation}
At very high
temperature, the favoured vacuum is perturbative, with $\ell_0 \approx 1$,
times $Z(3)$ rotations. One then chooses $b_3>0$ so that in the $Z(3)$ model,
there is always one vacuum with a real, positive expectation value for
$\ell_0$.
This produces a first-order deconfining transition, where $\ell_0$ jumps
from $0$ at $T_c^-$ to $\ell_c = 2b_3/3$ at $T_c^+$~\cite{Dumitru:2000in,Dumitru:2001bf,Dumitru:2001xa};
$T_c$ is given by $b_2(T_c) = -2b_3^2/9$.
The $\ell$ field has two masses, from its 
real ($m_\ell$) and imaginary ($\widetilde{m}_\ell$) parts.
At $T_c^+$, $\sqrt{Z_s} m_\ell/T = \ell_c$.
The mass for the imaginary part of $\ell$ is
$\sqrt{Z_s} \widetilde{m}_\ell(T)/ T \propto \sqrt{b_3 \ell}$;
at $T_c^+$, $\widetilde{m}_\ell/m_\ell = 3$, twice the value expected
from a perturbative analysis of the loop-loop correlation function,
obtained by expanding $\ell$ from Eq.~(\ref{PloopDef}) to order
$A_0^3$~\cite{Dumitru:2002cf,Dumitru:2001vc}. 
This mass ratio receives corrections if
five-point and six-point couplings are included in the effective
Lagrangian~\cite{Dumitru:2002cf,Dumitru:2001vc} but those 
are not crucial for the following
discussion. One notes that, in principle, all of the above coupling
constants could be determined on the lattice.
The lattice regularization requires non-perturbative
renormalization of the Polyakov-loop in order to define the proper
continuum limit of $\ell$~\cite{Kaczmarek:2002mc,Petreczky:2004pz,Zantow:2003uh,Dumitru:2003hp}.

Within the above mean-field theory, dynamical quarks act like a
``background magnetic field'' which breaks the $Z(3)$ symmetry
explicitly, and a term linear in $\ell$
also appears in $V(\ell)$~\cite{Banks:1983me,Green:1983sd,
Meisinger:1995qr,Alexandrou:1998wv}:
\begin{equation}
V(\ell) = - b_1 \frac{\ell + \ell^*}{2} 
- \frac{b_2}{2} |\ell|^2 
- \frac{b_3}{3}\frac{\ell^3 + \ell^*\!^3}{2}
+ \frac{1}{4} \left( |\ell|^2 \right)^2, \; \quad\quad(N_c=3,~m_\pi<\infty)~.
\end{equation}
Hence, as $m_\pi$ decreases from infinity, $b_1(m_\pi)$ turns on. The
normalisation of $b_2(T)$ for $T\to\infty$ is such that
$\ell_0\to 1$, i.e., $b_2(T=\infty)=1-b_1-b_3$. 

In Chapter II, the Polyakov-loop model is used
to study the QCD phase transition.

\section{The Cornwall-Jackiw-Tomboulis formalism}\label{seccjt}
At nonzero temperature $T$, ordinary perturbation theory in terms of 
the coupling constant $g$ breaks down in the sense that one
can no longer order different contributions
according to powers of $g$ \cite{Dolan:1974qd}. 
This is because the new energy scale introduced by the 
temperature can conspire with the typical momentum scale $p$
of a process so that $gT/p$ is no longer of order $g$, but can be of
order $1$ \cite{Braaten:1989kk,Braaten:1989mz}.
Consequently, all terms of order $gT/p$ have to be taken into account
which requires the resummation of certain classes of diagrams.

A convenient resummation method is provided by the 
extension of the Cornwall-Jackiw-Tomboulis (CJT) formalism 
\cite{Cornwall:1974vz} to nonzero temperatures and chemical potentials.
The CJT formalism is equivalent to the $\Phi$-functional 
approach of Luttinger and
Ward \cite{Luttinger:1960ua} and Baym \cite{Baym:1962sx}.
It generalises the concept of the 
effective action $\Gamma[\bar{\phi}]$ for the expectation value
$\bar{\phi}$ of the one-point function in the presence of
external sources to that for the effective action
$\Gamma[\bar{\phi},\bar{G}]$ for one and 
two-point functions, $\bar{\phi}$ and $\bar{G}$,
in the presence of external sources.
(For an extension of this approach to three- and more-point functions, see
\cite{Norton:1974bm,Kleinert:1982ki,Carrington:2004sn,Berges:2004pu}.)

The starting point for the CJT formalism is the generating
functional for Green's functions \cite{Cornwall:1974vz}
\bea
Z[J,K]&=&e^{W[J,K]}=\int{\cal D}\phi 
\exp\left\{I[\phi]+\phi J+\frac{1}{2}\phi K\phi\right\},
\eea
where $I[\phi]\equiv\int_X {\cal L}$ is the classical action,
$W[J,K]$ the generating functional for connected Green's functions, 
and with the short-hand notations
\bse\bea
\phi J&\equiv&\int_X \phi(X)J(X),\\
\phi K\phi&\equiv&\int_{X,Y} \phi(X)K(X,Y)\phi(Y).
\eea\ese
The expectation values for the
one- and two-point functions in the presence
of the external sources, $\bar\phi$ and $\bar G$, are given by
\bse\bea
\bar\phi(X)&\equiv&\frac{\delta W[J,K]}{\delta J(X)},\\
\bar G(X,Y)&\equiv&\frac{\delta^2 W[J,K]}{\delta J(X)\delta J(Y)}.
\eea\ese
Note that the derivative with respect to $K$ is
\be
\frac{\delta W[J,K]}{\delta K(X,Y)}
=\frac{1}{2}[\bar G(X,Y)+\bar\phi(X)\bar\phi(Y)].
\ee
The aim is now to eliminate the sources in favour
of the one- and two point functions with a double Legendre transformation
\be
\Gamma[\bar\phi,\bar G]=W[J,K]-\bar\phi J
-\frac{1}{2}\bar\phi K \bar\phi-\frac{1}{2}\bar GK,
\ee
where $\bar G K\equiv \int_{XY}\bar G(X,Y)K(X,Y)$. Minimisation
of this functional gives the expectation values of the one-
and two-point function in the absence of the external sources,
$\varphi$ and $\cal G$,
\bse\label{statcond1}\bea
\left. \frac{\delta \Gamma[\bar\phi,\bar G]}
{\delta\bar\phi(X)}\right| _{\bar\phi=\varphi
,\bar G={\cal G}}&=&0,\\
\left.\frac{\delta \Gamma[\bar\phi,\bar G]}
{\delta \bar G(X,Y)}\right|_{\bar\phi=\varphi,
\bar G={\cal G}}&=&0.
\eea\ese
The last equation corresponds to the Dyson-Schwinger equation for the
full dressed propagator. In the CJT formalism, one uses the following
effective action \cite{Cornwall:1974vz}
\be\label{effactcjt}
\Gamma[\bar\phi,\bar G]
=I[\bar\phi]-\frac{1}{2}\mbox{Tr ln}(\bar G^{-1})-\frac{1}{2}
\mbox{Tr}(G^{-1}\bar G-1)+\Gamma_2[\bar\phi,\bar G],
\ee
where,
\be
G^{-1}(X,Y;\bar\phi)
=-\left.\frac{\delta^2I[\phi]}{\delta\phi(X)\delta\phi(Y)}
\right|_{\phi=\bar\phi},
\ee
is the tree-level propagator, and $\Gamma_2[\bar{\phi},\bar{G}]$
the sum of all two-particle irreducible
(2PI) vacuum diagrams with internal lines given by $\bar{G}$. I'm
interested in the case where the fields are translationally invariant,
so that one may consider the effective potential $V$ instead of the effective
action. For translationally invariant systems, these two
quantities are related via
$\Gamma[\bar{\phi},\bar{G}] = - \Omega_3/T \, V [\bar{\phi},\bar{G}] \;,$
where $\Omega_3$ is the 3-volume of the system, thus
\be\label{effpotcjt}
V[\bar{\phi},\bar G] = 
U(\bar{\phi}) + 
\frac{1}{2} \int_k \ln \bar G^{-1}(k) + 
\frac{1}{2} \int_k [G^{-1}(k;\bar{\phi}) \bar G(k) - 1] 
+ V_2[\bar{\phi},\bar G],   
\ee
where
\begin{eqnarray}
V_2[\bar{\phi},\bar G] =-T\Gamma_2[\bar{\phi},\bar G]/\Omega_3.
\end{eqnarray}
The stationary conditions in Eqs.~(\ref{statcond1}),
can now be expressed in terms of the effective potential
\bse\bea\label{statcond}
\left. \frac{\delta V[\bar\phi,\bar G]}
{\delta\bar\phi}\right|_{\bar\phi={\varphi},\bar G=
{\cal G}}&=&0,\\
\left. \frac{\delta V[\bar\phi,\bar G]}
{\delta \bar G(K)}\right|_{\bar\phi={\varphi},\bar G=
{\cal G}}&=&0.
\eea\ese
Together with Eq.~(\ref{effpotcjt}), the last stationary condition
gives the Dyson-Schwinger equation
\be
{\cal G}^{-1}(k)=G^{-1}(k;\varphi)+\Pi(k),
\ee
where
\be\label{Pi}
\Pi(k)\equiv2\left.\frac{\delta V_2[\bar\phi,\bar G]}
{\delta \bar G(k)}\right|_{\bar\phi={\varphi},
\bar G={\cal G}},
\ee
is the self-energy. Note that the thermodynamic pressure is (up to
a sign) identical to the effective potential \cite{rivers}
\be
p=-V[\varphi,{\cal G}].
\ee

\section{The aim of this work}
The main focus of my thesis is QCD and its phase transitions as
discussed in the introduction. 

In chapter II, I present a study I've done with Adrian Dumitru and 
J\"org Ruppert \cite{Dumitru:2003cf} about the phase transition
temperature of QCD, and its dependence on the 
quark (or pion) mass. In the first part of this chapter, 
I use the linear $\sigma$-model with $O(N)$ symmetry (cf. Sec. 1.4),
within the CJT formalism (cf. Sec. 1.5), which
links the transition to chiral symmetry restoration. The parameters
of this model depend on the vacuum mass of the pion and
$\sigma$-meson and the vacuum value of the decay constant
of the pion. From lattice QCD calculations we know the
dependence of these values on the
quark mass, and hence one can determine the dependence of
the chiral phase transition (especially the transition temperature)
on the quark mass. In the second part of this chapter, I
use the Polyakov-loop model (cf. Sec. 1.4), which links the
transition to the restoration of $Z(N_c)$ symmetry.
The aim of this part is to find how ``strong'' one has to
break the $Z(3)$ symmetry in order to reproduce the transition 
temperature (given by lattice QCD calculations). 

In chapter III, I present a study I've done with my
Ph.D. advisor Dirk Rischke and J\"org Ruppert \cite{Roder:2005vt} 
about the improvement of the standard Hartree-Fock approximation by
including nonzero decay width effects.
In the standard Hartree-Fock approximation one treats 
all particles as stable quasiparticles,
which means that the spectral densities of these particles are
just delta-functions with zero decay width, $\Gamma=0$.
Obviously, this is a reasonable approximation for all particles with
small decay width, i.e., the pion with 
a decay with of $\Gamma_\pi\sim 8$ eV
in vacuum, but not for broad particles,
like the $\sigma$-meson with a vacuum decay width of 
$\Gamma_\sigma\sim (600-1000)$ MeV. The decay width is proportional
to the imaginary part of the self-energy of the particle 
$\Gamma\sim \Im\Pi$. In the standard Hartree-Fock approximation 
only real-valued tadpole diagrams (cf. Fig. 3.1 b, c, and 3.2 a, c)
are taken into account. I improve this scheme by taking additionally into
account the cut sunset diagrams (cf. Fig. 3.1 d, e, and 3.2 d),
which have a real {\it and} an imaginary part.
Beside the imaginary part, another major difference between these two
types of diagrams is that the cut sunset diagram is a functional
of an external four momentum vector $K=(\omega,{\bf k})$
(where $\omega$ is the energy and ${\bf k}$ the momentum of the particle).
Hence, the Dyson-Schwinger equations
for the full propagators and the condensate equations become
integral equations one has to solve on an energy-momentum grid.
In this chapter, my focus is to study the influence of the inclusion
of nonzero decay width effects,
therefore, as a first approximation, I simply neglect
the real part arising from these cut sunset diagrams.

Finally, in chapter IV, I present a work \cite{Roder:2005qy} I've done to
study the effects
of the (in chapter III neglected) 4-momentum dependent
real parts of the cut sunset
diagrams in the linear $\sigma$-model with $O(N)$ symmetry.
I do this not in the full Hartree-Fock but
in the Hartree approximation. The difference between these
both approximation is that all terms of order $\sim 1/N$
are neglected on the level of the condensate and Dyson-Schwinger
equations in the Hartree approximation.
This leads to a vanishing imaginary part of the pion 
self-energy, hence to a vanishing pion decay width. 
The decay width of the $\sigma$-meson remains nonzero. Besides
the effects of the real part of the cut sunset diagram, I
study in this chapter the influence of the ``choice'' of the vacuum
mass of the $\sigma$-meson. As mentioned above, the parameters 
depend on the vacuum mass of the $\sigma$-meson mass, which is 
not well defined, $m_\sigma=(400-1200)$ MeV, because of the very 
large (vacuum) decay width of the $\sigma$-meson. Therefore, I compare 
the results for $m_\sigma=400$, 600, and 800 MeV, in the case
of explicit chiral symmetry breaking $m_\pi\neq 0$ and the chiral
limit $m_\pi=0$.

\vfill

\newpage
\section*{Conventions}
I denote 4-vectors by capital letters, $X \equiv(x_0,{\bf x})$, with
${\bf x}$ being a 3-vector of modulus
$x\equiv |{\bf x}|$. 
The imaginary-time formalism is used to compute quantities at
nonzero temperature. Integrals over 4-momentum $K = (\omega, {\bf k})$ are
denoted as
\be
\int_K \, f(K) \equiv T \sum_{n=-\infty}^{\infty}
                       \int \frac{d^{3} k}{(2\pi)^{3}} \,
         f(-i\omega_n,{\bf k}) \;,
\ee
where $T$ is the temperature and
$\omega_n= 2 \pi n T$, $n= 0, \pm 1, \pm 2, \ldots$ 
are the bosonic Matsubara frequencies.
My units are $\hbar=c=k_{B}=1$.  The metric tensor is $g^{\mu \nu}
= {\rm diag}(+,-,-,-)$.  

%**********************************************************************
%*********Chapter II***************************************************
\chapter{The quark mass dependence of the transition temperature}
\section{Motivation}

Lattice QCD calculations at finite temperature and
with dynamical fermions are presently performed
for quark masses exceeding their physical values; for a recent review
see~\cite{Laermann:2003cv}. To date,
pion masses as low as $\approx400$~MeV are feasible~\cite{Karsch:2000kv},
about three times the physical pion mass. When comparing
effective theories to first-principles numerical data obtained on the
lattice it is therefore important to fix the parameters (coupling
constants, vacuum expectation values and so on) such as to match the
values of physical observables, e.g.\ of $m_\pi$, to those of the
lattice calculations. For example, the QCD equation of state in the
confined phase appears to be described reasonably well by that of a
hadron resonance gas model, {\em after extrapolating} the physical
spectrum of hadrons and resonances to that from the
lattice~\cite{Karsch:2003zq,Karsch:2000kv}.
Thus, lattice data on the dependence of various observables on the
quark (or pion) mass constrain effective theories for the QCD phase
transition at finite temperature and could provide relevant
information on the driving degrees of freedom.

In the following, I analyse the dependence of the {\em chiral} symmetry
restoration temperature on the vacuum mass of the pion using a linear
sigma model in section 2.2.
The linear sigma model provides an effective Lagrangian
approach to low-energy QCD near the
 chiral limit~\cite{Pisarski:1983ms,Rajagopal:1992qz}. It
incorporates the global flavour symmetry, assuming that ``colour'' can
be integrated out. For example, it allows one to discuss the
order of the $N_f=2+1$ chiral phase transition as a function of the quark
masses~\cite{Pisarski:1983ms,Rajagopal:1992qz,Lenaghan:2000ey,Metzger:1993cu,Meyer-Ortmanns:1992pj,Goldberg:1983ju,Gavin:1993yk}.

Instead of working up from zero quark mass, one could start with
the quark masses taken to infinity, that is, with a pure gauge theory.
Then, one can discuss the {\em deconfinement} transition at finite
temperature within an effective Lagrangian for the Polyakov
loop with global $Z(N_c)$
symmetry~\cite{Yaffe:1982qf,Pisarski:2002ji,Dumitru:2000in,Dumitru:2001bf,
Dumitru:2001xa,Scavenius:2001pa,Ogilvie:1999if,Meisinger:2001cq,Mocsy:2003qw,
Fukushima:2003fw} 
($N_c$ is the
number of colours). For finite pion mass, the
symmetry is broken explicitly, and the phase transition (or 
crossover) temperature is shifted, relative to the pure gauge theory where
pions are infinitely heavy. In section 2.3., I determine
the endpoint of the line of first-order transitions for three colours,
and extract the magnitude of the explicit $Z(3)$ breaking from
lattice data on $\Delta T_c$.

\section*{Pion Mass and Decay Constant in Vacuum}

As discussed in Sec. \ref{secQCD},
the Lagrangian of QCD with the quark mass matrices set to zero
is invariant under independent rotations of the $N_f$ right- and
left-handed quark fields. It exhibits a global $SU(N_f)_r \times
SU(N_f)_\ell$ symmetry, leading to $2(N_f^2-1)$ conserved currents. Those
are $N_f^2-1$ vector currents, $V_i^\mu=\bar{\psi}\gamma^\mu \lambda_i
\psi\;/2$, and $N_f^2-1$ axial currents, $A_i^\mu = \bar{\psi}\gamma^\mu
\gamma_5 \lambda_i \psi\;/2$, with $\lambda_i$ the generators of
$SU(N_f)$, normalised according to ${\rm tr}~\lambda_i
\lambda_j=2\delta_{ij}$. The $SU(N_f)_V$ subgroup of vector
transformations is preserved in the vacuum~\cite{Vafa:1984tf}, while the
$SU(N_f)_A$ is broken spontaneously by a non-vanishing chiral
condensate $\langle\bar{q}_R q_L\rangle\neq0$, leading to
non-conservation of the axial currents.

In reality, of course, even $SU(N_f)_V$ is broken explicitly by the
non-vanishing quark mass matrix. Nevertheless, since $m_u\simeq m_d$,
the $SU(2)_V$ symmetry is almost exact in QCD. The
small explicit breaking of $SU(2)_A$ is responsible for the
non-vanishing pion mass, as given by the Gell-Mann-Oakes-Renner relation
\be \label{GOR}
m_\pi^2 = \frac{1}{f_\pi^2} \; m_q \; \langle\bar{q} q\rangle~.
\ee
I neglect isospin breaking effects here, and so assume that $m_u=m_d
\equiv m_q$.
$\langle\bar{q} q\rangle$ denotes the sum of the vacuum expectation
values of the operators $\bar{u}_R u_L$ and $\bar{d}_R d_L$, and their
complex conjugates. The proportionality constant $f_\pi$ is the pion
decay constant. It should be noted that~(\ref{GOR}) is only
valid at tree level, and that loop effects induce an implicit
dependence of both $f_\pi$ and $\langle\bar{q} q\rangle$ on $m_q$.
For small $m_q$, this dependence can be computed in chiral
perturbation theory~\cite{Gerber:1988tt}. For example, at 
next-to-leading order,
\bse\bea
m_\pi^2 &=& M^2 \left[1- \frac{1}{2} \left(\frac{M}{4\pi F}\right)^2\log
\frac{\Lambda_3^2}{M^2}\right]~, \label{cPT_mpi}\\
f_\pi &=& F \left[ 1+ \left(\frac{M}{4\pi F}\right)^2\log
\frac{\Lambda_4^2}{M^2}\right]~, \label{cPT_fpi}
\eea\ese
where $M$ and $F$ are the couplings of the effective theory
(equivalent to $m_q$ and $\langle\bar{q} q\rangle$), and
$\Lambda_3$ and $\Lambda_4$ are two renormalization-group invariant
scales. These relations link the behaviour of $f_\pi$ to that of
$m_\pi$, the mass of a physical state. (In what follows, I use $m_\pi$
to vary the strength of explicit symmetry breaking rather than using directly
the scale-dependent quark masses). 

More accurate results than Eqs.~(\ref{cPT_mpi},\ref{cPT_fpi}) can
perhaps be obtained by computing quark propagators for various quark
masses on the lattice. Ref.~\cite{Chiu:2003iw} analysed the
propagators for gauge-field
configurations generated with the standard Wilson gauge action
(``quenched QCD''), using overlap fermions with exact chiral symmetry.
They obtained a parametrisation of both $m_\pi$
and $f_\pi$ in terms of the mass $m_q$ of $u$ and $d$ quarks (see
section~2 in~\cite{Chiu:2003iw}) which allows to express $f_\pi$
as a function of $m_\pi$.
Their data covers an interval of
0.4~GeV$\lton m_\pi\lton1$~GeV, and 0.15~GeV$\lton \sqrt{2}
f_\pi\lton0.22$~GeV.

\section{Results: The linear $\sigma$-model with $O(N)$ symmetry}
In this section, I discuss chiral symmetry restoration at nonzero
temperature, and in particular the dependence of the symmetry
restoration temperature on the pion mass.
For simplicity, I restrict myselfe here to the two-flavour case.
My emphasis is not on the order of the transition as the
strange quark mass is varied but rather on how the temperature at
which the transition occurs (be it either a true phase transition or
just a crossover) depends on the pion mass. 
Such dependence arises from two effects. First, of course, due to
explicit symmetry breaking occurring when $m_\pi>0$. Second, due to
the ``indirect'' dependence of spontaneous symmetry breaking, i.e., of
the condensate $\langle\bar{q}q\rangle$ resp.\ $f_\pi$, on the pion
mass (through pion loops, see previous section).

In the following I use the counter-term renormalization scheme as discussed in
section V.B in \cite{Lenaghan:1999si}. In this scheme
a mass renormalization scale $\mu$ is
introduced and the couplings then depend on that scale. However, choosing
\be
\mu_{ren}^2
=\exp\left[\frac{m_\sigma^2(\ln\,m_\sigma^2-1)-m_\pi^2(\ln\,m_\pi^2-1)}
{m_\sigma^2-m_\pi^2}\right]~,
\ee
the four-point coupling $\lambda(\mu)=\lambda_{\rm tree}$ retains its
tree-level (classical) value~\cite{Lenaghan:1999si}. In other words, this
renormalization prescription evolves the renormalization scale $\mu_{ren}$
in such a way as to keep $\lambda$ constant.

\begin{figure}
\begin{center}
\includegraphics[height=6.3cm]{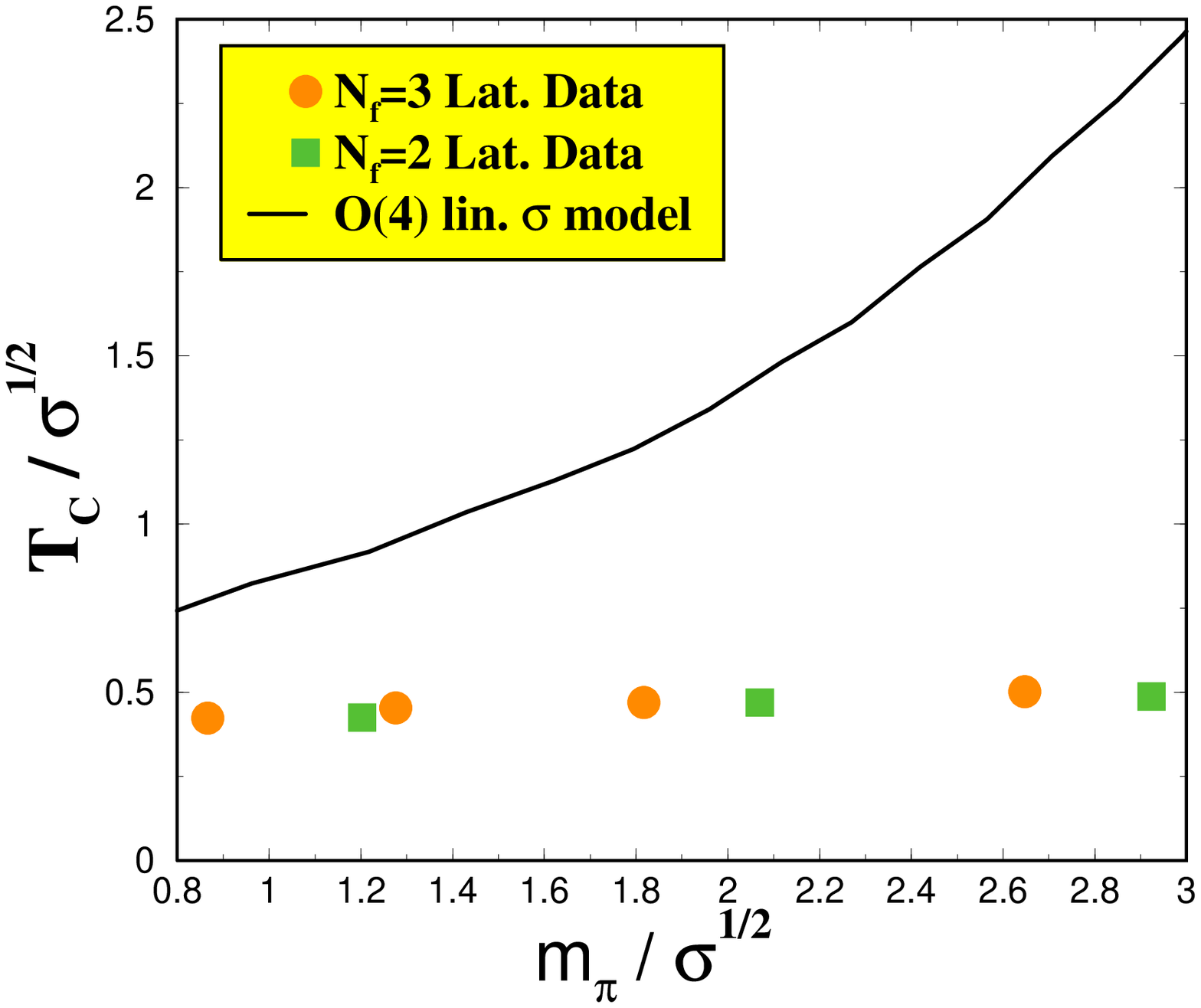}
\includegraphics[height=6.3cm]{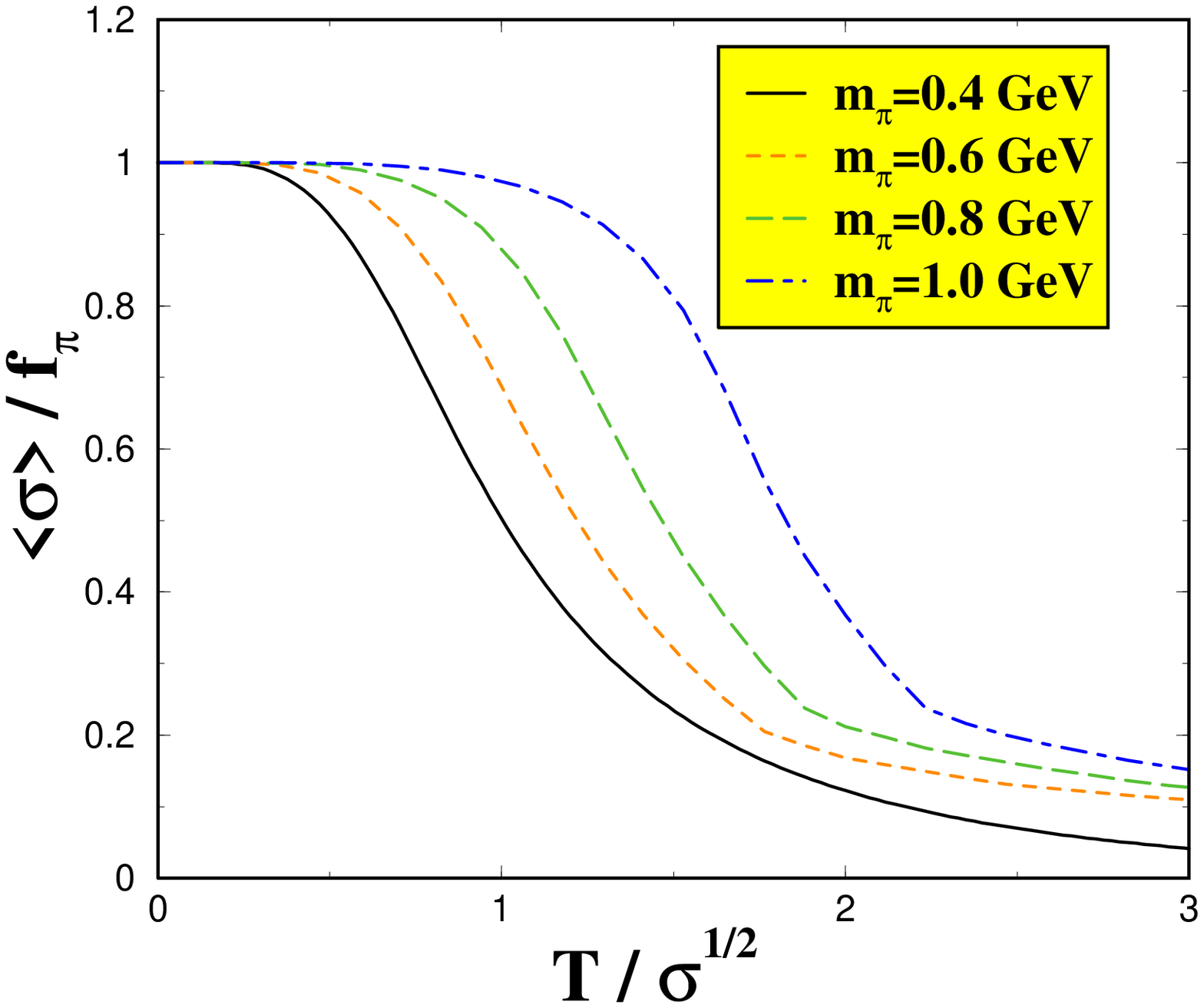}
\caption[The crossover temperature and the scalar condensate.]
{Left: The crossover temperature $T_c$ as a function of the
(vacuum) pion mass as obtained from the linear sigma model
with $O(4)$ symmetry in comparison to lattice
data~\protect\cite{Karsch:2000kv} for two and three flavours.
The scale for both $T_c$ and $m_\pi$ is set by the zero-temperature
string tension in the pure gauge theory, $\surd\sigma\simeq0.425$~GeV.
Right: The scalar condensate,
$\langle\sigma\rangle$, as a function of temperature for
various pion masses.}
\label{Fig_sigma}
\end{center}
\end{figure}

Explicitly, this leads to the following expressions for the
couplings~\cite{Lenaghan:1999si}:
\be
\lambda = \frac{1}{2} \frac{m_\sigma^2-m_\pi^2}{f_\pi^2},\quad
 H=f_\pi\left(m_\sigma^2-2\lambda f_\pi^2 \right),\quad
m^2 = -\frac{1}{2} \left( m_\sigma^2-3 m_\pi^2\right)
-6\lambda Q_\mu(m_\pi),
\ee
where
\be
Q_\mu(M)\equiv \frac{1}{(4\pi)^2}\left[M^2\ln\frac{M^2}{\mu_{ren}^2}
-M^2+\mu_{ren}^2\right].
\ee
These equations determine the couplings in vacuum in terms of $m_\pi$,
$f_\pi$ and $m_\sigma$. The dependence of $f_\pi$ and $m_\pi$ on
the quark mass is
taken from the data of ref.~\cite{Chiu:2003iw}.
Roughly, for $m_\pi:0.4~{\rm GeV}\to 1$~GeV, $f_\pi$
increases by about 50~\%, leading to an increase of the explicit
symmetry breaking term $H$ by a factor of 10.
I also require the dependence of $m_\sigma$
on $m_\pi$, which I take from a computation with standard
Wilson fermions~\cite{Kunihiro:2003yj}. Those authors
find that $m_\sigma$ is essentially a linear function of
$m_\pi^2$. I checked how my results in Fig.~\ref{Fig_sigma} depend
on this assumption by using, alternatively, a linear dependence
$m_\sigma= m_\pi+{\rm const.}$, with
$m_\sigma=0.6$~GeV for $m_\pi=0.14$~GeV. I found essentially the same
dependence of $T_c$ on $m_\pi$.

At nonzero temperature, one uses the 
effective potential for composite operators, as discussed
in Sec.~\ref{seccjt}, to determine
the masses and the scalar condensate in the Hartree-Fock approximation.
This approximation is defined by only taking into account the
double-bubble diagrams shown in Figs.~\ref{paper1} d and e in the effective
potential $V$. The gap equations for
the condensate and the masses are given by 
minimisation of this potential.
The expectation values of the one- and two-point functions in 
the absence of external sources, $\sigma$ and
${\cal S}$, ${\cal P}$,
are determined from the stationary points of $V$,
\bea
\left.\frac{\delta V}{\delta \bar\sigma}\right
|_{\bar\sigma=\sigma,\bar{S}={\cal S},\bar{P}={\cal P}}=0,
\left.\frac{\delta V}{\delta \bar{S}}\right
|_{\bar\sigma=\sigma,\bar{S}={\cal S},\bar{P}={\cal P}}=0,
\left.\frac{\delta V}{\delta \bar{P}}\right
|_{\bar\sigma=\sigma,\bar{S}={\cal S},\bar{P}={\cal P}}=0,
\eea
leading to the gap equations for the chiral condensate $\sigma$, and the 
effective masses of the $\sigma$-meson and the pions,
$M_\sigma$ and $M_\pi$,
\bse\bea
 H &=& \sigma \left[M_\sigma^2-2\lambda \sigma^2 \right], \label{eq9}\\
 M_\sigma^2 &=& m^2+3 \lambda \left\{\sigma^2 + 
[Q_T(M_\sigma)+Q_\mu(M_\sigma)]+
[Q_T(M_\pi)+Q_\mu(M_\pi)]\right\}, \hspace{.5cm} \label{eq10}\\
 M_\pi^2 &=& m^2+ \lambda \left\{\sigma^2 + 
[Q_T(M_\sigma)+Q_\mu(M_\sigma)]
+5[Q_T(M_\pi)+Q_\mu(M_\pi)]\right\}, \label{eq11}
\eea\ese
where the nonzero-temperature contribution of the tadpole diagram
is given by
\be
 Q_T(m) \equiv 
\frac{1}{2\pi^2}\int_0^\infty d q[\omega (q)]^{-1}
f[\omega(q)],
\ee
where $f(\omega)\equiv 1/[\exp(\omega/T)-1]$ is the Bose-Einstein
distribution function and $\omega(q)\equiv\sqrt{q^2+m^2}$ the
quasiparticle energy.
The self-consistent solution of the above
gap equations for a given vacuum pion mass determines
the temperature dependence of the scalar condensate as the order
parameter of chiral symmetry restoration. For explicitly broken chiral
symmetry, $H>0$, the transition in this approach is a crossover. I
define the crossover temperature $T_c$ by the peak of
$\partial  \sigma  / \partial T$.
The dependence of $T_c$ on $m_\pi$ is depicted in Fig.~\ref{Fig_sigma}
(left), where I have also shown lattice results obtained with two and
three degenerate quark flavours, respectively~\cite{Karsch:2000kv} (the
$N_f=2$ data with standard action, the $N_f=3$ data with improved
p4-action). Driven by the increase of both $f_\pi$ and $H$ with
$m_\pi$, the linear sigma model predicts a rather rapid rise of $T_c$
with the pion mass, as compared to the data which is nearly flat on the
scale of the figure. While lattice data indicate a rather weak
dependence of $T_c$ on the quark mass (see also ref.~\cite{Bernard:1996cs}),
models with spontaneous symmetry breaking in the vacuum
naturally predict a rather steep rise of $T_c$
with the VEV $\sigma_{\rm vac}=f_\pi$, which itself
increases with the quark (or pion) mass. My findings here are in
qualitative agreement with those from ref.~\cite{Berges:1997eu} who employed
nonperturbative flow equations to compute the effective potential
for two-flavour QCD within the linear sigma model. They also find a
steeper slope of $T_c(m_\pi)$ than indicated by the lattice, even though their
analysis appears to predict a somewhat weaker increase of $f_\pi$
with $m_\pi$ than the data of~\cite{Chiu:2003iw}, which I employ here.

Fig.~\ref{Fig_sigma} also shows the temperature dependence of the
$\sigma$ condensate (right). With $m_\sigma$ a linear function of
$m_\pi^2$~\cite{Kunihiro:2003yj}, the width of the crossover is
approximately independent of 
the pion mass for 0.4~GeV$\lton m_\pi\lton1$~GeV, while I found 
considerable broadening when $m_\sigma$ is linear in $m_\pi$ (not shown).
The chiral susceptibility $\partial \sigma/\partial T$ at
its maximum is $\approx 0.25$, i.e., the crossover is in fact
quite broad for the range of $m_\pi$ considered. Since this is
at variance with lattice data on QCD thermodynamics (pressure
and energy density as functions of temperature, see e.g.\ the review
in~\cite{Laermann:2003cv}), one might argue that the crossover is in
fact not driven by the order parameter field but by heavier degrees of
freedom~\cite{Karsch:2003zq,Karsch:2000kv,Gerber:1988tt}. Such degrees 
of freedom could reduce the pion-mass dependence of
the transition substantially: using
three-loop chiral perturbation theory (i.e., the non-linear model),
Gerber and Leutwyler find~\cite{Gerber:1988tt} that
$T_c$ increases rapidly from
$\approx 190$~MeV in the chiral limit
(using their set of couplings) to $\approx 240$~MeV for
physical pion mass. However, when heavy states are included (in the
dilute gas approximation), then $T_c$ increases less rapidly, from
$\approx 170$~MeV in the chiral limit to $\approx 190$~MeV for
physical pion mass. 

\vfill
\section{Results: The Polyakov-loop model}
In this section I consider the Polyakov-loop model as introduced in
Sec. 1.4. As discussed, the effective potential for three
colours with explicit breaking of the $Z(3)$ symmetry is given by, 
\begin{equation}
V(\ell) = - b_1 \frac{\ell + \ell^*}{2} 
- \frac{b_2}{2} |\ell|^2 
- \frac{b_3}{3}\frac{\ell^3 + \ell^*\!^3}{2}
+ \frac{1}{4} \left( |\ell|^2 \right)^2,
 \; \quad\quad(N_c=3,~m_\pi<\infty)~.
\label{e3_Q}
\end{equation}

First, I consider the case where $b_1$ is very small, and take
the term linear in $\ell$ as a perturbation; then
the weakly first-order phase transition of the
pure gauge theory persists (in what follows, the critical temperature
in the pure gauge theory with $b_1=0$ will be denoted $T_c^*$). 
The critical temperature is determined from
\be \label{b2Tc}
b_2(T_c) = -\frac{2}{9} b_3^2 \left(1+\frac{27}{2}\frac{b_1}{b_3^3}
\right) + {\cal O}(b_1^2)~.
\ee
The order parameter jumps at $T_c$,
from 
\be \label{ellTc-}
\ell_0(T_c^-) = \frac{9}{2} \frac{b_1}{b_3^2} + {\cal O}(b_1^2)~,
\ee
to
\be \label{ellTc+}
\ell_0(T_c^+) = \frac{2}{3}b_3 - \frac{9}{2} \frac{b_1}{b_3^2}
                   + {\cal O}(b_1^2)~. 
\ee
Note that numerically $\ell_0(T_c^-)$ could be much larger than $b_1$
if the phase transition in the pure gauge theory is weak and so the
correlation length $\xi=1/m_\ell$ near $T_c$ is large (i.e., if
$b_3$ is small), as indeed appears to be the case for $N_c=3$
colours~\cite{Kaczmarek:1999mm}. In other words, it could be that on the lattice
$\ell$ quickly develops a non-vanishing expectation value at $T_c^-$
already for rather large quark (or pion) masses, but this does not
automatically imply a large explicit symmetry breaking (see also
Fig.~\ref{Fig_ell} below).

From Eq.~(\ref{b2Tc}) one can estimate the shift of $T_c$ induced by
letting $m_\pi<\infty$.
Writing the argument of $b_2$ in that equation as
$T_c^* + \Delta T_c$ and expanding to first-order in $\Delta T_c$ one
obtains
\bea \label{DelTc}
\frac{\Delta T_c}{T_c^*} &=& -3\; \frac{b_1}{b_3}~\left(
T\frac{\partial b_2}{\partial T}\right)^{-1}_{T=T_c^*} + {\cal O}(b_1^2)\nn\\
&=&-\frac{2}{3}~\ell_0(T_c^-)~b_3\left(
T\frac{\partial b_2}{\partial T}\right)^{-1}_{T=T_c^*}+ {\cal O}(b_1^2)~.
\eea
The shift in $T_c$ with decreasing pion mass is proportional to
the expectation value of the Polyakov-loop just below $T_c$; all other
factors on the right-hand side of Eq.~(\ref{DelTc}) do not depend on
$b_1$ or $m_\pi$.
Numerical values for $b_3$ and for $b_2(T)$ were obtained
in~\cite{Dumitru:2000in,Dumitru:2001bf,Scavenius:2001pa,Scavenius:2002ru} 
by fitting the effective potential~(\ref{e3_N=3})
to the pressure and energy density of the pure gauge theory with three
colours; those are $b_3\approx0.9$ and $b_3^2/[T_c^* \partial
b_2(T_c^*)/\partial T]\approx 1$, to within 10\%. I therefore expect
that numerically $\Delta T_c/T_c^*$ is roughly equal to $\ell_0(T_c^-)$.

The Eqs.~(\ref{ellTc-},\ref{ellTc+}) seem  to indicate that
the discontinuity of $\ell_0$ at $T_c$ vanishes, i.e., that the phase
transition turns into a crossover, at a pion mass such that 
$b_1(m_\pi)=2b_3^3/27$. However, one cannot really extend the ${\cal
O}(b_1)$ estimates to the endpoint of the line of first-order
transitions because it applies, near $T_c$, only if $-4b_2(T_c) \ll
b_3^2$, which translates into $b_1 \ll b_3^3/108$, see Eq.~(\ref{b2Tc}).
To find the endpoint of the line of first-order transitions I
solve numerically for the global minimum of~(\ref{e3_Q}) as a function
of $b_2$, for given $b_1$, see Fig.~\ref{Fig_ell} (left). The
numerical solution is ``exact'' and does not assume
that $b_1$ is small. I employ $b_3=0.9$ to properly account for the
small latent heat of the pure gauge
theory~\cite{Dumitru:2000in,Dumitru:2001bf,Dumitru:2001xa,Scavenius:2001pa,Dumitru:2001vc,Dumitru:2002cf,Scavenius:2002ru}. 
Also, for $b_1=0$, this $b_3$ corresponds to $\ell_c=0.6$, which is close
to the expectation value of the renormalised (fundamental) loop for
the $N_c=3$ pure gauge theory~\cite{Kaczmarek:2002mc,Petreczky:2004pz,Zantow:2003uh,Dumitru:2003hp}.

Clearly, for very small $b_1$ the order parameter $\ell_0$ jumps at
some $b_2^c\equiv b_2(T_c)$, i.e., the first-order phase
transition persists. (The abscissa is normalised by
$|b_2(T_c^*)| = 2b_3^2/9$.)
I find that the discontinuity vanishes at $b_1^c=0.026(1)$, so
there is no true phase transition for $b_1 > b_1^c$. Nevertheless,
I define $b_2^c$ even in the crossover regime via
the peak of $\partial\ell_0(b_2)/\partial b_2$.
The shift of $b_2^c$ with increasing $b_1$ can now be converted
into the shift of $T_c$ itself by expanding about $T_c^*$:
\be \label{DelTc2}
\frac{\Delta T_c}{T_c^*} = {\Delta b_2^c}~\left(
T\frac{\partial b_2}{\partial T}\right)^{-1}_{T=T_c^*}~,
\ee
as already discussed above. I also note that from Fig.~\ref{Fig_ell}
(left) the susceptibility for the Polyakov-loop at its maximum is
$\partial \ell_0/\partial b_2\simeq 3.5$, 2, 1.5 for $b_1=0.06$, 0.1,
and 0.126, respectively. That is, the crossover is rather sharp for
the values of $b_1$ shown in the figure.

Explicit breaking of the Z(3) symmetry of the gauge theory
has previously been studied
in~\cite{Green:1983sd,Meisinger:1995qr,Alexandrou:1998wv}, and has been 
identified as the essential factor in determining the endpoint of
deconfining phase transitions. Moreover, while the term $\sim b_1$
quickly washes out the transition, those studies showed that along the
line of first-order transitions the shift of $T_c$ (or, alternatively, of
the critical coupling $\beta_c$) is moderate, which agrees with
my findings. However, the numerical values for the critical ``external field"
at the endpoint obtained in~\cite{Meisinger:1995qr,Alexandrou:1998wv} 
from actual
Monte-Carlo simulations can not be compared directly to my estimate for
$b_1^c$ because I work here with the renormalised (continuum-limit) loop,
not the bare loop.
Ref.~\cite{Karsch:2000kv} studied finite-temperature
QCD with $N_f=3$ flavours and various quark masses on the lattice (with
improved p4-action), and determined the critical (or crossover)
temperature as a
function of the pion mass. Using Eq.~(\ref{DelTc2}) one can
match $\Delta T_c/T_c^*$ to the data from~\cite{Karsch:2000kv} to
determine $b_1(m_\pi)$. In other words, I extract the
function $b_1(m_\pi)$ required to match the effective
Lagrangian~(\ref{e3_Q}) to $T_c(m_\pi)$ found on the lattice.
The result is shown in Fig.~\ref{Fig_ell} on the right. (Again, the pion mass
is normalised to the zero-temperature string tension in the pure gauge
theory, $\surd\sigma\simeq0.425$~GeV.)
\begin{figure}
\begin{center}
\includegraphics[height=6.2cm]{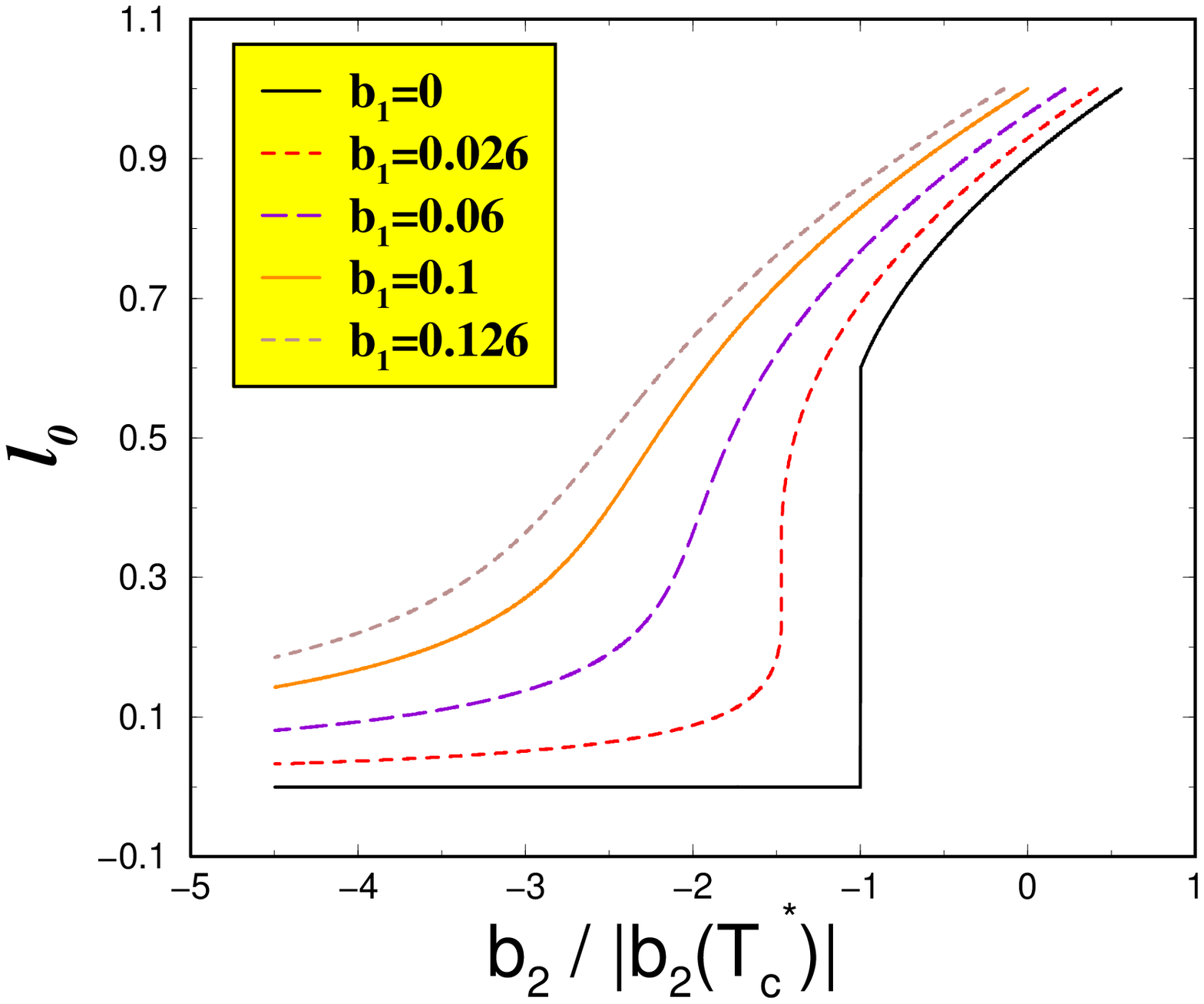}
\includegraphics[height=6.2cm]{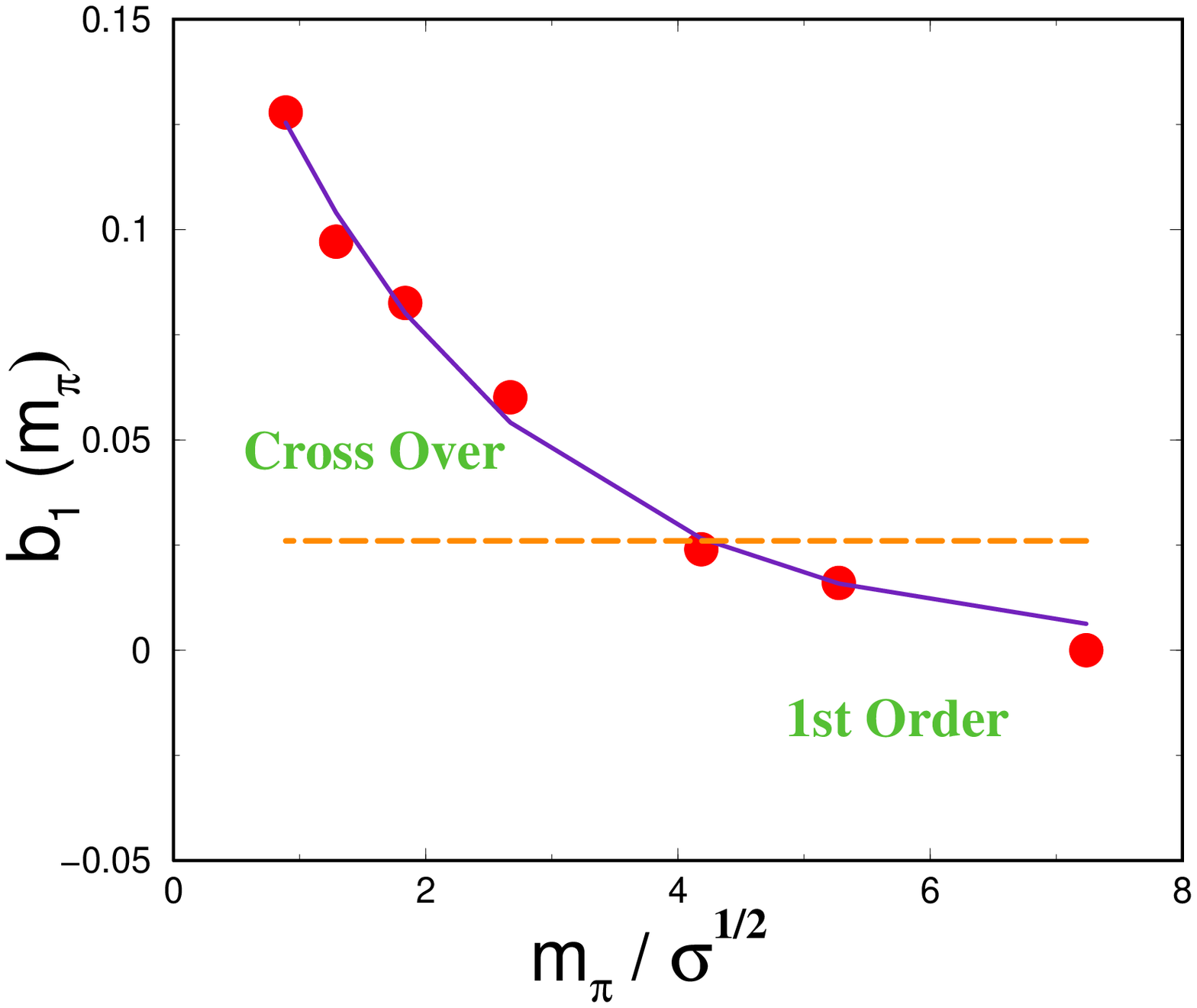}
\caption[The expectation value for the Polyakov-loop
and the explicit symmetry breaking term.]
{Left: The expectation value for the Polyakov-loop,
$\ell_0(b_2(T))$, for various values of the explicit
symmetry breaking coupling, $b_1$. All curves terminate at
$\ell_0=1\Leftrightarrow T=\infty$.
Right: $b_1$ as a function of $m_\pi$, obtained by matching to three-flavour
lattice data for $T_c(m_\pi)$. The solid line corresponds to an exponential
increase of $b_1$ with decreasing $m_\pi$, see text.
The broken horizontal line displays the
endpoint of the line of first-order phase transitions in terms of
$b_1$; the intersection with the $b_1(m_\pi)$ curve then gives the
corresponding pion mass.}  \label{Fig_ell}
\end{center}
\end{figure}
Evidently, the $\approx33\%$ reduction of $T_c$ from $m_\pi=\infty$
(pure gauge theory) to $m_\pi/\surd\sigma\approx 1$ requires only
small explicit breaking of the $Z(3)$ symmetry for the Polyakov-loop
$\ell$: I find that $b_1<0.15$ even for the smallest pion
masses available on the lattice. This is due to the rather weak first-order
phase transition of the pure gauge theory with $N_c=3$ colours,
reflected by the strong dip of the string tension 
in the confined phase near $T_c^-$ and of the Polyakov-loop screening
mass $m_\ell$ in the deconfined phase near $T_c^+$~\cite{Kaczmarek:1999mm}; 
cf.\
also the discussion in~\cite{Pisarski:2000eq,Pisarski:2002ji,
Dumitru:2000in,Dumitru:2001bf,Dumitru:2001xa,Dumitru:2001vc,Dumitru:2002cf}.

Moreover, $b_1(m_\pi)$ appears to follow the
expected behaviour $\sim\exp(-m_\pi)$. The exponential fit shown by the
solid line corresponds to $b_1(m_\pi) = a \exp(-b \; m_\pi/\surd\sigma)$,
with $a=0.19$ and $b=0.47$~. Surprisingly, by naive extrapolation
one obtains a pretty small
explicit symmetry breaking even in the chiral limit,
$b_1\approx 0.2$.

Finally, the endpoint of the line of
first-order transitions at $b_1^c=0.026$ (indicated by the dashed
horizontal line) intersects the curve $b_1(m_\pi)$ at
$m_\pi/\surd\sigma\approx 4.2$. For heavier pions the theory exhibits a
first-order deconfining phase transition, which then turns into a
crossover for $m_\pi\lton4.2\;\surd\sigma\approx 1.8$~GeV
($\hat{=}\,m_q\approx 0.9$~GeV). According to my estimate, the
endpoint of the line of first-order transitions occurs at $\Delta
T_c/T_c^*\approx 12\% $, which is slightly less than a previous
(qualitative) estimate of 26\% from ref.~\cite{Ogilvie:1999if}.

Of course, so far my analysis is restricted to pion masses
$m_\pi/\surd\sigma\gsim1$. On the other hand,
one might cross a chiral critical point for some pion
mass $m_\pi/\surd\sigma<1$~\cite{Gavin:1993yk}. Attempting a fit
with the model~(\ref{e3_Q}) beyond that point would then lead to 
deviations from $b_1\sim\exp(-m_\pi)$.

%**********************************************************************
%*********Chapter II***************************************************
\chapter{The improved Hartree-Fock approximation}
\section{Motivation}
In this section I consider the linear $\sigma$-model with $O(N)$ symmetry
(cf. Sec. \ref{introon}) within the CJT formalism (cf. Sec. \ref{seccjt}).
Of course, it is practically impossible to solve the exact theory (i.e. to
include all possible diagrams in the effective potential), therefore one 
has to solve the model within a many-body approximation scheme.
The most popular among these many-body approximation
schemes is the Hartree-Fock approximation.
In this case, $\Gamma_2$ [cf. Eq. (\ref{effactcjt})] solely contains
diagrams of so-called
``double-bubble'' topology, cf.\ Fig.\ \ref{paper1} a--c. 
Note that, in Refs.\ 
\cite{Petropoulos:1998gt,Lenaghan:1999si,Lenaghan:2000ey,Roder:2003uz}, this
scheme has been termed ``Hartree'' approximation.
However, since the exchange contributions with respect to internal
indices are in fact included, it is more
appropriate to call it ``Hartree-Fock'' approximation. Neglecting
the exchange contributions leads to the actual Hartree approximation
which, in the case of the $O(N)$ model, has also been termed 
``large-$N$'' approximation \cite{Petropoulos:1998gt,Lenaghan:1999si}.
In the Hartree-Fock approximation, 
the self-energies, which according to Eq.\ (\ref{Pi})
are obtained by cutting lines in the diagrams for $\Gamma_2$,
only consist of ``tadpole'' diagrams, cf.\ Figs.\ \ref{selfenergy_sigma} 
b,c, and \ref{selfenergy_pion} a,c.
For the chiral effective theories, such as the $U(N_f)_r \times
U(N_f)_\ell$ and $O(N)$ models, this approximation scheme
has been studied in great detail 
\cite{Baym:1977qb,Bochkarev:1995gi,Roh:1996ek,Amelino-Camelia:1997dd,Petropoulos:1998gt,Lenaghan:1999si,Lenaghan:2000ey,Roder:2003uz}.
The Hartree-Fock approximation is a very simple approximation scheme,
since tadpole self-energies do not have an imaginary part and,
consequently, all particles are stable quasiparticles. Moreover, since
tadpoles are independent of energy and momentum,
the Dyson-Schwinger equations for the propagators reduce to 
fix-point equations for the in-medium masses.

There are, however, several problems with the Hartree-Fock approximation
related to the truncation of $\Gamma_2$.
For instance, in the case of the $O(N)$ model,
it does not correctly reproduce the order of the chiral
symmetry restoring phase transition 
and it violates Goldstone's theorem in the sense
that the Goldstone bosons do not remain massless for nonzero temperatures
$0 <T< T_c$.
Several ways to remedy this shortcoming have been suggested.
The simplest one is to neglect subleading contributions in $1/N$,
leading to the Hartree (or large-$N$, see discussion above) approximation 
\cite{Petropoulos:1998gt,Lenaghan:1999si}.
Here, the Goldstone bosons remain massless for $T< T_c$
and the transition is of second-order, as expected from universality
class arguments \cite{Pisarski:1984ms}.
Another possibility to restore Goldstone's theorem is via
so-called ``external'' propagators 
\cite{Aouissat:1997nu,vanHees:2001ik,VanHees:2001pf,vanHees:2002bv,Aarts:2002dj}. 
These objects are obtained from the Hartree-Fock propagators
by additionally resumming all diagrams pertaining to the
Random Phase Approximation (RPA), with internal lines given
by the Hartree-Fock propagators.
Another way to restore the second-order nature of the chiral 
phase transition is to include so-called ``two-particle point-irreducible'' 
(2PPI) contributions to $\Gamma_2$ 
\cite{Verschelde:1992bs,Verschelde:2000dz,Baacke:2002pi}.

In this work, I am not concerned with these formal shortcomings
of the Hartree-Fock approximation: I focus exclusively on
the case realized in nature where chiral symmetry is
already explicitly broken by (small) nonzero quark masses, such that
the pion is no longer a Goldstone boson and assumes a mass 
$m_\pi \simeq 139.5~{\rm MeV}$ \cite{PDBook}.
My goal in this work is to include the nonzero decay width
of the particles in a self-consistent many-body approximation
scheme. To this end, one has to go beyond the Hartree-Fock
approximation and add other diagrams to $\Gamma_2$ which, upon cutting
lines according to Eq.\ (\ref{Pi}), yield self-energies with
nonzero imaginary part. The most simple of such diagrams, and the
ones considered in the following,
are those of the so-called ``sunset'' topology, cf.\ Fig.\ \ref{paper1} d,e,
leading to the self-energy diagrams shown in Figs.\ \ref{selfenergy_sigma} d,e
and \ref{selfenergy_pion} d. 

The self-energies arising from the sunset diagrams in $\Gamma_2$
depend on the energy and the momentum of the incoming particle.
Therefore, the Dyson-Schwinger equations no longer reduce to
fix-point equations for the in-medium masses. Since
the self-energies now have a nonzero imaginary part, implying
a finite lifetime of the corresponding particles, the spectral densities are
no longer delta-functions. It is therefore convenient to rewrite
the Dyson-Schwinger equations for the propagators into 
a set of self-consistent integral equations for the spectral densities 
which has to be solved numerically.

\section{The Dyson-Schwinger and the condensate equations}

In this section I discuss the application of the improved
Hartree-Fock approximation to the $O(N)$ model. The linear
$\sigma$ model with $O(N)$ symmetry and its effective potential
is already introduced in Chapter \ref{introon}. As mentioned above,
in the standard Hartree-Fock approximation one takes into account 
only the three possible double-bubble diagrams, $V_2^a$, $V_2^b$, and
$V_2^c$ of Eqs. \ref{V2db}, in the effective potential. As explained,
in order to include the nonzero decay width
of the particles one has to go beyond the standard Hartree-Fock
approximation by additionally including the two sunset diagrams,
$V_2^d$ and $V_2^e$ of Eqs. \ref{V2sunset}.
The complete expression for $V_2$ is obtained by the sum of
all contributions,
\be
V_2=V_{2}^a+V_{2}^b+V_{2}^c+V_2^d+V_2^e\;.
\ee
The expectation values of the one- and two-point functions in 
the absence of external sources, $\sigma$ and
${\cal S}$, ${\cal P}$, are determined from the stationary points of $V$,
\bea
\left.\frac{\delta V}{\delta \bar\sigma}\right
|_{\bar\sigma=\sigma,\bar{S}={\cal S},\bar{P}={\cal P}}=0\;,\;\;\;
\left.\frac{\delta V}{\delta \bar{S}}\right
|_{\bar\sigma=\sigma,\bar{S}={\cal S},\bar{P}={\cal P}}=0\;,\;\;\;
\left.\frac{\delta V}{\delta \bar{P}}\right
|_{\bar\sigma=\sigma,\bar{S}={\cal S},\bar{P}={\cal P}}=0\;,
\eea
leading to an equation for the scalar condensate $\sigma$,
\bse\label{gap_equations}
\be\label{condensate} 
H  =  \mu^2 \, \sigma + \frac{4 \lambda}{N}\, \sigma^3
 +  \frac{4\lambda}{N}\, \sigma \int_Q\, 
\left[ 3 \,{{\cal S}}(Q) + (N-1) \, {{\cal P}}(Q)\right] 
+ \left. \frac{\delta V_2}{\delta \bar\sigma}\right
|_{\bar\sigma=\sigma,\bar{S}={\cal S},\bar{P}={\cal P}}\; ,
\ee
where
\bea
\left. \frac{\delta V_2}{\delta \bar\sigma}\right
|_{\bar\sigma=\sigma,\bar{S}={\cal S},\bar{P}={\cal P}} & =&
\left(\frac{4\lambda}{N}\right)^2\sigma
\left[2(N-1)
\int_L\int_Q
{\cal S}(L){\cal P}(Q){\cal P}(L+Q) \right. \nn \\
&  & \hspace*{1.8cm}+ \left. 3!\int_L\int_Q
{\cal S}(L){\cal S}(Q){\cal S}(L+Q)\right]\; , \label{dv2ds}
\eea
and to the Dyson-Schwinger equations for the scalar and pseudoscalar
propagators,
\bea 
{\cal S}^{-1}(K;\sigma)&=&S^{-1}(K;\sigma)
+\Sigma[K;\sigma]\label{schwinger_dyson_1}\;,\\
{\cal P}^{-1}(K;\sigma)&=&P^{-1}(K;\sigma)
+\Pi[K;\sigma]\label{schwinger_dyson_2}\;.
\eea
\ese
Here I introduced the self-energies of the scalar and pseudoscalar fields,
\be
\Sigma = \Sigma^a+\Sigma^b+\Sigma^c+\Sigma^d+\Sigma^e\;, \;\;\;\;
\Pi=\Pi^a+\Pi^b+\Pi^c+\Pi^d+\Pi^e\;,
\ee
with
\begin{figure}
\begin{center}
\includegraphics[height=2cm]{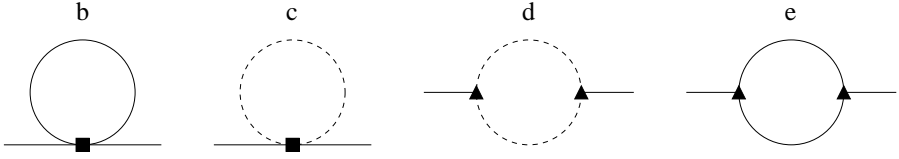}
\caption[The self-energy of the $\sigma$-meson.]
{The self-energy of the $\sigma$-meson. The diagrams b and c
are tadpole contributions generated by cutting an internal $\sigma$-line in the
double-bubble diagrams b and c in Fig.\ \ref{paper1}. The diagrams 
d and e are one-loop contributions generated by cutting an internal
$\sigma$-meson line in the sunset diagrams d and e of Fig.\ \ref{paper1}. 
Lines and vertices as in Fig.\ \ref{paper1}.}
\label{selfenergy_sigma}
\end{center}
\end{figure}
\bse\label{self_energy_sigma}
\bea
\label{self_sigmaa}
\Sigma^a&\equiv&2\left.\frac{\delta V_2^a}{\delta \bar{S}}\right
|_{\bar\sigma=\sigma,\bar{S}={\cal S},\bar{P}={\cal P}}
=0\; ,\\
\label{self_sigmab}
\Sigma^b&\equiv&2\left.\frac{\delta V_2^b}{\delta \bar{S}}\right
|_{\bar\sigma=\sigma,\bar{S}={\cal S},\bar{P}={\cal P}}
=\frac{4\lambda}{N} \,  3\, \int_Q \, {\cal S}(Q)\; ,\\
\label{self_sigmac}
\Sigma^c&\equiv&2\left.\frac{\delta V_2^c}{\delta \bar{S}}\right
|_{\bar\sigma=\sigma,\bar{S}={\cal S},\bar{P}={\cal P}}
=\frac{4\lambda}{N} (N-1)\, \int_Q \, {\cal P}(Q)\; ,\\
\label{self_sigmad}
\Sigma^d[K;\sigma]&\equiv&2\left.\frac{\delta V_2^d}{\delta \bar{S}}\right
|_{\bar\sigma=\sigma,\bar{S}={\cal S},\bar{P}={\cal P}}
=\left(\frac{4\lambda \sigma}{N}\right)^2
2(N-1)\int_Q{\cal P}(K-Q){\cal P}(Q)\; ,\hspace{1.3cm}\\
\label{self_sigmae}
\Sigma^e[K;\sigma]&\equiv&2\left.\frac{\delta V_2^e}{\delta \bar{S}}\right
|_{\bar\sigma=\sigma,\bar{S}={\cal S},\bar{P}={\cal P}}
=\left(\frac{4\lambda \sigma}{N}\right)^2
3\cdot 3!\int_Q{\cal S}(K-Q){\cal S}(Q)
\eea\ese
and
\bse\label{self_energy_pion}
\bea
\label{self_pia}
\Pi^a&\equiv&\frac{2}{N-1}\left.\frac{\delta V_2^a}{\delta \bar{P}}\right
|_{\bar\sigma=\sigma,\bar{S}={\cal S},\bar{P}={\cal P}}
=\frac{4\lambda}{N}\, (N+1)\,  \int_Q \, {\cal P}(Q)\; ,\\
\label{self_pib}
\Pi^b&\equiv&\frac{2}{N-1}\left.\frac{\delta V_2^b}{\delta \bar{P}}\right
|_{\bar\sigma=\sigma,\bar{S}={\cal S},\bar{P}={\cal P}}
=0\; ,\\
\label{self_pic}
\Pi^c&\equiv&\frac{2}{N-1}\left.\frac{\delta V_2^c}{\delta \bar{P}}\right
|_{\bar\sigma=\sigma,\bar{S}={\cal S},\bar{P}={\cal P}}
=\frac{4\lambda}{N} \int_Q \, {\cal S}(Q)\; ,\\
\label{self_pid}
\Pi^d[K;\sigma]&\equiv&\frac{2}{N-1}\left.\frac{\delta V_2^d}{\delta 
\bar{P}}\right
|_{\bar\sigma=\sigma,\bar{S}={\cal S},\bar{P}={\cal P}}
=\left(\frac{4\lambda \sigma}{N}\right)^2
4\int_Q{\cal P}(K-Q){\cal S}(Q)\;,\hspace{1.3cm}\\
\label{self_pie}
\Pi^e&\equiv&\frac{2}{N-1}\left.\frac{\delta V_2^e}{\delta \bar{P}}\right
|_{\bar\sigma=\sigma,\bar{S}={\cal S},\bar{P}={\cal P}}=0\; .
\eea
\ese
The calculation of the self-energy contributions  
Eqs.~(\ref{self_energy_sigma}) and (\ref{self_energy_pion}) corresponds to
opening internal lines in the diagrams of Fig.~\ref{paper1}. Employing this
procedure to the double-bubble diagrams leads to the tadpole contributions
for the self-energies of the
$\sigma$-meson (see Figs.~\ref{selfenergy_sigma} b, c)
and the pion (see Figs.~\ref{selfenergy_pion} a, c).
\begin{figure}
\begin{center}
\includegraphics[height=2cm]{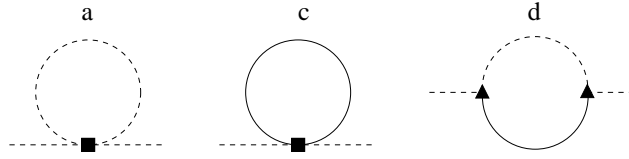}
\caption[The self-energy of the pion.]
{The self-energy of the pion. The diagrams a and c
are tadpole contributions generated by cutting an internal pion line in the
double-bubble diagrams a and c in Fig.\ \ref{paper1}. The diagram 
d is the one-loop contribution generated by cutting an internal pion
line in the sunset diagram d in Fig.~\ref{paper1}. 
Lines and vertices as in Fig.\ \ref{paper1}.}
\label{selfenergy_pion}
\end{center}
\end{figure}
This defines the standard Hartree-Fock approximation.
Additionally, the sunset diagrams of 
Figs.~\ref{paper1} d, e lead to the one-loop contributions shown
in Figs.~\ref{selfenergy_sigma} d, e and \ref{selfenergy_pion} d. 
The latter contributions depend on the energy and the momentum 
of the particles and lead to nonvanishing imaginary parts for the 
self-energies.
The explicit calculation of the self-energies (\ref{self_energy_sigma})
and (\ref{self_energy_pion})
is discussed in Appendix A.1.
The next step is to rewrite the set of Dyson-Schwinger
equations (\ref{schwinger_dyson_1}) and (\ref{schwinger_dyson_2}) in
terms of a set of integral equations for the spectral densities.
In general, the spectral densities for the $\sigma$-meson and the pion
are defined as
\bse\bea
\rho_\sigma (\omega, {\bf k}) &\equiv&2\, 
\Im\;{\cal S}(\omega+i\eta,{\bf k};\sigma)\;,\\
\rho_\pi (\omega, {\bf k}) &\equiv&2\, 
\Im\; {\cal P}(\omega+i\eta,{\bf k};\sigma)\;.
\eea\ese
If the imaginary parts of the self-energies are zero, as in
the standard Hartree-Fock approximation, all particles are
stable quasiparticles,
i.e., their spectral densities become delta-functions with support
on the quasiparticle mass shell,
\bse \label{specdens1} \bea
\rho_\sigma (\omega, {\bf k}) &=& 2 \pi\, 
Z_\sigma[\omega_\sigma({\bf k}),{\bf k}]\, 
 \left\{ \delta [ \omega - \omega_\sigma({\bf k})] - \delta [ \omega + 
\omega_\sigma({\bf k})] \right\}\;,\\
\rho_\pi (\omega, {\bf k}) &=& 2\pi \, 
Z_\pi[\omega_\pi({\bf k}),{\bf k}]\, 
 \left\{ \delta [ \omega - \omega_\pi({\bf k})] - \delta [ \omega + 
\omega_\pi({\bf k})] \right\}\;,
\eea \ese
where $\omega_{\sigma,\pi}({\bf k})$ are the quasiparticle energies
for $\sigma$-meson and pions, defined by the positive solutions of
\bse \label{qpe} \bea
\omega_\sigma^2({\bf k}) & = & k^2 + m_\sigma^2(\sigma) + 
\Re\, \Sigma[\omega_\sigma({\bf k}),{\bf k};\sigma]\;,\\
\omega_\pi^2({\bf k}) & = & k^2 + m_\pi^2(\sigma) + 
\Re\, \Pi[\omega_\pi({\bf k}),{\bf k};\sigma]\;,
\eea \ese
and 
\bse \bea
Z_{\sigma}[\omega_\sigma({\bf k}),{\bf k}] & \equiv & \left| 
\frac{\partial {\cal S}^{-1}(K;\sigma)}{\partial k_0} 
\right|_{k_0 = \omega_\sigma({\bf k})}^{-1}\;,\\
Z_{\pi}[\omega_\pi({\bf k}),{\bf k}] & \equiv & \left| 
\frac{\partial {\cal P}^{-1}(K;\sigma)}{\partial k_0} 
\right|_{k_0 = \omega_\pi({\bf k})}^{-1}\;,
\eea \ese
are the wave-function renormalization factors on the quasi-particle
mass shell. Since the real parts of the self-energies are even
functions of energy, these factors are also even in 
$\omega_{\sigma,\pi}({\bf k})$.
In the standard Hartree-Fock approximation, the 
(real parts of the) self-energies do not depend on $K^\mu$, such that
\be \label{qpe2}
\omega_\sigma(k) = \sqrt{k^2 + M_\sigma^2(\sigma)}\;,
\;\;\;\;
\omega_\pi(k) = \sqrt{k^2 + M_\pi^2(\sigma)}\;,
\ee
where 
\be \label{effmass}
M_\sigma^2 (\sigma) \equiv m_\sigma^2(\sigma) + \Re\, \Sigma(\sigma)\;, 
\;\;\;\;
M_\pi^2 (\sigma) \equiv m_\pi^2(\sigma) + \Re\, \Pi(\sigma)\;,
\ee
are the effective $\sigma$-meson and pion masses, and
\be 
Z_{\sigma}[\omega_\sigma(k)] = \frac{1}{2 \omega_\sigma(k)}\;,
\;\;\;\;
Z_{\pi}[\omega_\pi(k)] = \frac{1}{2 \omega_\pi (k)}\;.
\ee
For nonvanishing imaginary parts of the self-energies the spectral
densities assume the following form:
\bse\label{spectral_density}
\bea
\rho_\sigma(\omega, {\bf k})&=&-\frac{2\,\Im\,\Sigma(\omega, {\bf k};\sigma)}
{[\omega^2-k^2-m_\sigma^2(\sigma)-\Re\, \Sigma(\omega, {\bf k};\sigma)]^2
+[\Im \,\Sigma(\omega, {\bf k};\sigma)]^2},\hspace{0.5cm} \\
\rho_\pi(\omega, {\bf k})&=&-\frac{2\,\Im\, \Pi(\omega, {\bf k};\sigma)}
{[\omega^2-k^2-m_\pi^2(\sigma)-\Re\, \Pi(\omega, {\bf k};\sigma)]^2
+[\Im\, \Pi(\omega, {\bf k};\sigma)]^2} .
\eea
\ese
The general calculation of the imaginary parts of the self-energies 
is discussed in Appendix A.1. It turns out that
they do not depend on the direction of 3-momentum ${\bf k}$, but
only on the modulus $k$. The tadpole diagrams of 
Figs.~\ref{selfenergy_sigma} b, c (for the $\sigma$-meson) and
Figs.~\ref{selfenergy_pion} a, c (for the pion) do not have an
imaginary part. Thus [see Appendix, Eq.\ (\ref{general_im})],
\bse\label{im}\bea
\Im\,\Sigma(\omega, k ;\sigma)
&=&\Im \,\Sigma^d(\omega,k;\sigma)+\Im\, \Sigma^e(\omega,k;\sigma)\nn\\
&=&\left(\frac{4\lambda \sigma}{N}\right)^2
\frac{1}{2(2\pi)^3}\frac{1}{k}
\int_{-\infty}^{\infty} d \omega_1\, d \omega_2\,
[1+f(\omega_1)+f(\omega_2)]\delta(\omega-\omega_1-\omega_2)\nn \\
&\times& 
\int_0^{\infty}   d q_1 \, q_1 \, d q_2 \, q_2\;
\Theta\left(|q_1-q_2|\leq k\leq q_1+q_2\right)
\nn\\
&\times& 
\left[2(N-1)\, \rho_\pi(\omega_1,q_1)\rho_\pi(\omega_2,q_2)
+3\cdot 3!\,\rho_\sigma(\omega_1,q_1)\rho_\sigma(\omega_2,q_2)
\right],\hspace{0.5cm}\\
\Im\, \Pi(\omega,k;\sigma)&=&\Im\, \Pi^d(\omega,k;\sigma)\nn\\
&=&\left(\frac{4\lambda \sigma}{N}\right)^2
\frac{1}{2(2\pi)^3}\frac{1}{k}
\int_{-\infty}^{\infty} d \omega_1\, d \omega_2\,
[1+f(\omega_1)+f(\omega_2)]\delta(\omega-\omega_1-\omega_2)\nn \\
&\times& 
\int_0^{\infty}   d q_1 \, q_1 \, d q_2\, q_2\; 
\Theta\left(|q_1-q_2|\leq k\leq q_1+q_2\right)
\nn\\
&\times&
4\,\rho_\sigma(\omega_1,q_1)\rho_\pi(\omega_2,q_2)
\,\,,
\eea\ese
where $f(\omega)= [ \exp(\omega/T)-1 ]^{-1}$ is the Bose-Einstein
distribution function.
These expressions are not ultraviolet divergent and thus do not
need to be renormalised. 

For the real parts, I employ the following approximation. One only
takes into account the Hartree-Fock contributions, arising 
from the tadpole diagrams Figs.~\ref{selfenergy_sigma} b, c (for the
$\sigma$-meson) and Figs.~\ref{selfenergy_pion} a, c (for the pion).
As discussed above, these contributions do not depend on energy and
momentum. The real parts of the diagrams in Figs.~\ref{selfenergy_sigma}
d, e and \ref{selfenergy_pion} d are functions of energy and momentum
and are much harder to compute, involving an additional numerical integration
as compared to the respective imaginary parts (\ref{im}).
Therefore, in the present first step where I focus exclusively on
the effects of a nonzero decay width, I neglect them.
Consequently [see Appendix, Eq.~(\ref{general_eq_tadpole})],
\bse \label{re} \bea
\Re\, \Sigma(\sigma)&=&\Re\,\Sigma^b(\sigma)+\Re\, \Sigma^c(\sigma)\nn \\
&=&\frac{4\lambda}{N} \, 
\frac{4}{(2\pi)^3}
\int_{0}^{\infty} d \omega\, d q\, q^2\,f(\omega)\,
\left[ 3\, \rho_\sigma(\omega,q) + (N-1) \, \rho_\pi(\omega,q)\right]
\,,\hspace{1cm}\\
\Re\, \Pi(\sigma)&=&\Re\, \Pi^a(\sigma)+\Re\, \Pi^c(\sigma)\nn \\
&=&\frac{4\lambda}{N} 
\frac{4}{(2\pi)^3}
\int_{0}^{\infty} d \omega\, d q\, q^2\,
f(\omega)\, \left[ \rho_\sigma(\omega,q)
+ (N+1)\, \rho_\pi(\omega,q)\right]\,.
\eea\ese
Since the real parts do not depend on energy and momentum,
they only modify the masses of the $\sigma$-meson and the pion, 
cf.\ Eq.\ (\ref{effmass}). Note that I only consider the
temperature-dependent contributions in Eqs.\ (\ref{re}).
The omitted vacuum parts are ultraviolet divergent and have to be 
renormalised. The proper way to perform renormalization 
within the CJT-formalism was first presented in Refs.\
\cite{vanHees:2001ik,VanHees:2001pf,vanHees:2002bv}.
In the standard Hartree-Fock approximation for the $O(N)$ model, 
this leads to the same equations that one obtains applying the 
renormalization procedure of Ref.\ \cite{Lenaghan:1999si}. 
In that paper, the dependence of the
results on the renormalization scale was studied in detail.
It was found that simply neglecting the vacuum parts instead of
applying the proper renormalization procedure
leads only to quantitative,
but not to qualitative changes in the results.
Since I am interested in the effect of nonzero imaginary parts for
the self-energies and not in issues of renormalization, I
simply neglect the vacuum parts.
This prescription is also used to compute the tadpole and the
sunset contributions in Eq.~(\ref{condensate}) for the condensate.

After obtaining real and imaginary parts (\ref{re}) and (\ref{im}), 
respectively, one inserts them in the expressions
(\ref{specdens1}) (if the imaginary part is zero) or
(\ref{spectral_density}) (if the imaginary part is nonzero) 
for the spectral densities. These spectral densities can then
be used to again evaluate the real and imaginary parts of the
self-energies. This defines an iterative scheme to self-consistently
solve for the spectral densities as a function of energy and momentum. 
A convenient starting point
for this scheme is the standard Hartree-Fock approximation, i.e., neglecting
the imaginary parts altogether.

The self-consistently computed spectral densities obey a sum rule
\cite{leBellac},
\be \label{sumrule}
\int_{-\infty}^\infty \frac{d \omega}{2 \pi} \, \omega\,
\rho_{\sigma,\pi}(\omega,{\bf k}) = 1\;.
\ee
In my calculation, there are two reasons why this
sum rule may be violated. One is because I neglected the
real parts of the self-energy diagrams in Figs.~\ref{selfenergy_sigma}
d, e and \ref{selfenergy_pion} d.
The other is due to the numerical realization of the
above described iterative scheme. 
Numerically, one has to solve for the spectral density on a finite,
discretized grid in energy-momentum space. If the imaginary part
of the self-energy becomes very small, the spectral density converges
towards a delta function. The support of the delta function may 
be located between grid sites, which causes a loss of spectral strength,
which in turn violates the sum rule.

My prescription to restore the validity of the sum rule is the
following. One first checks whether
the imaginary part is smaller than the grid spacing (in my calculations,
10 MeV in both energy and momentum direction) at the
location of the quasiparticle mass shell, $\omega = \omega_{\sigma,\pi}(k)$.
If this is the case, one adds a sufficiently wide numerical 
realization of the delta function, $\delta_{\rm num}$, 
to the original spectral density,
$\rho(\omega,k) \rightarrow \rho (\omega, k)
+ c \cdot \delta_{\rm num}[\omega - \omega_{\sigma,\pi}(k)]$,
where $c$ is a constant that is adjusted
so that the sum rule is fulfilled on our energy-momentum grid,
\be
\label{sumrule2}
\int_{-\omega_{\rm max}}^{\omega_{\rm max}} 
\frac{d \omega}{2 \pi} \, \omega\,
\rho_{\sigma,\pi}(\omega,{\bf k}) = 1\;,
\ee
where $\omega_{\rm max}$ is the maximum energy on the energy-momentum grid.

On the other hand, if the imaginary part turns out to be sufficiently
large, I presume that a possible violation
of the sum rule (\ref{sumrule2}) 
arises from neglecting the real parts of the
self-energy diagrams in Figs.~\ref{selfenergy_sigma}
d, e and \ref{selfenergy_pion} d. In this case, one
multiplies the spectral density by a constant, $\rho \rightarrow c' \cdot
\rho$, where $c'$ is adjusted so that the sum rule (\ref{sumrule2}) is
fulfilled. (By comparing the results to the case $c'=1$, I
found that this somewhat {\em ad hoc\/} correction procedure does not 
lead to major quantitative changes.)

Restricting the sum rule to a finite range in energy as
in Eq.\ (\ref{sumrule2}), however, causes the following problem.
If the decay width of the particles is very large and
consequently the spectral density a rather broad distribution
around the quasiparticle mass shell, there will be parts 
which lie outside the energy-momentum grid. I could estimate the
magnitude of this physical effect if I knew the behaviour of the 
spectral density at energies $\omega > \omega_{\rm max}$. 
This is possible at zero temperature (with the help of Weinberg's sum rules),
but not at nonzero temperature. Here, I simply
assume that this effect is sufficiently small to be neglected, i.e.,
I assume that a possible violation of the sum rule (\ref{sumrule2}) is
due to the two above mentioned artifacts and accordingly perform
the correction of the spectral density.

Another physical effect which causes a loss of spectral strength
is if the quasiparticle energy 
$\omega_{\sigma,\pi}(k) > \omega_{\rm max}$. 
In this case, I do not perform the correction of the 
spectral density as described above. This occurs at
large momenta $k$, close to the edge of the energy-momentum grid.
I checked that, 
in the numerical calculation of the integrals in Eqs.\ (\ref{im}),
the integrands become sufficiently small in this region, so that 
the imaginary parts are not sensitive to this effect.

This concludes the discussion of the improvement of the
standard Hartree-Fock approximation by self-consistently including 
nonzero decay widths.

\section{Results}\label{HFresults}

In this section I present the numerical results for the $O(4)$
linear sigma model obtained in the 
Hartree-Fock approximation improved by including nonzero decay widths,
as discussed in Sec. 3.2.
\begin{figure}
\begin{center}
\vspace{-1cm}
\includegraphics[width=15.5cm]{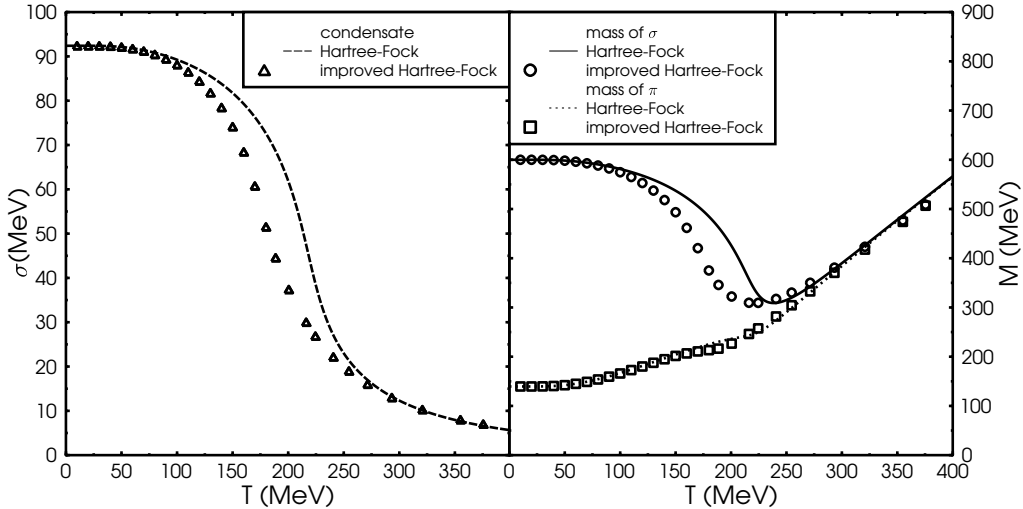}
\vspace{-1cm}
\caption[The condensate 
and the effective masses of the $\sigma$-meson and the pion.]
{The values for the condensate (left panel) 
and the effective masses of the $\sigma$-meson and the pion, (right panel)
as functions of temperature. The values are calculated in the standard
Hartree-Fock approximation (dashed and solid lines) 
and in the improved Hartree-Fock approximation (symbols)
as discussed in Sec. 3.2.}
\label{Hartree_values}
\end{center}
\end{figure}
The condensate $\sigma$ is shown in the left part of Fig.~\ref{Hartree_values}
as a function of temperature for the standard 
and the improved Hartree-Fock approximation. 
The qualitative behaviour is similar in the two approximations:
the condensate drops significantly with temperature indicating the
restoration of chiral symmetry. Since I consider the case of explicit
$O(4)$ symmetry breaking by taking $H \neq 0$ in Eq.\ (\ref{U}),
the chiral phase transition is a crossover transition.
Nevertheless, one can define a transition temperature, $T_\chi$,
as the temperature where the chiral susceptibility 
$\partial \sigma / \partial T$ assumes a maximum. 
Quantitatively, the inclusion of a nonzero
decay width lowers $T_\chi$ by about 
20\% as compared to the standard Hartree-Fock approximation. 
The transition temperature $T_\chi \simeq 175$~MeV
agrees within error bars
with recent lattice results, $T_\chi \simeq 172 \pm 5$~MeV 
for the two-flavour case \cite{Laermann:2003cv}.
(Note, however, that the latter value is extracted from an
extrapolation to the chiral limit, while my results are for the
case of explicit symmetry breaking, i.e., for nonzero quark masses.)

For the solution of the condensate equation (\ref{condensate})
the relative magnitude of the
contribution from the sunset diagrams, Eq.\ (\ref{dv2ds}),
is negligibly small, of order $\sim 10^{-4}$, and thus can be safely neglected.
In turn, not having to compute the integrals pertaining to
the double loop, cf.\ Eq.\ (\ref{sunset}), 
considerably speeds up the computation.

In the right part of Fig.~\ref{Hartree_values}, I show the effective masses 
$M_{\sigma}\equiv m_\sigma^2+\Re\, \Sigma$
and $M_{\pi} \equiv m_\pi^2+\Re\, \Pi$ of the $\sigma$-meson and 
the pion as functions of temperature in both approximation schemes. 
Since the decay width of the pion remains comparatively small, cf.\
\begin{figure}
\begin{center}
\vspace{-1cm}
\includegraphics[height=9.5cm]{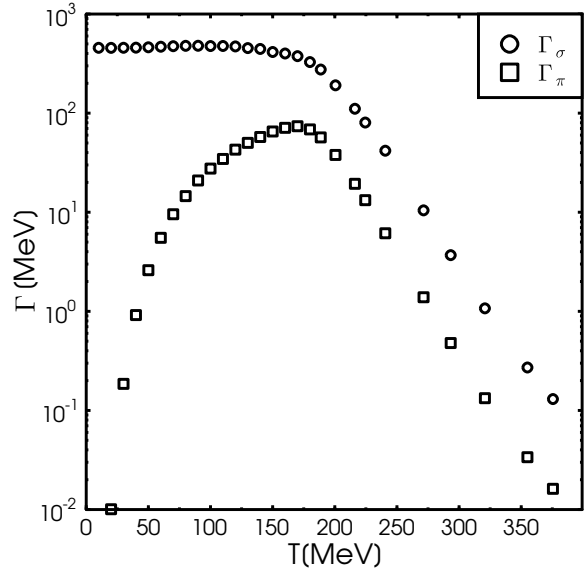}
\vspace{-1cm}
\caption[The decay width of the $\sigma$-meson and the pion.]
{The decay width of the $\sigma$-meson,
$\Gamma_\sigma=\Im \Sigma(\omega_\sigma)/\omega_\sigma$, and the pion, 
$\Gamma_\pi=\Im \Pi(\omega_\pi) /\omega_\pi$, in the
improved Hartree-Fock approximation as a function of the temperature 
at momentum $k=325~{\rm MeV}$.}
\label{decay_width}
\end{center}
\end{figure}
Fig.~\ref{decay_width}, the mass of the pion does not change
appreciably when taking the nonzero decay width into account.
On the other hand, the large decay width of the $\sigma$-meson at
temperatures below $T_\chi$, cf.\ Fig.~\ref{decay_width}, 
does influence the mass: in this range of temperatures
the mass exhibits a stronger decrease with temperature in the
improved Hartree-Fock approximation.
At large temperatures $T > T_\chi$, the decay width of the $\sigma$-meson
becomes negligibly small, cf.\ Fig.~\ref{decay_width}, 
and the mass approaches the value computed in the standard
Hartree-Fock approximation.
Both $\sigma$-meson and pion masses approach each other
above $T_\chi$, indicating the restoration of chiral symmetry.

In Fig.~\ref{decay_width} I show the decay widths of $\sigma$-mesons
and pions, defined as \cite{Weldon:1983jn,leBellac}
\be
\Gamma_\sigma (k) \equiv
\frac{\Im\,\Sigma[\omega_\sigma(k),k;\sigma]}{
\omega_\sigma(k)}\;,\;\;\;\;
\Gamma_\pi (k) \equiv \frac{\Im\,\Pi[\omega_\pi(k),k;\sigma]}{
\omega_\pi(k)}\;,
\ee 
where $\omega_{\sigma,\pi}(k)$ is the energy on
the quasi-particle mass shell, cf.\ Eq.~(\ref{qpe}).
At small temperatures, due to the possible decay of a $\sigma$-meson
into two pions, the decay width of the former
is large, of the order of its mass.
Note that the value $\Gamma_\sigma \simeq 460$~MeV obtained here
at $T=0$ is completely determined by the parameters $m^2, \lambda,$ and $H$
of the $O(4)$ model, i.e., without adjusting any additional
parameter. It is reasonably close to the experimentally measured
value $\Gamma_\sigma \sim (600 - 1000)$~MeV \cite{PDBook}.
The $\sigma$ decay width increases with temperature up to a maximum
of about $500$~MeV at a temperature $T\simeq 100$ MeV and then
decreases rapidly. The decay width of the pion vanishes at
$T=0$. It increases at nonzero temperature
and assumes a maximum at about $T\simeq 180$~MeV. At this temperature
the decay width is about half of the corresponding pion mass. It
decreases rapidly at higher temperature. Although the decay
widths of both particles decrease at high temperatures,
they do not become degenerate, the decay width of the $\sigma$-meson 
remains about a factor of 8 larger than that
of the pion. This difference can be traced to the symmetry factors
multiplying the one-loop self-energies for the $\sigma$-meson
and pion, cf.\ Eqs.\ (\ref{im}). For the $\sigma$-meson, there is a
factor $2 (N-1)$ in front of the pion loop and a factor 
$3 \cdot 3!$ in front of the $\sigma$-loop, so that the overall
factor is $\sim 2 (N-1) + 3 \cdot 3! = 24$. For the pion,
there is only a mixed $\sigma$-pion loop with a symmetry factor
$4$. From this simple argument
one already expects that the decay width of the $\sigma$-meson is about
a factor of 6 larger than that of the pion. The remaining difference
comes from the fact that the self-consistently computed
spectral densities of $\sigma$-meson and pion
under the integrals in Eqs.\ (\ref{im}) are also different, cf.\
Fig.\ \ref{rho_sigma_pion}. 

One might argue that, at asymptotically large temperatures,
effects from chiral symmetry breaking can no longer play a role
and the decay widths, as well as the spectral densities, 
of $\sigma$-meson and pion should become degenerate.
This is true in the chiral limit, where there is a 
thermodynamic phase transition between the phases of broken and
restored chiral symmetry, and where $\sigma \equiv 0$ in the latter phase.
Here, however, I consider the case of explicitly broken chiral symmetry,
where $\sigma >0 $, even when the temperature is very 
large. Since the decay widths are
proportional to $\sigma^2$, they also do not vanish at large temperature.
The difference in the symmetry factors for the self-energies
of $\sigma$-mesons and pions then leads to different
values for the decay widths and spectral densities.

\begin{figure}
\begin{center}
\includegraphics[height=15cm]{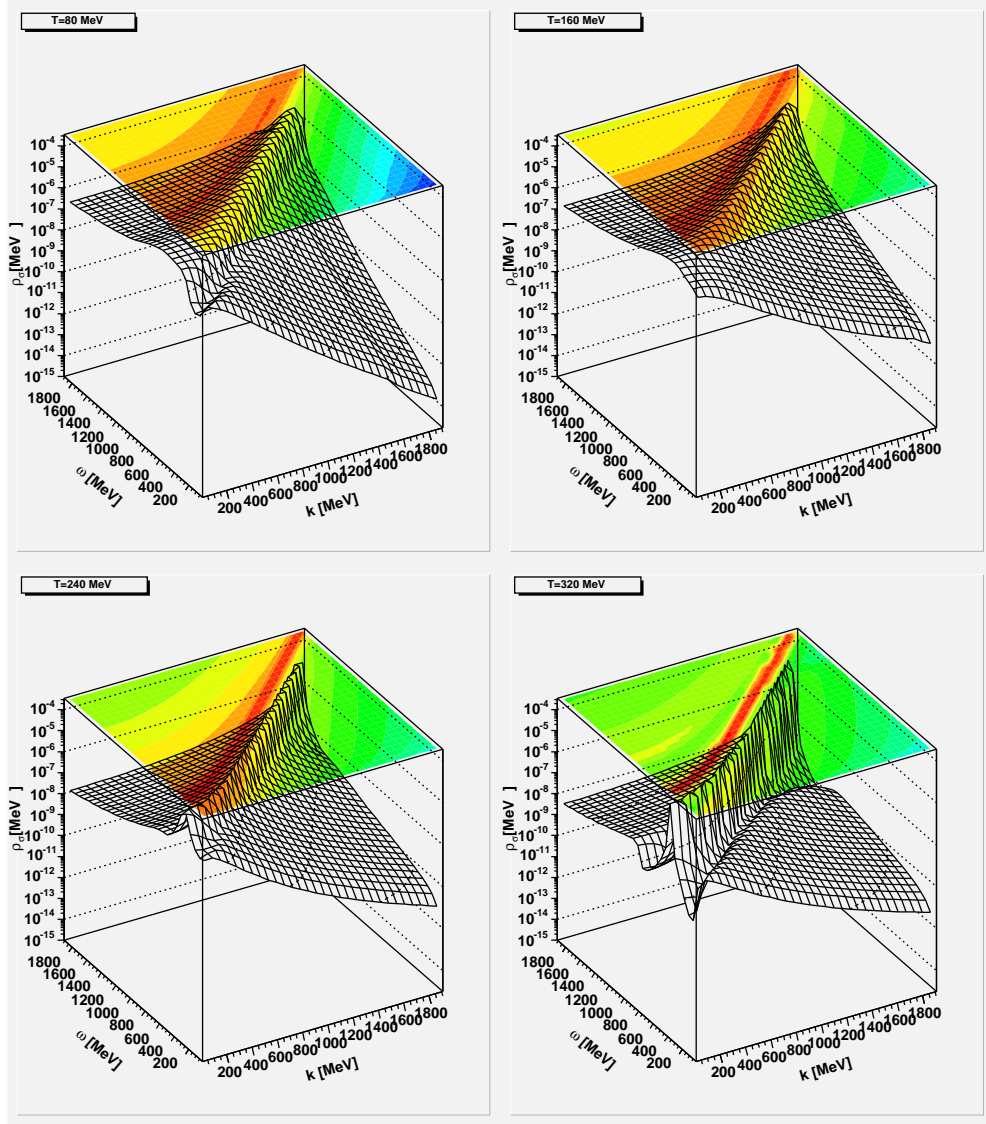}
\caption[The spectral density of the $\sigma$-meson.]
{The spectral density of the $\sigma$-meson
as a function of energy $\omega$ and momentum $k$
at temperatures 80, 160, 240, and 320 MeV.}
\label{rho_sigma}
\end{center}
\end{figure}

The self-consistently calculated spectral densities of 
the $\sigma$-meson and the pion as functions 
of the external energy $\omega$ and 
momentum $k$ are shown in Fig.~\ref{rho_sigma} 
\begin{figure}
\begin{center}
\includegraphics[height=15cm]{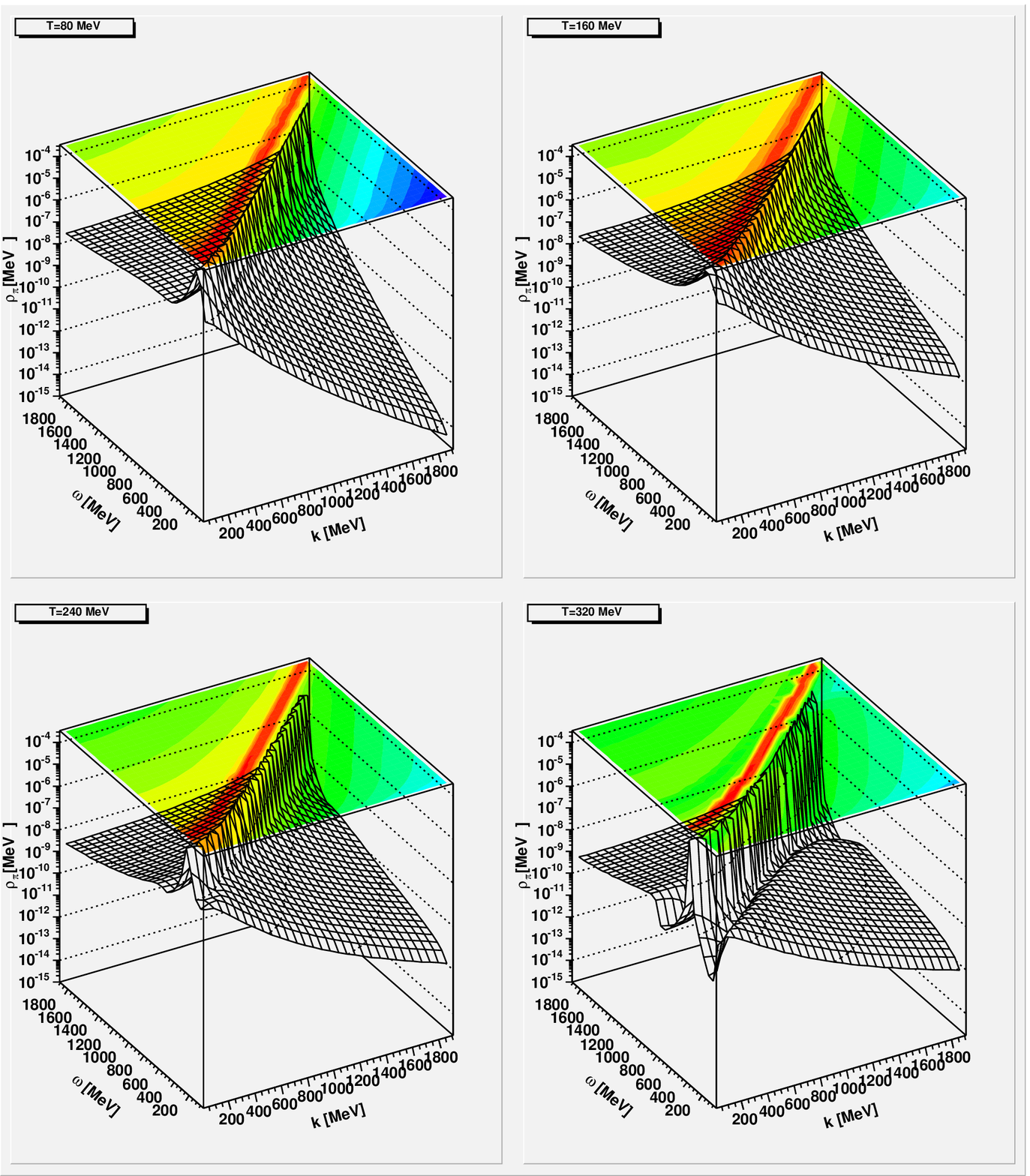}
\caption[The spectral density of the pion.]
{The spectral density of the pion as a 
function of  energy $\omega$ and momentum $k$
at temperatures 
80, 160, 240, and 320 MeV.}
\label{rho_pion}
\end{center}
\end{figure}
\begin{figure}
\begin{center}
\includegraphics[height=15cm]{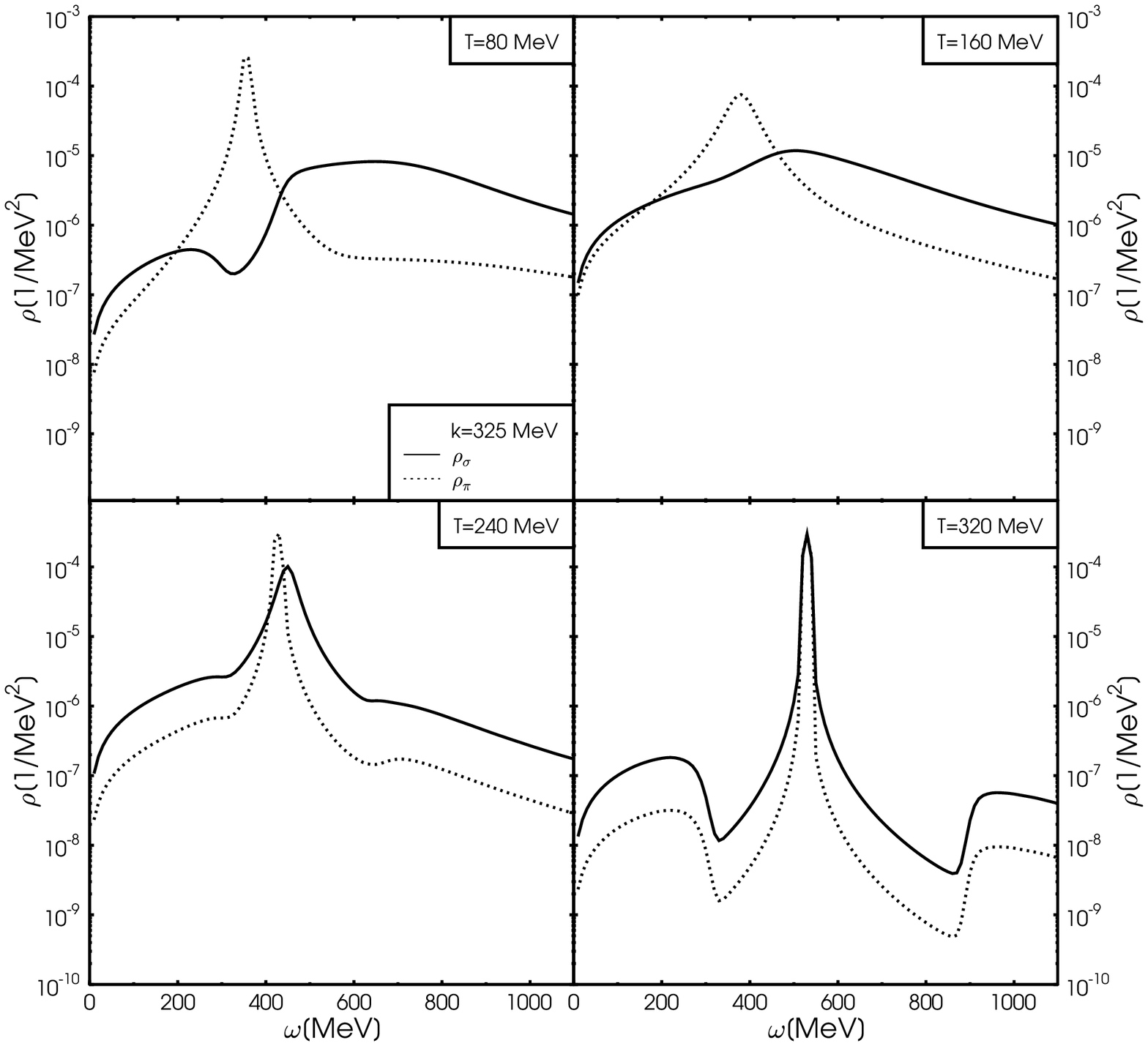}
\caption[The spectral density of the $\sigma$-meson and pion at a
fixed momentum.]
{The spectral density of the $\sigma$-meson and pion as a function
of energy $\omega$ at temperatures 80, 160,
240, and 320 MeV. The momentum is $k~=~325~{\rm MeV}$.}
\label{rho_sigma_pion}
\end{center}
\end{figure}
and Fig.~\ref{rho_pion} for different temperatures,
$T=80,\, 160,\, 240,$ and $320~\,{\rm MeV}$. 
For a detailed discussion, let us fix the momentum
at $k = 325$ MeV and consider the spectral densities
as functions of energy $\omega$ for different temperatures,
cf.\ Fig.~\ref{rho_sigma_pion}.

At all temperatures, the pion spectral density (dotted line) 
exhibits a pronounced peak on
the mass shell, at $\omega_\pi(k) = \sqrt{k^2 + M_\pi^2(\sigma)}$.
When the temperature is below $200$ MeV, such that 
$M_\pi \simeq m_\pi = 139.5$ MeV, cf.\ Fig.~\ref{Hartree_values},
the peak is located at $\omega_\pi(325\, {\rm MeV}) \simeq 350$ MeV. 
Above $T \sim 200$ MeV, $M_\pi$ increases significantly with temperature, 
and the position of the peak shifts towards larger energies, 
$\omega_\pi(325\, {\rm MeV}) \simeq 500$ MeV.
The broadening of the peak is due to scattering of the pion
off $\sigma$-mesons in the medium.

In contrast to the pion spectral density, for temperatures
below $\sim 170$ MeV the $\sigma$
spectral density (full line) does not exhibit a peak at the
mass shell, $\omega_\sigma (k) = \sqrt{k^2 + M_\sigma^2(\sigma)}$.
The reason is that $\omega_\sigma(k)$ is still 
sufficiently large to allow for the decay into two pions.
Consequently, in this temperature range the $\sigma$ spectral
density is very broad. On the other hand,
for temperatures above $\sim 170$ MeV, where $\omega_\sigma(k)$
drops below $2 \, M_\pi$, the two-pion decay channel is
closed and the $\sigma$ spectral density develops a distinct peak,
whose width is due to scattering of the $\sigma$-mesons off pions and
other $\sigma$-mesons in the medium.

Two other features of the spectral densities shown in
Fig.~\ref{rho_sigma_pion} are noteworthy. The first is
the region below the light-cone, $K^2 =\omega^2- k^2 <0$, where 
the mesons are Landau-damped. The second is the 
two-particle decay threshold. For $\sigma \rightarrow \pi \pi$,
this threshold is located at $\omega \sim 2\, M_{\pi}$, for 
$\sigma \rightarrow \sigma \sigma$, it is at $\omega \sim 2\, M_{\sigma}$,
and for $\pi \rightarrow \sigma \pi$ at $\omega \sim M_\sigma + M_\pi$.
The threshold is most clearly seen at large temperatures, e.g.\ $T= 320$ MeV,
when both particles become degenerate in mass,
$M_\sigma \simeq M_\pi \sim 400$ MeV, and the threshold is at
$\omega \sim 900$ MeV.

%**********************************************************************
%*********Chapter III**************************************************
\chapter{The improved Hartree approximation}
\section{Motivation}
In the standard Hartree (or Hartree-Fock) approximation of the $O(N)$ model,
one only takes into account the double-bubble diagrams. In these approximations
the self-energies of the particles are only real valued, therefore no
decay width effects are included. The difference between
the Hartree and the Hartree-Fock approximation is, that 
in the Hartree approximation all terms of order $\sim 1/N$
are neglected on the level of the
Dyson-Schwinger and the condensate equations.

In the last chapter I presented the so-called improved Hartree-Fock
approximation, which takes additionally into account sunset type diagrams.
This leads to
4-momentum dependent real and imaginary parts 
in the Dyson-Schwinger equations. Indeed, in Chapter III, I
neglected the 4-momentum dependent {\em real} parts of the Dyson-Schwinger
equations for simplicity. In the present Chapter,
I include them,
although in the Hartree approximation \cite{Roder:2005qy}.
This leads to a vanishing imaginary
part of the pion self-energy (because it is of order $\sim 1/N$),
i.e., the spectral density of the pion is just a delta function.
As shown in chapter III, the inclusion of the pion decay
width does not lead to a major change of the results, therefore
this is still a good approximation for the pion. 
The contributions from the $\sigma\to 2\sigma$ decay, in the self-energy
of the $\sigma$-meson, is also of order $\sim 1/N$, and vanishes, 
but (a part of) the contribution from the $\sigma\to 2\pi$ decay remains.
Therefore, its spectral density has a nonzero width, as
expected for the $\sigma$-meson with a very
large (vacuum) decay width, $\Gamma_\sigma=(600-1000)~{\rm MeV}$.
In the following, I call this the improved Hartree approximation. 

As discussed in Sec. 1.4, the parameter $H$ is given as a function
of the vacuum mass $m_\pi$,
and the vacuum decay constant $f_\pi$ of the pion : $H=m_\pi^2f_\pi$. 
In this chapter I compare the chiral limit, 
$m_\pi=0$ and $f_\pi=90~{\rm MeV}$, 
with the case of explicit chiral symmetry breaking, $m_\pi=139.5~{\rm MeV}$ 
and $f_\pi=92.4~{\rm MeV}$. The bare mass $\mu^2$ and the four-point coupling
$\lambda$ depends additionally on the vacuum mass of the $\sigma$-meson
$m_\sigma$: $\mu^2=-(m_\sigma^2-3\,m_\pi^2)/2$, and
$\lambda=N(m_\sigma^2-m_\pi^2)/(8f_\pi^2)$.
As mentioned above, the decay width of the $\sigma$-meson in
vacuum is very large, therefore its vacuum mass is not
well defined, $m_\sigma=(400-1200)~{\rm MeV}$ \cite{PDBook}. I compare
the results for $m_\sigma=400,\;600,$ and $800~{\rm MeV}$. The parameter
sets of the model, for $N=4$, are summarised in table \ref{parameter}.
\begin{table}
\begin{center}
\begin{tabular}{|l|l|l|} \hline
$\begin{array}{c}
\mbox{$\sigma$-meson}\\
\mbox{vacuum mass}\\
\end{array}$
& 
$\begin{array}{c}
\mbox{explicit chiral symmetry breaking}\\
m_\pi=139.5\,{\rm MeV},f_\pi=92.4\,{\rm MeV}\\
H=(121.6\,{\rm MeV})^3
\end{array}$
& $\begin{array}{c}
\mbox{chiral limit}\\
m_\pi=0\,{\rm MeV},f_\pi=90\,{\rm MeV}\\
H=0
\end{array}$\\ \hline
$ m_{\sigma}=400\,{\rm MeV}$ & $\lambda=8.230 $ & $\lambda=9.877 $\\
                             & $\mu^2= -(225.41\,{\rm MeV})^2$ &
                               $\mu^2= -(282.84\,{\rm MeV})^2$ \\\hline
$ m_{\sigma}=600\,{\rm MeV}$ & $\lambda=19.043 $ & $\lambda=22.222 $\\
                             & $\mu^2= -(388.34\,{\rm MeV})^2$ &
                               $\mu^2= -(424.264\,{\rm MeV})^2$  \\\hline
$ m_{\sigma}=800\,{\rm MeV}$ & $\lambda=36.341 $& $\lambda=39.506 $\\
                             & $\mu^2= -(539.27\,{\rm MeV})^2$ &
                               $\mu^2= -(565.685\,{\rm MeV})^2$ \\\hline
\end{tabular}
\end{center}
\vspace{3mm} 
\caption{ The masses and decay constants at vanishing
temperature and the corresponding parameter sets for the
$O(4)$ linear sigma model for the two symmetry breaking patterns
studied here.}
\label{parameter}
\end{table}
\section{The Dyson-Schwinger and the condensate equations}

In the following I repeat the derivation of the
condensate and the Dyson-Schwinger equations in the CJT 
formalism, as discussed in chapter III, and show
how the equations simplify in the improved Hartree 
approximation. The diagrams included in the effective potential
are the three double-bubble diagrams [$\sim(\int\bar{P})^2$,
$\sim(\int\bar{S})^2$, and $\sim\int\bar{P}\int\bar{S}$]
shown in Figs.~\ref{paper1} a, b, c, and
the two sunset diagrams ($\sim\int\int\bar{S}\bar{P}\bar{P}$ and
$\sim\int\int\bar{S}\bar{S}\bar{S}$) shown in Figs.~\ref{paper1} d, e.
To get the expectation values for the one- and two-point functions in the 
absence of external sources, $\sigma$, ${\cal S}$, and $\cal P$, one has
to find the stationary points of the effective potential 
(\ref{CJT-potential}).
Minimisation of the effective potential with respect to the one-point function
\be
\left.\frac{\delta V}{\delta \bar\sigma}\right
|_{\bar\sigma=\sigma,\bar{S}={\cal S},\bar{P}={\cal P}}=0,
\ee
leads to an equation for the scalar condensate $\sigma$, 
\bea
H  &=&  \mu^2 \, \sigma + \frac{4 \lambda}{N}\, \sigma^3
 +  \frac{4\lambda}{N}\, \sigma \int_Q\, 
\left[ 3 \,{{\cal S}}(Q) + (N-1) \, {{\cal P}}(Q)\right] \nn\\
&+&
\left(\frac{4\lambda}{N}\right)^2\sigma
\left[2(N-1)
\int_L\int_Q
{\cal S}(L){\cal P}(Q){\cal P}(L+Q) \right. \nn\\
&&\hspace{2cm}+ \left. 3!\int_L\int_Q
{\cal S}(L){\cal S}(Q){\cal S}(L+Q)\right]\; . \label{dv2ds_ln}
\eea
Using the fact that $\sigma^2\sim N$ [cf. Eq. (\ref{vev_tree})],
this equation becomes in the large-$N$ limit,
\be\label{condensate_largen} 
H  =  \sigma\left\{\mu^2 + \frac{4 \lambda}{N}\left[\sigma^2
 + N\, \int_Q\, {{\cal P}}(Q)\right]\right\}.
\ee
The minimisation of the effective potential with respect to
the two-point functions,
\bea
\left.\frac{\delta V}{\delta \bar{S}}\right
|_{\bar\sigma=\sigma,\bar{S}={\cal S},\bar{P}={\cal P}}=0\;,\;\;\;
\left.\frac{\delta V}{\delta \bar{P}}\right
|_{\bar\sigma=\sigma,\bar{S}={\cal S},\bar{P}={\cal P}}=0\;,
\eea
leads to the Dyson-Schwinger equations for the scalar and pseudoscalar
propagators, ${\cal S}$ and ${\cal G}$,
\bse\label{schwinger_dyson_ln}
\bea
{\cal S}^{-1}(K;\sigma)&=&S^{-1}(K;\sigma)
+\Sigma(K;\sigma)\label{schwinger_dyson_1_ln}\;,\\
{\cal P}^{-1}(K;\sigma)&=&P^{-1}(K;\sigma)
+\Pi(K;\sigma)\label{schwinger_dyson_2_ln}\;.
\eea\ese 	
Here I introduced the self-energy of the scalar 
\bse
\bea\label{self_energy_sigma_ln}
\Sigma(K;\sigma)&\equiv&
2\left.\frac{\delta V_2}{\delta \bar{S}}
\right|_{\bar\sigma=\sigma,\bar{S}={\cal S},\bar{P}={\cal P}}\nn\\
&=&\frac{4\lambda}{N}\left[ 3\, \int_Q \, {\cal S}(Q)\;
+(N-1)\, \int_Q \, {\cal P}(Q)\right]\; \nn\\
&+&\left(\frac{4\lambda \sigma}{N}\right)^2
\left[2(N-1)\int_Q{\cal P}(K-Q){\cal P}(Q)\; \right.\nn\\
&&\hspace{2cm}\left.+3\cdot 3!\int_Q{\cal S}(K-Q){\cal S}(Q)\right]\; 
\eea
and pseudoscalar fields
\bea\label{self_energy_pion_ln}
\Pi(K;\sigma)&\equiv&
\frac{2}{N-1}\left.\frac{\delta V_2}{\delta \bar{P}}\right
|_{\bar\sigma=\sigma,\bar{S}={\cal S},\bar{P}={\cal P}}\nn\\
&=&\frac{4\lambda}{N}\left[ \int_Q \, {\cal S}(Q)\; 
+(N+1)\,  \int_Q \, {\cal P}(Q)\right]\nn\\
&+&\left(\frac{4\lambda \sigma}{N}\right)^2
4\int_Q{\cal P}(K-Q){\cal S}(Q)\,\,.
\eea\ese
In the large-$N$ limit, all $\sigma$-meson tadpoles and a
part of the pion tadpole
($\sim\int{\cal S}$ and $\sim\int{\cal P}$) vanish, 
and only a part of the pion term ($\sim\int{\cal P}{\cal P}$) remains,
\bse
\bea\label{self_energy_sigma_large_N}
\Sigma(K;\sigma)
&=&\frac{4\lambda}{N}N\, \int_Q \, {\cal P}(Q)\; 
+\left(\frac{4\lambda \sigma}{N}\right)^2
2\,N\int_Q{\cal P}(K-Q){\cal P}(Q)\; ,\\
\Pi(K;\sigma)&=&\frac{4\lambda}{N}
N\,  \int_Q \, {\cal P}(Q).
\eea\ese
The tadpole contributions have no imaginary parts, therefore
\bse\label{im_largen}
\bea
\label{im_sigma_largen}
\Im\,\Sigma(K;\sigma)&=&
\left(\frac{4\lambda \sigma}{N}\right)^2
2\,N\,\Im\,\int_Q{\cal P}(K-Q){\cal P}(Q)\;,\\
\Im\,\Pi&=&0\label{im_pion_largen}.
\eea\ese
The imaginary part of the $\sigma$-meson self-energy depends
on the 4-momentum vector, $K=({\bf k},\omega)$, but the real part
can be split into terms which do
not depend on $K$, $[\Re\, \Sigma]_1$, arising from the tadpole
contribution, and terms which are 4-momentum 
dependent, $[\Re\, \Sigma(K,\sigma)]_2$,
arising from the sunset diagram,
\bea
\Re\, \Sigma(K;\sigma)&=&
[\Re\, \Sigma]_1+[\Re\, \Sigma(K;\sigma)]_2,
\eea
where 
\bse\bea
[\Re\, \Sigma]_1&=&
\frac{4\lambda}{N}\,N\,\int_Q \, {\cal P}(Q)\;,\\
\mbox{[}\Re\, \Sigma(K;\sigma)]_2&=&\left(\frac{4\lambda \sigma}{N}\right)^2
2\,N\,\Re \int_Q{\cal P}(K-Q){\cal P}(Q)\label{re_largen_2}.
\eea\ese
As mentioned above, the 4-momentum dependent terms in the pion self-energy 
vanish in
the large-$N$ limit. The 4-momentum independent term of the real part
of the pion self-energy is the same as for the $\sigma$-meson,
\be\label{re_largen}
\Re\, \Pi=[\Re\, \Sigma]_1=
\frac{4\lambda}{{\it N}}\,{\it N}\,\int_Q \, {\cal P}(Q)\;.
\ee
I want to calculate the spectral density of the
$\sigma$-meson, $\rho_\sigma$. To this aim, I rewrite
the Dyson-Schwinger equations (\ref{schwinger_dyson_ln}) and the equation 
for the chiral condensate (\ref{condensate_largen}) as functions of the
spectral density of the pion $\rho_\pi$. In the large-$N$ limit, the
imaginary part of the pion vanishes, and therefore the spectral density is
just a delta function 
\be
\label{spectral_density_pion_ln}
\rho_\pi (\omega, k) = \frac{\pi}{\omega_\pi (k)}
 \left\{ \delta [ \omega - \omega_\pi(k)] 
- \delta [ \omega + \omega_\pi(k)] \right\}\;,
\ee
with support on the quasiparticle energy for the pion $\omega_\pi(k)$,
\be
\omega_\pi(k) = \sqrt{k^2 + M_\pi^2(\sigma)}\;,
\ee
where I defined an effective mass for the pion $M_\pi$,
\be\label{Mpi_ln}
M_\pi^2 (\sigma) \equiv m_\pi^2(\sigma) + \Re\, \Pi\;.
\ee
In the chirally broken phase ($\sigma\ne 0$) the imaginary part of the
$\sigma$-meson is nonzero, therefore the spectral density assumes
the following form: 
\be\label{spectral_density_sigma_1_ln}
\rho_\sigma(\omega, {\bf k})=-\frac{2\,\Im\,\Sigma(\omega, {\bf k};\sigma)}
{[\omega^2-k^2-m_\sigma^2(\sigma)-\Re\,\Sigma(\omega,{\bf k};\sigma)]^2
+[\Im \,\Sigma(\omega, {\bf k};\sigma)]^2}\,\, .
\ee
In the chirally restored phase ($\sigma=0$) the 4-momentum dependent 
parts of the $\sigma$-meson self-energy vanish, hence
also the spectral density of the $\sigma$-meson becomes a delta function, 
\be
\label{spectral_density_sigma_2_ln}
\rho_\sigma (\omega, k) = \frac{\pi}{\omega_\sigma (k)}
 \left\{ \delta [ \omega - \omega_\sigma(k)] 
- \delta [ \omega + \omega_\sigma(k)] \right\}\;,
\ee
with support on the quasiparticle energy for the $\sigma$-meson
$\omega_\sigma (k)$,
\be
\label{qpe1sigma_ln}
\omega_\sigma(k) = \sqrt{k^2 + M_\sigma^2(\sigma)}\;,
\ee
where I defined an effective mass for the $\sigma$-meson $M_\sigma$,
\be\label{Msigma_ln}
M_\sigma^2 (\sigma) \equiv m_\sigma^2(\sigma) + [\Re\, \Sigma]_1\;.
\ee
Note that in the chirally broken phase ($\sigma\ne 0$), the quasiparticle
energy for the $\sigma$-meson $\omega_\sigma ({\bf k})$ is given
by the solution of
\bea
\label{qpe2sigma_ln}
\omega_\sigma^2({\bf k})-k^2-M_\sigma^2(\sigma)
-[\Re\,\Sigma(\omega_\sigma({\bf k}),{\bf k};\sigma)]_2=0.
\eea
The imaginary  and the real parts of the self-energies,
Eqs. (\ref{im_sigma_largen}), (\ref{re_largen_2}), and (\ref{re_largen}),
can be written as functions of the pion spectral
density. Note that for the real
parts, I only consider temperature-dependent contributions and neglect
the (ultraviolet divergent) vacuum parts, which is a simple way to
renormalize the integrals. The imaginary parts are not ultraviolet divergent
and thus do not need to be renormalised. The
4-momentum independent term is just
the standard tadpole integral, cf. Eq.~(\ref{ap2_tp}),
\be
\Re\, \Pi=[\Re\, \Sigma]_1=\frac{1}{2\pi^2}
\int_0^\infty {\it d q}\,{\it q}^2[\omega_\pi ({\it q})]^{-1}
f[\omega_\pi({\it q})].
\ee
The derivation of the equations for the 4-momentum dependent terms is
given in appendix A.2, 
\bse\label{4md_ln}
\bea
\Im\,\Sigma(\omega,k)
&=&\left(\frac{4\lambda \sigma}{N}\right)^2
\frac{N}{8\pi}\frac{1}{k}
\int_0^{\infty}  \, d q \;q\,[\omega_\pi (q)]^{-1}
\Theta\left(|q_0-q|\leq k\leq q_0+q\right)
\nn\\&\times&
\{1+f[\omega_\pi(q_0)]+f[\omega_\pi(q)]\},\\
\mbox{[}\Re\,\Sigma(\omega,k)]_2&=&
\left(\frac{4\lambda \sigma}{N}\right)^2
\frac{N}{8\pi^2}\frac{1}{k}
\mbox{P}\int_0^{\infty}   d q_1 \, q_1 \, d q_2 \, q_2\nn\\
&\times&
\Theta\left(|q_1-q_2|\leq k\leq q_1+q_2\right)
[\omega_\pi(q_1)\omega_\pi(q_2)]^{-1}\nn\\
&\times& \left\{ 
\frac{f[\omega_\pi(q_1)]+f[\omega_\pi(q_2)]}
{\omega_\pi(q_1)+\omega_\pi(q_2)-\omega}
+\frac{f[\omega_\pi(q_1)]+f[\omega_\pi(q_2)]}
{\omega_\pi(q_1)+\omega_\pi(q_2)+\omega}\right.\\
&&\left.+\frac{-f[\omega_\pi(q_1)]+f[\omega_\pi(q_2)]}
{\omega_\pi(q_1)-\omega_\pi(q_2)-\omega}
+\frac{-f[\omega_\pi(q_1)]+f[\omega_\pi(q_2)]}
{\omega_\pi(q_1)-\omega_\pi(q_2)+\omega}\right\}\;,\nn
\eea\ese
where $f(\omega)\equiv 1/[\exp(\omega/T)-1]$ is the Bose-Einstein
distribution function, $q_0\equiv\sqrt{[\omega-\omega_\pi(q)]^2-M_\pi^2}$,
and $\mbox{P}\int\dots$ denotes the principal value of the integral. Note that 
$\Theta\left(|q_1-q_2|\leq k\leq q_1+q_2\right)/k=2\delta(q_1-q_2)$ 
in the limit $k\to 0$, which can be used to perform the $q$-integration,
cf. Eqs.~(\ref{im2k0_ln}), and (\ref{re2k0_ln}). 
An appropriate way to perform 
the principal value numerically is discussed in the appendix,
cf. Eqs.~(\ref{re2k_ln}), and (\ref{re2k0_ln}).

The spectral densities have to obey a sum rule \cite{leBellac},
\be \label{sumrule_ln}
\int_{-\infty}^\infty \frac{\d \omega}{2 \pi} \, \omega\,
\rho_{\sigma,\pi}(\omega,{\bf k}) = 1\;.
\ee
In the improved Hartree approximation, the spectral density of the pion
(\ref{spectral_density_pion_ln}) is just a delta function, 
and normalised in such way that this sum rule is {\it{a priori}}
fulfilled. As mentioned above, this is not the case for the 
$\sigma$-meson spectral density in 
the chirally broken phase. The main reason for a
possible violation of the sum rule is due to
neglecting terms of the order $\sim 1/N$ in the self-energy,
which leads to a loss of spectral strength. I found that the inclusion of the 
4-momentum dependent real part $[\Re\,\Sigma]_2$ is close to negligible
for the validity of the sum rule. Other possible problems
arise from the numerical realisation of the spectral density on a finite
$\omega_{max}\times k_{max}$ energy-momentum grid, where 
$\omega_{max}\ge |\omega|$ and $k_{max}\ge |k|$ are the boundaries of the 
grid. If the imaginary 
part of the self-energy becomes very small, the spectral density becomes
more and more a delta function, which one has to realise numerically. On
the other side, if this imaginary part is very large, i.e., the width of
the spectral density becomes very large, one has to use a large $\omega_{max}$,
otherwise one loses too much spectral strength for energies
$\omega>\omega_{max}$. To minimise these numerical problems, I
use a rather large and fine (quadratic) energy-momentum grid, with
$\omega_{max}=k_{max}=2$ GeV and a lattice spacing of 5 MeV.

As discussed in Chapter III, if the sum rule is not fulfilled, 
I use the following way to restore it. If the imaginary part
of the $\sigma$-meson self-energy is very small,
I add a numerical realisation of a delta function $\delta_{\rm num}$ 
to the spectral density:
$\rho_\sigma'(\omega,k)\rightarrow\rho_\sigma(\omega,k)
+c_1\cdot\delta_{\rm num}[\omega-\omega_\sigma(k)]$.
If the imaginary part of the self-energy is large enough (compared
with the lattice spacing) I just multiply the
spectral density by a factor:
$\rho_\sigma'(\omega,k)\rightarrow c_2\cdot\rho_\sigma(\omega,k)$.
The constants $c_1$ and $c_2$ are adjusted in a way that 
$\rho_\sigma'$ fulfills the sumrule on the energy-momentum grid 
\be
\int_{-\omega_{max}}^{\omega_{max}} \frac{\d \omega}{2 \pi} \, \omega\,
\rho'_\sigma(\omega,{\bf k}) = 1\;.
\ee

The numerically scheme for the improved Hartree approximation is the
following. At first, one solves the standard Hartree approximation,
i.e., the condensate
and the Dyson-Schwinger equations, Eqs.~(\ref{condensate_largen})
and (\ref{schwinger_dyson_ln}) with $[\Re\,\Sigma]_2=[\Im\,\Sigma]_2=0$, to
obtain the chiral condensate and
the effective mass of the pion, $\sigma$ and $M_\pi$.
With these results, one calculates the
4-momentum dependent real and imaginary parts (\ref{4md_ln}) 
of the $\sigma$-meson self-energy, and finally 
the spectral density, Eq.~(\ref{spectral_density_sigma_1_ln}) or
(\ref{spectral_density_sigma_2_ln}). The decay width 
of the $\sigma$-meson $\Gamma_\sigma$ is defined as 
\cite{Weldon:1983jn,leBellac}
\be\label{decayw_ln}
\Gamma_\sigma (k) = 
\frac{\Im\,\Sigma[\omega_\sigma(k),k;\sigma]}{
\omega_\sigma(k)}\;.
\ee

\label{ImprovedH}
\section{Results}\label{Hresults}

In this section I present the numerical results for the linear
sigma model with $O(4)$ symmetry in the improved Hartree approximation
as discussed in the last section, for the parameter sets given in 
table \ref{parameter}. 

\subsection*{The mass}\label{res_1_ln}

\begin{figure}
\begin{center}
\vspace{-1.5cm}
\includegraphics[width=13.5cm]{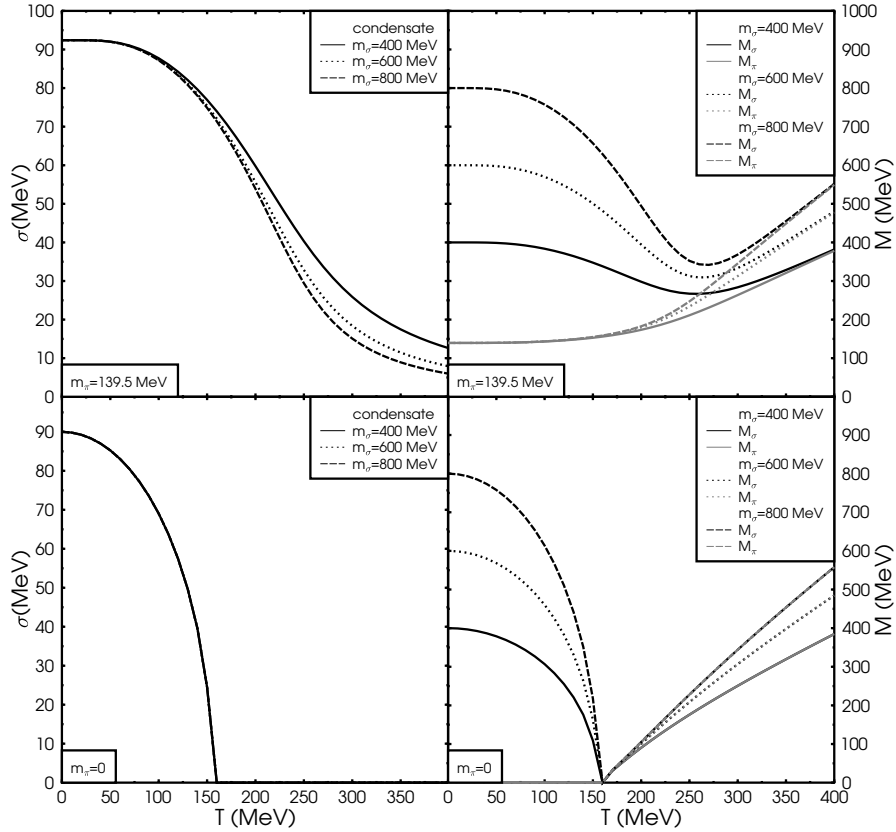}
\vspace{-0.5cm}
\caption[The chiral condensate 
and the effective masses of the $\sigma$-meson and the pion.]
{The chiral condensate $\sigma$ (left column) 
and the effective masses of the $\sigma$-meson and the pion,
$M_\sigma$ and $M_\pi$ (right column),
as functions of temperature $T$ and $\sigma$-meson vacuum mass $m_\sigma$. 
In the upper row the results with explicit chiral symmetry breaking 
are shown and in the lower row the results in the chiral limit.}
\label{Hartree_values_ln}
\end{center}
\end{figure}

\begin{figure}
\begin{center}
\includegraphics[height=7cm]{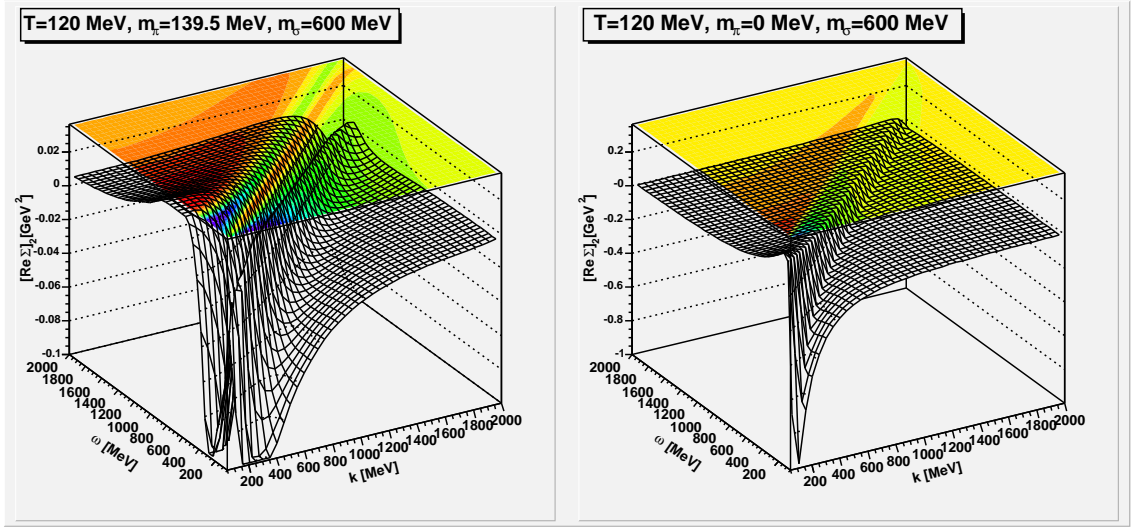}
\caption[The 4-momentum dependent real part
of the $\sigma$-meson self-energy.]{The 4-momentum
dependent real part of the $\sigma$-meson self-energy $[\Re\,\Sigma]_2$,
in the energy-momentum plane, for the case with explicit chiral
symmetry breaking (left) and the corresponding result
for the chiral limit (right) at a temperature of $T=120$ MeV.}
\label{re_t100_ln}
\end{center}
\end{figure}
I start the discussion of the results 
with the chiral condensate $\sigma$, given
by the solution of Eq.(\ref{condensate_largen}), and the
4-momentum independent effective
masses of the $\sigma$-meson and the pion, $M_\sigma$ and $M_\pi$,
as defined in Eqs.~(\ref{Msigma_ln}) and (\ref{Mpi_ln}). Note, however,
that (in the chirally broken phase, $\sigma\neq 0$) the mass of
the $\sigma$-meson (the energy where the $\sigma$-meson spectral density
assumes its maximum) is additionally modified by the 4-momentum dependent
real part of the $\sigma$-meson self-energy $[\Re\,\Sigma(K;\sigma)]_2$
in Eq.~(\ref{re_largen_2}), which is discussed later.

In the upper row of Fig.~\ref{Hartree_values_ln} the results for the chiral
condensate $\sigma$ (left column) and the effective mass for the 
$\sigma$-meson and the pion, $M_\sigma$ and $M_\pi$
(right column), are shown as functions of $T$
and $m_\sigma$, in the case with explicit chiral symmetry 
breaking. The results show
the behaviour of a crossover transition. Neither
the chiral condensate nor the mass of the scalar particle $M_\sigma$
become equal to zero, even for high temperatures. Thus, $M_\sigma$
and $M_\pi$ become only approximatively degenerate for large temperatures.
In the lower row of Fig.~\ref{Hartree_values_ln} the corresponding results for 
the chiral limit are presented. The results show the behaviour of 
a second-order phase transition. The condensate $\sigma$ and the mass
of the scalar particle $M_\sigma$ becomes equal to zero at a
critical temperature $T_\chi$, therefore $M_\sigma=M_\pi$ 
for $T\ge T_\chi$. The
condensate (accordingly $T_\chi$) does not depend on the vacuum mass
$m_\sigma$. This can be understood  as a consequence of the condensate
equation (\ref{condensate_largen}) in the chiral limit ($H=0$),
\be
0  =  \mu^2 + \frac{4 \lambda}{N}\left[\sigma^2
 + N\, \int_Q\, {{\cal P}}(Q)\right].
\ee
For $m_\pi=0$ the integral can be performed 
analytically \cite{Lenaghan:1999si},
\be
0  =- \frac{m_{\sigma}^2}{2} + \frac{4 \lambda}{N}\sigma^2
 + 4\lambda \frac{T^2}{12}.
\ee
Using (\ref{tree_level_masses}), this leads to an
($m_\sigma$-independent) equation for the condensate:
\be
\sigma=\sqrt{f_\pi^2-\frac{T^2N}{12}}\;
\ee
that determines the critical temperature (where $\sigma=0$)
to be $T_\chi=\sqrt{12/N}f_\pi$. This agrees with the numerical results
for the $O(4)$ model, $T_\chi=\sqrt{3}f_\pi\approx 160~{\rm MeV}$.

\begin{figure}\begin{center}
\includegraphics[height=9cm]{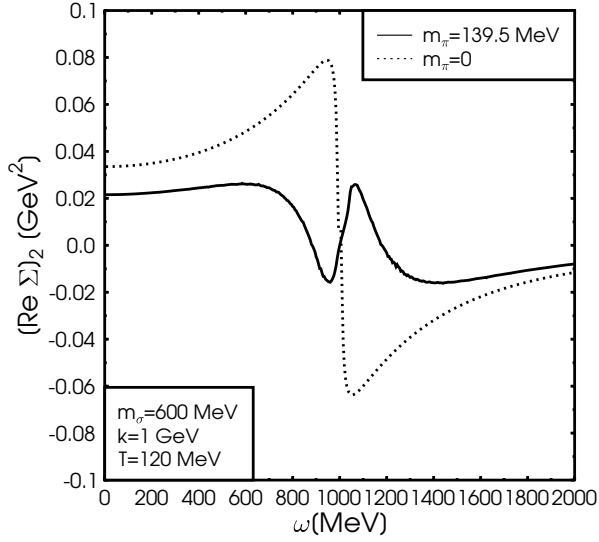}
\caption[The 4-momentum dependent real part
of the $\sigma$-meson self-energy at fixed momentum.]
{The 4-momentum dependent real part
of the $\sigma$-meson self-energy $[\Re\,\Sigma]_2$, at a
fixed momentum of $k=1$ GeV, for the case with  explicit chiral
symmetry breaking (full line) and the corresponding result
for the chiral limit (dotted line) at a temperature of $T=120$ MeV.}
\label{re_t100_k1000_ln}
\end{center}\end{figure}
The 4-momentum dependent real part of the $\sigma$-meson self-energy
$[\Re\,\Sigma]_2$ is shown in Fig.~\ref{re_t100_ln} on the whole
energy-momentum plane, and in Fig.~\ref{re_t100_k1000_ln} at fixed momentum
(in the middle of the grid) of $k=1$ GeV as a function of energy. It 
is larger in the chiral limit as 
in the case with explicitly broken chiral symmetry, because 
$[\Re\,\Sigma]_2\sim \lambda^2\sim (m_\sigma^2-m_\pi^2)^2$ is maximal
for $m_\pi^2=0$. 
Note that $[\Re\Sigma]_2$ is small compared to the (squared) 
4-momentum independent mass of the $\sigma$-meson, shown in 
Fig.~\ref{Hartree_values_ln}. I come back to the influence
of $[\Re\Sigma]_2$ in the discussion of the spectral density.

\subsection*{The decay width}\label{res_2_ln}
After calculating the ($4-$momentum dependent)
imaginary part of the $\sigma$-meson self-energy
(\ref{im_sigma_largen}), the decay width $\Gamma_\sigma$ 
(\ref{decayw_ln}) is calculated at the
quasiparticle energy $\omega=\omega_\sigma$.
The imaginary part of the self-energy, and
therefore the decay width, of the pion is of order $\sim 1/N$,
and therefore neglected.

\begin{figure}\begin{center}
\includegraphics[height=9cm]
{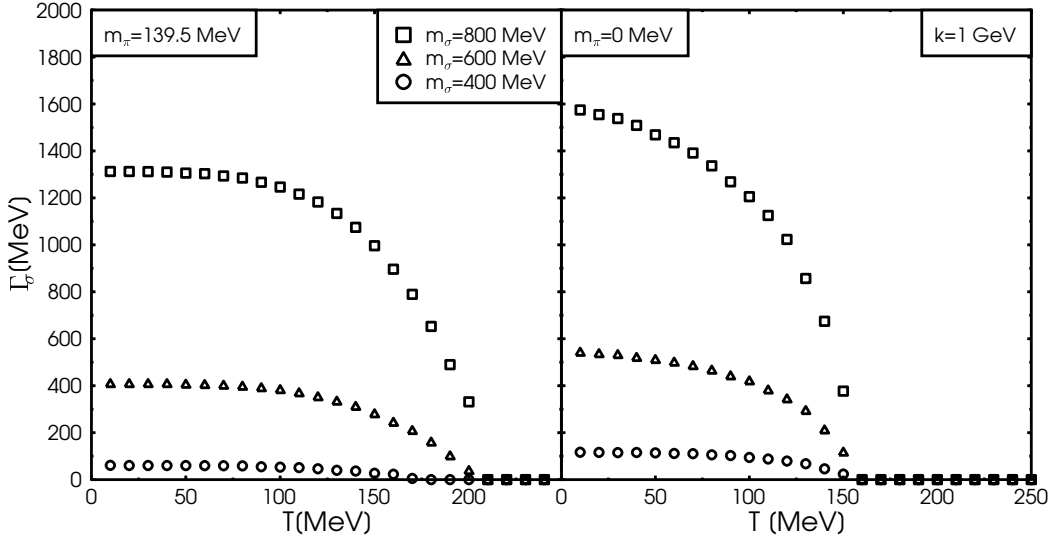}
\caption[The decay width of the $\sigma$-meson.]
{The decay width of the $\sigma$-meson $\Gamma_\sigma$,
as a function of temperature $T$ and $\sigma$-meson vacuum mass $m_\sigma$,
at momentum $k=1$ GeV, in the case with explicit chiral symmetry breaking 
(left) and the chiral limit (right).}
\label{decay_width_ln}
\end{center}\end{figure}
In Fig.~\ref{decay_width_ln} the decay width of the $\sigma$-meson
$\Gamma_\sigma$ is shown as a function of temperature $T$.
The qualitative behaviour is similar in all cases, but the decay width 
is larger in the chiral
limit ($m_\pi=0$) compared to the case with explicit
chiral symmetry breaking ($m_\pi\neq 0$), because
$\Gamma_\sigma\sim
\Im\,\Sigma\sim\lambda^2\sim (m_\sigma^2-m_\pi^2)^2\le m_\sigma^4$
[cf. Eqs. (\ref{im_sigma_largen}) and (\ref{def_para})]. 
The decay width is a strictly monotonic decreasing function with temperature,
and becomes approximately zero (equal zero)
for temperatures larger than $\sim 200$ MeV 
($T\ge T_\chi$) in the case with explicitly broken chiral symmetry 
(the chiral limit). This is a consequence of the (partial) restoration 
of the chiral symmetry, the masses of the chiral partners
become (approximatively) degenerate, and therefore the
phase space of the $\sigma\to 2 \pi$ decay is squeezed. 
The dependence on the vacuum mass of the $\sigma$-meson is significant. 
The reason for this is that $\Gamma_\sigma\sim\Im\,\Sigma\sim m_\sigma^4$
which agrees reasonably with the results, 
$\Im\,\Sigma=1:3:16$ for $m_\sigma=400:600:800$ MeV.

\begin{figure}\begin{center}
\includegraphics[height=9cm]
{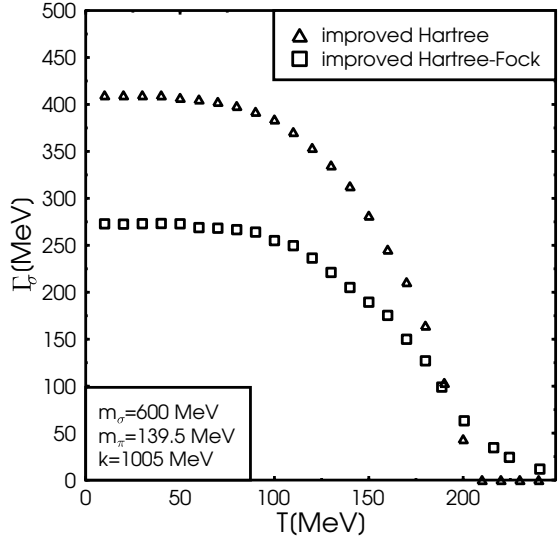}
\caption[The decay width of the $\sigma$-meson in the improved Hartree
and  Hartree-Fock approximation.]
{The decay width of the $\sigma$-meson $\Gamma_\sigma$,
in the improved Hartree
(triangles) and in the improved Hartree-Fock (squares) approximation, 
as a function of the temperature $T$, at momentum $k=1005$ MeV and 
$\sigma$-meson vacuum mass of $m_\sigma=600$ MeV,
in the case with explicit chiral symmetry breaking.}
\label{decay_width_cmp_hf_largen}
\end{center}\end{figure}
In Fig.~\ref{decay_width_cmp_hf_largen}
the decay width of the $\sigma$-meson in the improved Hartree
(triangles) is compared with the improved Hartree-Fock \cite{Roder:2005vt} 
(squares) approximation. 
The main difference comes from the combinatorial factor 
in front of the two-pion term ($\sim\int{\cal P}{\cal P}$) in 
the imaginary part of the self-energy [cf. Eqs.~(\ref{self_energy_sigma_ln})
with (\ref{im_sigma_largen})]. A part of this contribution is
of order $\sim 1/N$, and neglected in the Hartree approximation. Therefore,
this factor is $2\cdot N=8$ in the improved Hartree, and $2\cdot(N-1)=6$
in the improved Hartree-Fock approximation, which would lead to 
a decay width which is an factor $\approx 1.33$ larger
in the improved Hartree approximation.
The remaining difference, shown in the plot, comes from the $\sigma$-meson
term
($\sim3\cdot 3!\int{\cal S}{\cal S}$) in Eq. (\ref{im_sigma_largen}),
which vanishes in the large-$N$ limit.

\subsection*{The spectral density}\label{res_3_ln}
Finally, after solving the coupled condensate
and Dyson-Schwinger equations selfconsistently,
Eqs.~(\ref{condensate_largen})
and (\ref{schwinger_dyson_ln}), the spectral density of the $\sigma$-meson
$\rho_\sigma$
is given in the chiral broken phase ($\sigma\neq 0$) by 
Eq.~(\ref{spectral_density_sigma_1_ln}),
and in the restored phase ($\sigma=0$) by
Eq.~(\ref{spectral_density_sigma_2_ln}).

\begin{figure}\begin{center}
\includegraphics[height=7cm]
{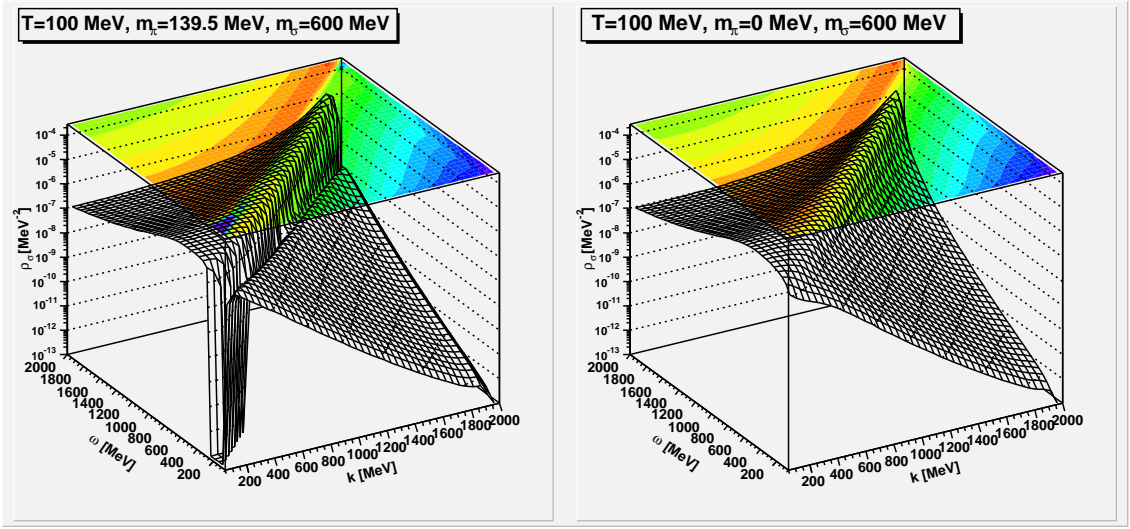}
\caption[The spectral density of the $\sigma$-meson.]
{The spectral density of the $\sigma$-meson $\rho_\sigma$, 
as a function of energy $\omega$, momentum $k$, 
and $\sigma$-meson vacuum mass $m_\sigma$ at a temperature of $T=100$ MeV,
in the case with explicit chiral symmetry
breaking (left) and in the chiral limit (right).}
\label{rho_sigma_cutp_xs_cl_3d_ln}
\end{center}\end{figure}
\begin{figure}\begin{center}
\includegraphics[height=8cm]{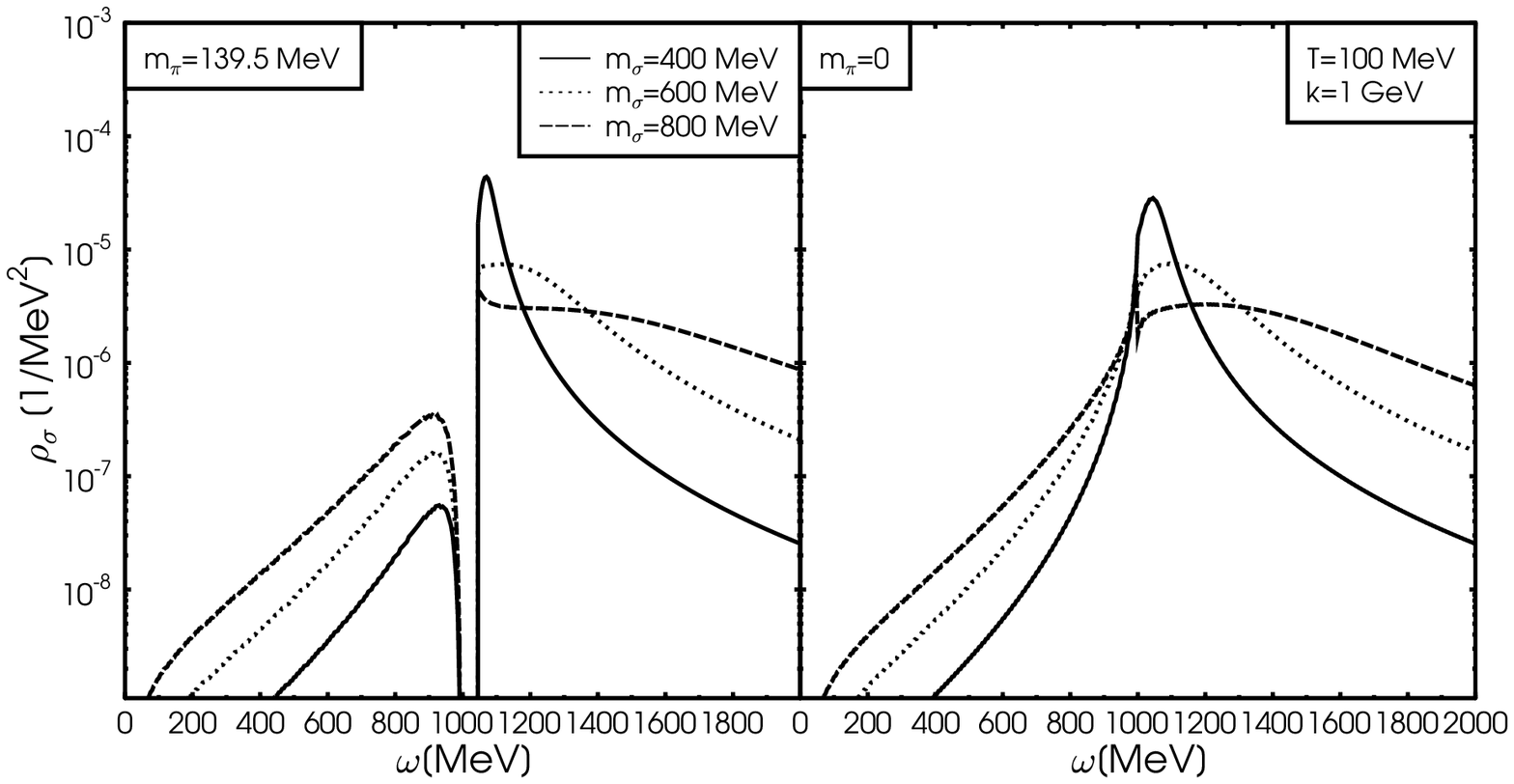}
\caption[The spectral density of the $\sigma$-meson at
fixed momentum.]
{The spectral density of the $\sigma$-meson $\rho_\sigma$, 
as a function of energy $\omega$, and $\sigma$-meson vacuum mass $m_\sigma$,
at a momentum of $k=1$ GeV and a temperature of $T=100$ MeV,
in the case with explicit chiral symmetry
breaking (left) and in the chiral limit (right).}
\label{rho_sigma_cutp_xs_cl_ln}
\includegraphics[height=8cm]{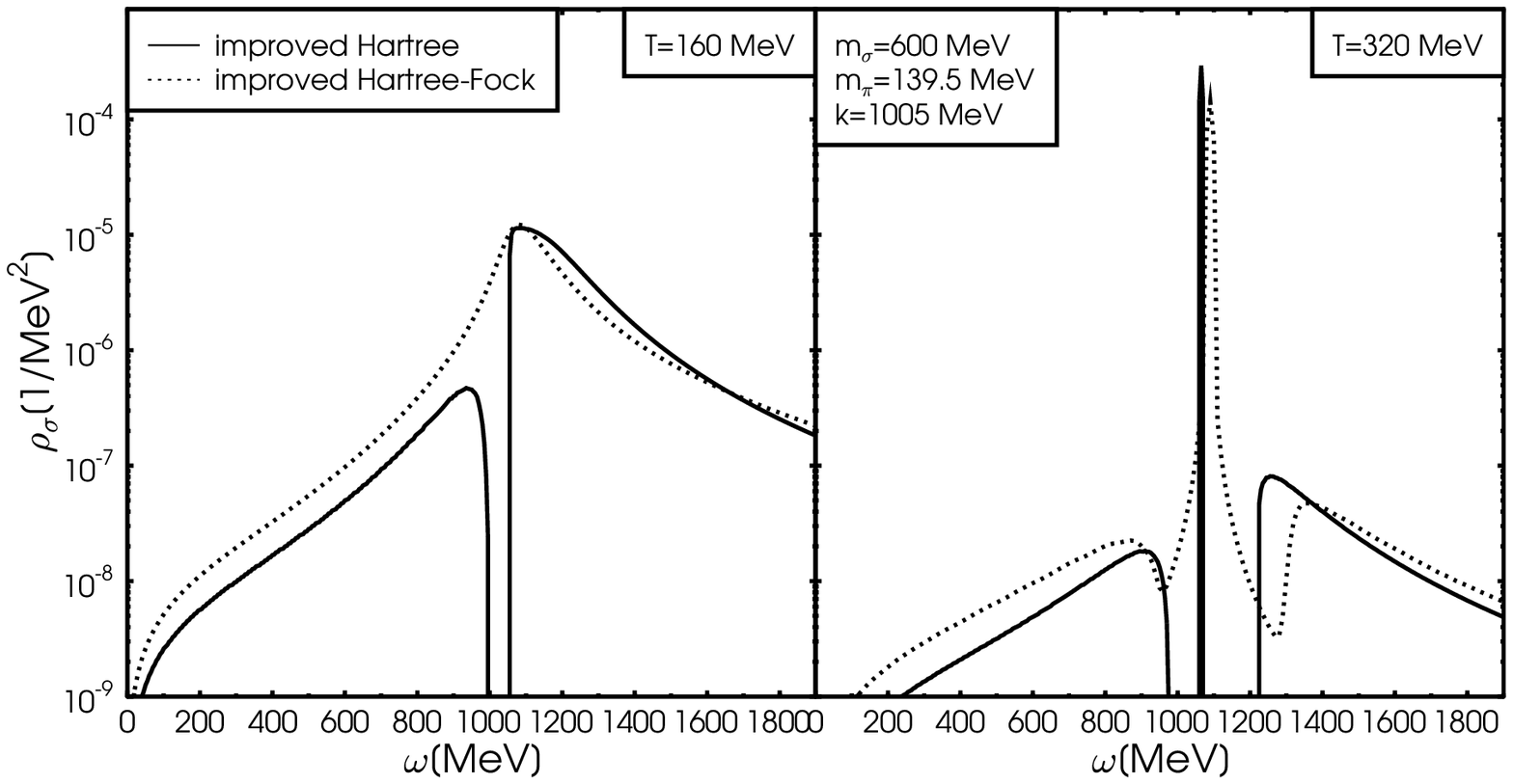}
\caption[The spectral density of the $\sigma$-meson in the improved Hartree 
and Hartree-Fock approximation.]
{The spectral density of the $\sigma$-meson $\rho_\sigma$,
as a function of energy $\omega$, at a momentum of $k=1$ GeV, 
and a vacuum $\sigma$-meson mass of $m_\sigma=600 $MeV, in
the case with explicit chiral symmetry breaking. The results are shown for 
the improved Hartree (full line) and the improved Hartree-Fock (dotted line)
approximation at temperatures of $T=160$ MeV (left), 
and 320 MeV (right).}
\label{rho_sigma_vgl_hfln}
\end{center}\end{figure}

In Fig.~\ref{rho_sigma_cutp_xs_cl_3d_ln} the spectral density
is plotted as a function of energy and momentum, and in 
Fig.~\ref{rho_sigma_cutp_xs_cl_ln} at fixed momentum (as for
the real part in the middle of the grid) of $k=1$ GeV. In 
Fig.~\ref{rho_sigma_cutp_xs_cl_ln} all possible parameter sets 
(cf. table 4.1) are compared. 
The spectral density does not exhibit a 
pronounced peak at the mass-shell
energy $\omega_\sigma=\sqrt{k^2+M_\sigma^2}$. The reason for this is
that the
energy of the $\sigma$-meson $\omega_\sigma$ is large enough
to decay into two pions. As discussed above, the decay width
of this process is large and becomes even larger for increasing $m_\sigma$,
as shown in the plot.
A remarkable difference between the chiral limit (right) and the
case with explicit chiral symmetry breaking (left) is that
$\rho_\sigma\approx 0$ around $\omega\approx 1$ GeV in
the case with explicit symmetry breaking but not in the chiral limit. This
effect
can be traced back to the threshold energy for $\sigma\rightarrow 2\pi$, which
is  $\omega=2M_\pi$, and therefore zero in the chiral limit but not in the
case with explicit chiral symmetry breaking. For $\omega<k$ 
the $\sigma$-meson is Landau-damped.

To illustrate the effects of neglecting terms of order
$\sim 1/N$, I compare in Fig.~\ref{rho_sigma_vgl_hfln}
the results from the improved Hartree (full lines) with the results 
from the improved
Hartree-Fock \cite{Roder:2005vt} (dotted lines) approximation. The results
are shown in the low-temperature regime, $T=160$ MeV (left), and in the
high-temperature regime, $T=320$ MeV (right).
In the improved Hartree-Fock approximation
one does not neglect the $\sim 1/N$ terms in the imaginary part of
the self-energy, which leads to a nonzero width of the pion spectral 
density, and therefore to a washed-out $\sigma$-meson spectral 
density.

Another major difference between the improved Hartree 
and Hartree-Fock approximation is that 
the 4-momentum dependent real parts are neglected in the latter one. However,
as mentioned, the differences in Fig.~\ref{rho_sigma_vgl_hfln} 
can be mainly traced back
to the the neglected terms of order $\sim 1/N$.
To quantify the difference arising from the 4-momentum dependent real parts,
I calculate the relative change between the spectral density 
with and without $[\Re\Sigma]_2$, averaged over
the energy-momentum grid, 
$\langle\mbox{diff}\rangle\equiv\langle|(\rho_{\mbox{w}}-\rho_{\mbox{wo}})
/\rho_{\mbox{w}}|\rangle_{k,\omega}$, where
$\rho_{\mbox{w}}$ is the standard spectral density given by
Eq.~(\ref{spectral_density_sigma_1_ln}), $\rho_{\mbox{wo}}$ 
the spectral density given by Eq.~(\ref{spectral_density_sigma_1_ln}) 
with $[\Re\Sigma]_2\equiv 0$, and $\langle\dots\rangle_{k,\omega}$ 
denotes the average over the energy-momentum grid. This mean value is
$\langle\mbox{diff}\rangle=5.61\pm 2.90\%$ for $T=160~{\rm MeV}$,
and $\langle\mbox{diff}\rangle=0.44\pm 0.01\%$ for $T=320~{\rm MeV}$.

%**********************************************************************
%*********Chapter IV***************************************************
\chapter{Conclusions \& Outlook}
In the following I briefly discuss the major results from the 
last three chapters and explain what is planned in the future.

\subsubsection*{Chapter II: The quark mass dependence of
the transition temperature \cite{Dumitru:2003cf}}
Three-colour QCD exhibits a (weakly) first-order deconfining phase
transition at a temperature $T_c/\surd\sigma\approx0.63$ in the limit
of infinitely heavy quarks ($\surd\sigma\approx0.425$~GeV denotes the
string tension at $T=0$ in this theory). Near $T_c$,
the screening mass for the fundamental Polyakov-loop $\ell$ drops
substantially~\cite{Kaczmarek:1999mm}, and so one might hope to capture the
physics of the phase transition using some effective Lagrangian for
$\ell$~\cite{Yaffe:1982qf,Pisarski:2002ji,Dumitru:2000in,Dumitru:2001bf,
Dumitru:2001xa,Scavenius:2001pa,Ogilvie:1999if,Meisinger:2001cq,Mocsy:2003qw,
Fukushima:2003fw}.

For finite quark masses, a term linear in $\ell$ appears which
breaks the $Z(3)$ center-symmetry explicitly. This reduces the
deconfinement temperature, with $\Delta T_c/T_c^*$ on the order of
the expectation value of the Polyakov-loop at $T_c^-$,
cf.\ Eq.~(\ref{DelTc}).

At some point then,
the line of first-order deconfinement phase transitions
ends~\cite{Gavin:1993yk,Green:1983sd,Meisinger:1995qr,
Alexandrou:1998wv,Brown:1990ev}. I have provided 
a quantitative estimate of this point, $m_\pi \simeq 4.2\;\surd\sigma
\approx 1.8$~GeV($\hat{=}\,m_q\approx 0.9$~GeV) and $T_c\simeq240$~MeV for
$N_f=3$ degenerate flavours, by matching the effective
Lagrangian for the Polyakov-loop to lattice data on
$T_c(m_\pi)$~\cite{Karsch:2000kv}. Assuming that $b_1\propto
N_f$~\cite{Alexandrou:1998wv} shifts ``D'' to $m_\pi \simeq 1.4$~GeV
for $N_f=2$ and to 0.8~GeV for $N_f=1$.

Going to even smaller quark (or pion) masses leaves a crossover
between the low-temperature and high-temperature regimes of QCD.
The dependence of the crossover temperature $T_c$
on the pion mass appears to be well described by a small explicit
breaking of the $Z(3)$ center symmetry, $b_1\sim\exp(-m_\pi)$, down to
$m_\pi/\surd\sigma\simeq1$, which is the smallest pion mass covered by
the lattice data of Ref.~\cite{Karsch:2000kv}. On the other hand, a
linear sigma model leads to a stronger dependence of $T_c$ on
$m_\pi$ than seen in the data. 

In turn, in the chiral limit, and for $N_f=3$ flavours, one
expects a first-order {\em chiral} phase
transition~\cite{Pisarski:1983ms,Gavin:1993yk,Brown:1990ev}. The linear sigma model should then
be an appropriate effective Lagrangian
for low-energy QCD~\cite{Pisarski:1983ms,Rajagopal:1992qz,Lenaghan:2000ey,Metzger:1993cu,Meyer-Ortmanns:1992pj,Goldberg:1983ju,Gavin:1993yk}. The first-order chiral
phase transition ends in a critical point ``C'' if either the mass of
the strange quark or those of all three quark flavours increase. Given
that the explicit symmetry breaking term for the Polyakov-loop remains
rather small when extrapolated to $m_\pi\to0$, that is $b_1\to0.2$,
I speculate that ``C'' might be rather close to the chiral
limit. Indeed, recent lattice estimates for $N_f=3$ place ``C'' at $m_\pi\simeq
290$~MeV~\cite{Karsch:2001nf} for standard staggered fermion action and
$N_t=4$ lattices; improved (p4) actions predict values as low as
$m_\pi\simeq67$~MeV~\cite{Karsch:2003va}.

Of course, the question arises why, for pion masses down to
$\simeq400$~MeV, the QCD crossover is described rather naturally by
a slight ``perturbation'' of the $m_\pi=\infty$ limit, in the form of
an explicit breaking of the global $Z(N_c)$ symmetry for the 
Polyakov-loop.
Physically, the reason is the flatness of the potential for $\ell$ in
the pure gauge theory at $T_c$, see e.g.\ the figures
in~\cite{Dumitru:2000in,Dumitru:2001bf,Scavenius:2002ru}, which causes the sharp drop of the screening mass
for $\ell$ near $T_c^+$~\cite{Kaczmarek:1999mm}. 
This is natural, given that finite-temperature expectation values of
Polyakov-loops at $N_c=3$ are close to those at
$N_c=\infty$~\cite{Dumitru:2003hp}, where the
potential at $T_c$ becomes entirely flat~\cite{Gross:1980he,Kogut:1981ez,Dumitru:2003hp}.
Hence, a rather small ``tilt'' of the
potential (due to explicit symmetry breaking) quickly washes out the
deconfining
phase transition of the pure gauge theory, and causes a significant
shift $\Delta T_c$ of the crossover temperature already for small $b_1$.
If so, then for $N_c\to\infty$, at the Gross-Witten
point, the endpoint ``D'' should be located at
$b_1=0$; the discontinuity for the Polyakov-loop at $T_c$,
which in a mean-field model for the pure gauge theory  is 1/2 at
$N_c=\infty$~\cite{Dumitru:2003hp,Gross:1980he,Kogut:1981ez},
then vanishes for arbitrarily
small explicit symmetry breaking. This has previously been noted by
Green and Karsch~\cite{Green:1983sd} within a mean-field model. If
confirmed by lattice Monte-Carlo studies, one might improve the
understanding of the degrees of freedom driving the QCD crossover for
pion masses above the chiral critical point ``C''.

\subsubsection*{Chapter III: The improved Hartree-Fock approximation
\cite{Roder:2005vt}}
In chapter III, I have studied the $O(N)$ linear sigma model 
at nonzero temperature within a self-consistent many-body resummation scheme.
This scheme extends the standard Hartree-Fock approximation by
including nonzero decay widths of the particles.
In the standard Hartree-Fock approximation, the self-energies of
the particles consist of tadpole diagrams which have no
imaginary part. Consequently, all particles are stable
quasi-particles. In order to obtain a nonzero decay width, 
one has to include diagrams in the self-energy, which have a
nonzero imaginary part corresponding to decay and, in a medium,
scattering processes.

In order to incorporate the nonzero decay width in a self-consistent
way, I apply the Cornwall-Jackiw-Tomboulis formalism. 
The standard Hartree-Fock approximation is obtained by
considering only double-bubble diagrams in the 2PI effective
action, leading to the (energy- and momentum-independent)
tadpole contributions in the 1PI self-energies.
In order to extend the Hartree-Fock approximation, I additionally
take into account diagrams of sunset topology in the 
2PI effective action. This has the
consequence that the 1PI self-energies obtain additional energy- and
momentum-dependent one-loop contributions which have a nonzero imaginary part.
The spectral densities of $\sigma$-mesons and pions are then
computed as solutions of a self-consistent set of Dyson-Schwinger equations 
for the $\sigma$-meson and pion two-point functions,
coupled to a fix-point equation for the chiral condensate. I
only take into account the imaginary parts of the new one-loop
contributions. I made sure that the spectral densities obey a standard
sum rule by adjusting their normalisation, if necessary.

%Results

I found that the temperature $T_\chi$ for chiral symmetry restoration 
is about $20\%$ smaller as compared to the Hartree-Fock approximation 
when including nonzero particle decay widths. 
My value $T_\chi \simeq 175$ MeV agrees reasonably well with 
lattice results \cite{Laermann:2003cv}. 
I computed the decay widths of $\sigma$-mesons and pions as
a function of temperature. The vacuum value for the $\sigma$-meson decay width
comes out to be in the experimentally observed range, without adjusting a 
parameter of the model. It stays approximately constant up
to temperatures $ \sim T_\chi$ and then decreases sharply with temperature.
The pion decay width grows from zero at $T=0$ to a value $\sim 100$ MeV at
$T \sim T_\chi$, and then also decreases with temperature.

I also investigated the spectral densities of $\sigma$-mesons and
pions as functions of energy and momentum for temperatures in the
range of $T=80$ to 320 MeV.
Below the chiral phase transition, the spectral density of the
$\sigma$-meson is broad, due to the possible decay into two pions.
It develops a peak at the quasi-particle mass shell above
the chiral phase transition, when this decay channel is closed.
On the other hand, the spectral density of the pion always exhibits a 
distinct peak at the quasi-particle mass shell. The width of this peak is 
due to scattering off $\sigma$-mesons in the medium.
Above the chiral phase transition, the spectral densities 
of $\sigma$-mesons and pions become degenerate in shape.

\subsubsection*{Chapter IV: The improved Hartree approximation
\cite{Roder:2005qy}}
In chapter IV, I systematically improved the standard Hartree
(or large-$N$)
approximation of the $O(N)$ linear sigma model by taking into
account, additionally to the
double-bubble diagrams, the sunset diagrams 
in the effective potential of the CJT formalism. 
This leads to 4-momentum dependent real and imaginary parts 
of the Dyson-Schwinger equations. In contrast
to chapter III, I didn't neglect the 4-momentum dependent real parts,
to study their influences to the results.
I solve these and the equation of the condensate 
selfconsistently in the chiral limit ($m_\pi=0$), and
in the case with explicitly broken chiral symmetry 
($m_\pi\neq 0$), to get the decay width
and the spectral density of the $\sigma$-meson. 

%real parts
First, I presented the results for the real parts
of the Dyson-Schwinger equations and the condensate equation.
The 4-momentum independent parts, i.e., the effective
masses and the chiral condensate exhibit a crossover transition
in the case with explicit chiral symmetry breaking and a second-order phase
transition in the chiral limit, which agrees with the predictions 
made by Pisarski and Wilczek \cite{Pisarski:1984ms}. 
I found, that the 4-momentum dependent real part of the self-energy
is rather small compared to the (squared) 4-momentum independent
effective masses. It is larger in the chiral limit, simply because
$[\Re]_2\sim\lambda^2\sim(m_\sigma^2-m_\pi^2)^2$ is 
maximal for $m_\pi^2=0$.

%imaginary part
The decay width shows qualitatively the same behaviour in all
cases. It is a decreasing function with temperature and becomes 
(approximatively) zero in the high-temperature regime.
Nevertheless, quantitatively it
depends strongly on the choice of the vacuum mass of the $\sigma$-meson,
by reason that $\Gamma_\sigma\sim\Im\,\Sigma\sim \lambda^2\sim m_\sigma^4$. 
Also, the decay widths are larger in the chiral limit, due to the fact
that $\Gamma_\sigma\sim\lambda^2\sim(m_\sigma^2-m_\pi^2)^2$.

%spectral density
In the low-temperature regime, the spectral density of the $\sigma$-meson
is a very broad function in energy 
(and becomes even broader for larger vacuum mass $m_\sigma$)
due to the $\sigma\to 2\pi$ decay. In the
high-temperature regime, the spectral density becomes more and more
a delta function. Remarkable is the fact
that $\rho_\sigma$ becomes zero in a certain energy interval,
in the case with explicit chiral
symmetry breaking but not in the chiral limit. This can be traced
back to the threshold energy for $\sigma\rightarrow 2\pi$, which is
$2m_\pi$, and therefore
zero in the chiral limit but not in the other case. For $\omega<k$ 
the $\sigma$-meson is Landau-damped. The influence of the 4-momentum
dependent
real part of the self-energy to the spectral density is rather small
and does not change the results qualitatively.

\newpage
\subsection*{Outlook}

The present studies can be continued along several lines:

\begin{itemize}

%new degrees of freedom
\item Throughout this thesis I used the linear
$\sigma$ model with $O(4)$ symmetry, which is
isomorphic to the linear $\sigma$ model with  
$SU(2)\times SU(2)$ symmetry. Both models contains only mesons
composed from the lightest two quarks (the up and down quarks), 
but the $O(4)$ model is the limit of the 
$SU(2)\times SU(2)$ model with maximally broken $U(1)_A$ symmetry, which
leads to the fact that the $\eta$ and the $a_0$ mesons
become infinitely heavy, and hence cannot be describted by
this theory, cf. \cite{Roder:2003uz}.
It would be exciting to study the influence of nonzero decay
width effects also for these mesons, and additionally 
for mesons with strange and maybe also charm degrees of freedom.
The scalar (s) and pseudoscalar (ps) mesons contained in the linear $\sigma$
models are summarized in the following table:

\begin{tabular}{|c|c|c|c|c|} \hline
&$O(4)$ & $SU(2)_r\times SU(2)_\ell$ 
& $SU(3)_r\times SU(3)_\ell$ & $SU(4)_r\times SU(4)_\ell$ \\\hline
s&$\sigma$ & $a_0^\pm,a_0^0$ &
    $\kappa^\pm,\kappa^0,\bar\kappa^0,f_0$ &
    $D_0^\pm,D_0^0,\bar D_0^0,D_{s,0}^\pm,\chi_{c,0}$\\\hline
ps&$\pi^\pm,\pi^0$ & $\eta$   &
     $K^\pm,K^0,\bar K^0,\eta'$ & $D^\pm,D^0,\bar D^0,D_s^\pm,\eta_c$\\\hline
\end{tabular}

\item In chapter II, the chiral critical point 
is at $m_q=0$, because only the 2-flavor
case is discussed. Extending the model by the strange degree of freedom 
would allow us to give an estimate for the position of this
point in the whole quark mass plane, discussed in Sec. 1.3, 
cf. Fig. \ref{s_ud_plane}. 

\item So far only the scalar and pseudoscalar mesons
are taking into account. The inclusion of baryonic degrees of freedom
into a framework, similar to that discussed in Chapters III or IV,
is still under investigation \cite{Beckmann:2005nm}.
Also, the inclusion of vector mesons
\cite{Ruppert:2004yg,strueberdipl} is under
investigation, and is of particular importance, since
in-medium changes in the spectral properties of vector mesons 
are reflected in the dilepton spectrum \cite{Ruppert:2004yg}
which, in turn, is  experimentally observable in heavy-ion collisions
at GSI-SIS, CERN-SPS and BNL-RHIC energies.

%finite mu
\item As disussed, a major advantage of effective models
over lattice QCD calculations is that no problems with a 
nonzero chemical potential occur 
\cite{VicenteVacas:2002se,Cabrera:2005wz}. Including the chemical potential in 
the framework of the improved Hartree or Hartree-Fock 
approximation would us allow to study, e.g.,
the influence of nonzero decay width effects on the chiral critical
endpoint in the $T-\mu-$plane shown in figure \ref{T_mu_plane}
(cf. e.g. \cite{Scavenius:2000qd}).

%renorm
\item Up to now, I neglected the ultraviolet divergent 
vacuum parts of the integrals, i.e., I just used trivial
renormalization. One should check what happens if one
use a ``real'' renormalization scheme 
(e.g. the cut-off or counter-term scheme,
cf. \cite{Lenaghan:1999si}).

%re_2 HF
\item The influence of the 4-momentum dependent real
parts of the self-energy in the
improved Hartree approximation is very small, nevertheless,
one should check its influence also in the full Hartree-Fock approximation.

\end{itemize}

%**********************************************************************
%*********Appendix*****************************************************
\begin{appendix}
\chapter{The calculation of the diagrams}
\section{In the improved Hartree-Fock approximation}
\begin{figure}
\includegraphics[height=3.5cm]{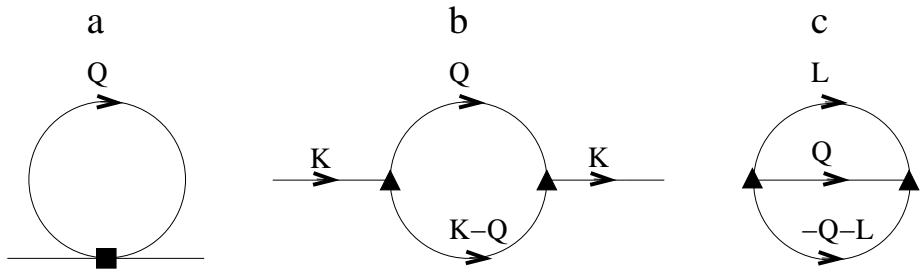}
\caption[The general topology of the tadpole diagram, the
cut sunset diagram, and the sunset diagram.]
{The general topology of the tadpole diagram a, the
cut sunset diagram b, and the sunset diagram c.}
\label{diagrams}
\end{figure}

In this Appendix I discuss the calculation of the three types of
diagrams (see Fig.~\ref{diagrams}) contributing to the self-energies and the 
condensate equation. To evaluate them explicitly, I use 
standard techniques of 
thermal field theory, see for example 
Refs.~\cite{das,kapusta,leBellac,Landshoff:1997jv,Landsman:1986uw,vanWeert}.

\subsection*{The tadpole diagram}\label{appendix1a}
In the following I discuss the calculation of the tadpole diagram
in Fig.~\ref{diagrams} a as a functional of the spectral density. 
In the imaginary-time formalism it is given by
\bea
{\cal T}\equiv T\sum_n \int \frac{d^3q}{(2\pi)^3} \Delta(-i\omega_n,{\bf q})
\eea
where $T$ is the temperature, and $\Delta$ is either the $\sigma$-meson
or pion propagator. 

The first step is to perform the sum over the Matsubara frequencies.
To this aim I introduce the mixed representation of the propagator
\be\label{mixed_representation}
\Delta(-i\omega_n,{\bf q})
=\int_0^{1/T} d\tau\exp(-i\omega_n\tau)\Delta(\tau,{\bf q}),
\ee
with
\bea\label{def_spectral_density}
\Delta(\tau,{\bf q})=\int_{-\infty}^{\infty}
\frac{d \omega}{2\pi}\, [\Theta(\tau)+f(\omega)]\,\rho(\omega,{\bf q})
\exp(-\omega\tau)\,\,,
\eea
where $f(\omega)=[\exp(\omega/T)-1]^{-1}$ is the Bose-Einstein distribution
function. With the identity
\be\label{delta_representation}
T\sum_n\exp(-i\omega_n\tau)=\sum_{m=-\infty}^{\infty}
\delta\left(\tau-\frac{m}{T}\right)\; ,
\ee
the Matsubara sum can be performed analytically
\bea
{\cal T}&=&T\sum_{n} 
\int \frac{d^3q}{(2\pi)^3}
\int_0^{1/T} d\tau \exp(-i\omega_n\tau)
\Delta(\tau,{\bf q}) \nn\\
&=&\int \frac{d^3q}{(2\pi)^3}
\int_0^{1/T} d\tau 
\Delta(\tau,{\bf q})\left[\delta({\tau})+ \delta\left(\tau - \frac{1}{T}
\right) \right] \nn \\
& = & \int \frac{d^3q}{(2\pi)^3} \frac{1}{2} \left[
\Delta(0,{\bf q}) + \Delta\left(\frac{1}{T},{\bf q} \right) \right]
\nn \\
& = & \int \frac{d^3q}{(2\pi)^3} \Delta(0,{\bf q})\,\,,
\eea
where use has been made of the KMS condition $\Delta(\tau,{\bf q})
\equiv \Delta(\tau- 1/T, {\bf q})$. With 
Eq.~(\ref{def_spectral_density}) I finally have
\be\label{rm_T_1}
{\cal T}=
\int_{-\infty}^{\infty} \frac{d \omega}{2 \pi}
\int \frac{d^3q}{(2\pi)^3}
\left[\frac 12+f(\omega)\right]\rho(\omega,{\bf q})\,\,.
\ee
Due to isotropy of space,
the spectral density of a scalar particle
cannot depend on the direction of ${\bf q}$, thus
the angular integration can be carried out:
\bea
{\cal T}=4\pi
\int_{-\infty}^{\infty} \frac{d \omega}{2 \pi}
\int_{0}^{\infty}\frac{d\,q}{(2\pi)^3}\,q^2
\left[\frac 12+f(\omega)\right]\rho(\omega,q)\,\,.
\eea
Using the fact that the spectral density for bosonic
degrees of freedom is an odd function of the energy, 
$\rho(\omega)=-\rho(-\omega)$,
\bea
{\cal T}&=&4\pi
\int_{0}^{\infty} \frac{d \omega}{2 \pi}\,
\frac{d q}{(2\pi)^3}\,q^2
[f(\omega)-f(-\omega)]\rho(\omega,q)\nn\\
&=&4\pi
\int_{0}^{\infty} \frac{d \omega}{2 \pi}\,
\frac{d q}{(2\pi)^3}\,q^2\,
[1+2f(\omega)]\,\rho(\omega,q)\,\,.
\eea
Subtracting the divergent vacuum contribution I obtain:
\bea\label{general_eq_tadpole}
{\cal T}=\frac{4}{(2\pi)^3}
\int_{0}^{\infty} d \omega\, d q\,q^2\,
f(\omega)\, \rho(\omega,q)\,\,.
\eea

\subsection*{The cut sunset diagram}\label{appendix1b}
In this section I calculate the imaginary part of the
cut sunset diagram of Fig.~\ref{diagrams} b. 
In the imaginary-time formalism 
the cut sunset diagram is given by
\bea
{\cal C}(-i\omega_m,{\bf k})&\equiv&
\int_Q \Delta_1 (K-Q)\, \Delta_2(Q)\nn\\
&=&T\sum_n \int \frac{d^3q}{(2\pi)^3}
\Delta_1(- i(\omega_m-\omega_n),\mathbf{k-q})
\Delta_2(-i\omega_n,{\bf q})\,\,,
\eea
where $\Delta_1,\Delta_2$ are the propagators of $\sigma$-mesons and/or
pions. The major difference
between the tadpole diagram, discussed in the last section, and this diagram
is the fact that it explicitly depends on the external four-momentum, 
$K^\mu\equiv(-i\omega_m,{\bf k})$.

Analogously to the last section one
introduces the mixed representation [see Eq.~(\ref{mixed_representation})] 
of the propagators $\Delta_1$ and $\Delta_2$ 
\bea
{\cal C}(-i\omega_m,{\bf k})
=T\sum_n \int \frac{d^3q}{(2\pi)^3}
&&\int_0^{1/T} d \tau_1\exp[-i(\omega_m-\omega_n)\tau_1]
\Delta_1(\tau_1,\mathbf{k-q})\nn\\
\times&&\int_0^{1/T} d \tau_2\exp(-i\omega_n\tau_2)
\Delta_2(\tau_2,{\bf q})\,\,.
\eea
The summation over the Matsubara frequencies leads to 
a delta function 
[cf.\ Eq.~(\ref{delta_representation})]
\bea
{\cal C}(-i\omega_m,{\bf k})
=\int\frac{d^3q}{(2\pi)^3}
\int_0^{1/T} d \tau\exp(-i\omega_m\tau)
\Delta_1(\tau,\mathbf{k-q})\Delta_2(\tau,{\bf q})
\,\,.
\eea
Introduction of the spectral densities
[see Eq.~(\ref{def_spectral_density})] for both
propagators in the mixed representation 
leads to
\bea
{\cal C}(-i\omega_m,{\bf k})
&=&\int \frac{d^3q}{(2\pi)^3}
\int_0^{1/T} d \tau\exp(-i\omega_m\tau)\nn\\
&\times&\int_{-\infty}^{\infty}
\frac{d \omega_1}{2\pi}[1+f(\omega_1)]\rho_1(\omega_1,\mathbf{k-q})
\exp(-\omega_1\tau)\nn\\
&\times&\int_{-\infty}^{\infty}
\frac{d \omega_2}{2\pi}[1+f(\omega_2)]\rho_2(\omega_2,{\bf q})
\exp(-\omega_2\tau)\;.
\eea
The integration over $\tau$ can be performed analytically. 
Employing the identity
\bea
\left[\exp\left(-\frac{\omega_1+\omega_2}{T}\right)-1\right]
\,[1+f(\omega_1)][1+f(\omega_2)]
=- 1 - f(\omega_1)- f(\omega_2)\;,
\eea
the expression can be rewritten as 
\be
{\cal C}(-i \omega_m,{\bf k})=
\int_{-\infty}^{\infty} \frac{d \omega_1}{2 \pi} \,\frac{d \omega_2}{2 \pi} 
\frac{1+f(\omega_1)+f(\omega_2)}
{i\omega_{\rm m}+\omega_1+\omega_2}
\int \frac{d^3 q }{(2 \pi)^3}\, \rho_1(\omega_1,\mathbf{k-q})
\rho_2(\omega_2,{\bf q})\;.
\ee
Using the fact that the 
spectral densities do not depend on the direction of momentum and
\bea\label{abs_ident_2}
\int \frac{d^3 x}{(2\pi)^3}
f(\left|\mathbf{y-x}\right|)g(x)
&=&\frac{2\pi}{y}\frac{1}{(2\pi)^3}
\int_0^{\infty} d x_1 \,  x_1 \, d x_2\, x_2\,
f(x_1)g(x_2)\nn\\
&\times& \Theta\left(|x_1-x_2|
\leq y\leq x_1+x_2\right)\,\,,
\eea
one can perform one angular integration, yielding
\bea\label{cut_sunset}
{\cal C}(-i \omega_m,{\bf k})&=&
\int_{-\infty}^{\infty} \frac{d \omega_1}{2 \pi} \,
\frac{d \omega_2}{2 \pi} \,
 \frac{1+f(\omega_1)+f(\omega_2)}{i\omega_{\rm m}+\omega_1+\omega_2}\nn \\
&\times&
\frac{2\pi}{k} \frac{1}{(2\pi)^3} 
\int_0^{\infty}  d q_1\, q_1 \, d q_2 \, q_2\,
\rho_1(\omega_1,q_1)\rho_2(\omega_2,q_2)\nn\\
&\times& \Theta\left(|q_1-q_2|
\leq k\leq q_1+q_2\right)\,\,.
\eea
The imaginary part of the retarded cut sunset diagram
can be extracted using analytical continuation, 
$-i \omega_m \rightarrow \omega + i \epsilon$, and
the Dirac identity $\Im 1/(x+i\epsilon)=-\pi\delta(x)$,
\bea\label{general_im}
\Im \,{\cal C}(\omega,{\bf k})&=&
\frac{1}{2(2\pi)^3}\frac{1}{ k}
\int_{-\infty}^{\infty} d \omega_1\, d \omega_2
[1+f(\omega_1)+f(\omega_2)]\, \delta(\omega-\omega_1-\omega_2)\\
&\times&
\int_0^{\infty}  d q_1 \, q_1 \, d q_2 \, q_2\, 
\Theta\left(|q_1-q_2|\leq k\leq q_1+q_2\right)\,
\rho_1(\omega_1,q_1)\, \rho_2(\omega_2,q_2).\nn
\eea
This integral is finite and therefore does not require renormalization.

\subsection*{The sunset diagram}\label{appendix1c}

In the imaginary-time formalism the sunset diagram shown in 
Fig.~\ref{diagrams} c is given by the following expression:
\bea
{\cal S}&=&T^2\sum_{n,m} 
\int \frac{d^3l}{(2\pi)^3}\, \frac{d^3q}{(2\pi)^3}
\Delta_1(-i\omega_m,{\bf l})
\Delta_2(-i\omega_n,{\bf q})
\Delta_3(-i(-\omega_n-\omega_m)
,-\mathbf{q-l})\;.\nn\\
\eea
I introduce the mixed
representation, see Eq.~(\ref{mixed_representation}), for the three
propagators $\Delta_1,\Delta_2$, and $\Delta_3$. One can perform
the Matsubara sums employing Eq.~(\ref{delta_representation}). 
In the mixed representation one introduces the spectral densities 
$\rho_1,\rho_2$, and $\rho_3$
in order to perform the $\tau$ integration analytically.
Employing the identity 
\bea
&&\left[\exp\left(-\frac{\omega_1+\omega_2+\omega_3}{T} \right) -1\right]
\left[1+f(\omega_1)\right]
\left[1+f(\omega_2)\right]\left[1+f(\omega_3)\right]\nn\\
&=&- 1-f(\omega_1)-f(\omega_2)-f(\omega_3)
-f(\omega_1)f(\omega_2)-f(\omega_1)f(\omega_3)-f(\omega_2)f(\omega_3),
\hspace{1.2cm}
\eea
for the Bose-Einstein distribution functions I obtain 
\bea 
{\cal S}&=&
\int_{-\infty}^{\infty} \frac{d \omega_1}{2 \pi}\,
\frac{d \omega_2}{2 \pi} \,
\frac{d \omega_3}{2 \pi}\nn\\
&\times&\, \frac{1+f(\omega_1)+f(\omega_2)+f(\omega_3) 
+f(\omega_1)f(\omega_2)
+f(\omega_1)f(\omega_3)
+f(\omega_2)f(\omega_3)}
{\omega_1+\omega_2+\omega_3}  \nn \\
&\times& 
\int \frac{d^3 l }{(2 \pi)^3} \, \frac{\d^3 q }{(2 \pi)^3}
\, \rho_1(\omega_1,{\bf l}) \rho_2(\omega_2,{\bf q})
\rho_3(\omega_3,\mathbf{-q-l})\,\,.
\eea
Using the fact that the spectral densities do not depend on the
direction of momentum, the angular integrals
can be carried out with the result
\bea 
{\cal S}&=&\frac{2}{(2\pi)^7}
\int_{-\infty}^{\infty} d \omega_1\, d \omega_2\, d \omega_3
\int_0^{\infty} dq_1\, q_1\, dq_2\, q_2\, dq_3\, q_3\,
\Theta\left(|q_1-q_2|\leq q_3\leq q_1+q_2\right)\nn\\
&\times&
\frac{1+f(\omega_1)+f(\omega_2)+f(\omega_3) 
+f(\omega_1)f(\omega_2)+f(\omega_1)f(\omega_3)+f(\omega_2)f(\omega_3)
}{\omega_1+\omega_2+\omega_3}  \nn \\
&\times& 
\rho_1(\omega_1,q_1)
\rho_2(\omega_2,q_2)
\rho_3(\omega_3,q_3)
\,\,.
\eea
Using the antisymmetry of the spectral densities, I obtain 
\bea 
{\cal S}&=&\frac{4}{(2\pi)^7}
\int_{0}^{\infty} d \omega_1\,d \omega_2\,d \omega_3
\, dq_1\, q_1\, dq_2\, q_2\,dq_3\, q_3\,
\Theta\left(|q_1-q_2|\leq q_3\leq q_1+q_2\right)\nn\\
&\times&\left[\frac{f(\omega_1)f(\omega_2)+f(\omega_1)f(\omega_3)
          +f(\omega_3)f(\omega_2)+f(\omega_1)+f(\omega_2)
          +f(\omega_3)+1}{\omega_1+\omega_2+\omega_3}\right.\nn\\
&&\quad+\frac{f(\omega_1)f(\omega_3)
             -f(\omega_1)f(\omega_2)
             +f(\omega_2)f(\omega_3)
             +f(\omega_3)}{\omega_1+\omega_2-\omega_3}\nn\\
&&\quad+\frac{f(\omega_1)f(\omega_2)
             -f(\omega_1)f(\omega_3)
             +f(\omega_2)f(\omega_3)
             +f(\omega_2)}{\omega_1-\omega_2+\omega_3}\nn\\
&&\quad+ \left.
        \frac{f(\omega_1)f(\omega_2)
             +f(\omega_1)f(\omega_3)
             -f(\omega_2)f(\omega_3)
             +f(\omega_1)}{-\omega_1+\omega_2+\omega_3}\right]\nn\\
&\times&
\rho_1(\omega_1,q_1) \rho_2(\omega_2,q_2)\rho_3(\omega_3,q_3)
\,\,.
\eea
\newpage
Finally, I discard the ultraviolet-divergent parts, and relabel the
integration variables
\bea 
{\cal S}&=&\frac{4}{(2\pi)^7}
\int_{0}^{\infty} d \omega_1\,d \omega_2\,d \omega_3
\, dq_1\, q_1\, dq_2\, q_2\,dq_3\, q_3\,
\Theta\left(|q_1-q_2|\leq q_3\leq q_1+q_2\right)\nn\\
&\times&
f(\omega_1) f(\omega_2) \left(
\frac{1}{\omega_1+\omega_2+\omega_3}
+ \frac{1}{-\omega_1+\omega_2+\omega_3}
+ \frac{1}{\omega_1-\omega_2+\omega_3}
+ \frac{1}{-\omega_1-\omega_2+\omega_3} \right) \nn\\
&\times&[
\rho_1(\omega_1,q_1) \rho_2(\omega_2,q_2)
\rho_3(\omega_3,q_3)
+\rho_1(\omega_1,q_1)
\rho_2(\omega_3,q_2)
\rho_3(\omega_2,q_3)\nn\\
&&+\vspace{0.5cm}\rho_1(\omega_2,q_1)
\rho_2(\omega_3,q_2)
\rho_3(\omega_1,q_3)]\,\,. \label{sunset}
\eea

\section{In the improved Hartree approximation}
In contrast to the improved Hartree-Fock approximation, the spectral
density of the pion, $\rho_\pi$, in the improved Hartree approximation
is just a delta-function, because the imaginary part of the
pion self-energy if of order $\sim 1/N$ and thus neglected.
Therefore the equations for the tadpole and the cut sunset diagrams,
Eqs.~(\ref{general_eq_tadpole}) and
(\ref{cut_sunset}) simplify, but one has to calculate additionally
the real part of the cut sunset diagram. The full
sunset diagram, discussed in \ref{appendix1c}, is of order $\sim 1/N$ and 
is therefore neglected in the large-$N$ limit. 

\subsection*{The tadpole diagram}
The Dyson-Schwinger and the condensate equations of the improved 
Hartree approximation contain {\it exclusively} the spectral 
density of the pion. Therefore one has to solve equations of  the form 
\bea
{\cal T}=\frac{4}{(2\pi)^3}
\int_{0}^{\infty} d \omega\, d q\,q^2\,
f(\omega)\, \rho_\pi(\omega,q)\,\,,
\eea
with different combinatorial factors in front. (Note that this is the
version where the divergent term is already neglected, i.e.,
the equation in the trivial renormalisation scheme.) Using the fact that 
the spectral density of the pion is just a delta function, 
$\rho_\pi (\omega, k) = \pi/\omega_\pi (k)
 \{ \delta [ \omega - \omega_\pi(k)] 
- \delta [ \omega + \omega_\pi(k)]\}$, one gets 
\be\label{ap2_tp}
{\cal T}=\frac{1}{2\pi^2}
\int_0^\infty d q\,q^2[\omega_\pi (q)]^{-1}
f[\omega_\pi(q)].
\ee

\subsection*{The cut sunset diagram}

In this section I derive the equations for the 4-momentum dependent
imaginary and real part of the self-energy (arising from the sunset 
diagram). As discussed in Sec. A.1
this diagram can be expressed as a function of the spectral density
$\rho_\pi$ in the imaginary time formalism
\bea\label{cut_sunset_1_ln}
\Sigma(-i\omega_m,k)&=&
\left(\frac{4\lambda \sigma}{N}\right)^2
\frac{2N}{(2\pi)^4}\frac{1}{k}
\int_{-\infty}^{\infty} d \omega_1\,d \omega_2
\frac{1+f(\omega_1)+f(\omega_2)}{i\omega_m+\omega_1+\omega_2}\\
&\times& 
\int_0^{\infty}   d q_1 \, q_1 \,d q_2 \, q_2\;
\Theta\left(|q_1-q_2|\leq k\leq q_1+q_2\right)
 \rho_\pi(\omega_1,q_1)\rho_\pi(\omega_2,q_2),\nn
\eea
where $f(\omega)=1/[\exp(\omega/T)-1]$ is the Bose-Einstein 
distribution function, and $\omega_m=2\pi m T$ are the Matsubara
frequencies. To extract the imaginary part of the retarded self-energy, 
one uses the Dirac identity, $\Im 1/(x+i\epsilon)=-\pi\delta(x)$, and
analytic continuation, $-i\omega_m\rightarrow \omega+i\epsilon$,
\bea
\Im\,\Sigma(\omega,k)&=&
\left(\frac{4\lambda \sigma}{N}\right)^2
\frac{N}{2(2\pi)^3}\frac{1}{k}
\int_{-\infty}^{\infty} d \omega_1\,d \omega_2
[1+f(\omega_1)+f(\omega_2)]\,\delta(\omega-\omega_1-\omega_2)\nn \\
&\times &
\int_0^{\infty}   d q_1 \, q_1 \,d q_2 \, q_2\;
\Theta\left(|q_1-q_2|\leq k\leq q_1+q_2\right)
\rho_\pi(\omega_1,q_1)\rho_\pi(\omega_2,q_2).\hspace{1.2cm}
\eea
As discussed in the improved Hartree approximation 
of the $O(N)$ model the spectral density of the pion 
is just a delta function (\ref{spectral_density_pion_ln}), 
\bea
\Im\,\Sigma(\omega,k)&=&
\left(\frac{4\lambda \sigma}{N}\right)^2
\frac{N}{8\pi}\frac{1}{k}
\int_{-\infty}^{\infty} d \omega_1\, d \omega_2
[1+f(\omega_1)+f(\omega_2)]\,\delta(\omega-\omega_1-\omega_2)\nn \\
&\times& 
\int_0^{\infty}   d q_1 \, q_1 \,d q_2 \, q_2\;
\Theta\left(|q_1-q_2|\leq k\leq q_1+q_2\right)
[\omega_\pi (q_1)\omega_\pi (q_2)]^{-1}\nn\\
&\times&
 \left\{ 
\delta [ \omega_1 - \omega_\pi(q_1)] 
\,\delta [ \omega_2 - \omega_\pi(q_2)]+
\delta [ \omega_1 + \omega_\pi(q_1)]
\,\delta [ \omega_2 + \omega_\pi(q_2)]\right.\nn\\
&&-\left.
\delta [ \omega_1 - \omega_\pi(q_1)] 
\,\delta [ \omega_2 + \omega_\pi(q_2)]-
\delta [ \omega_1 + \omega_\pi(q_1)] 
\,\delta [ \omega_2 - \omega_\pi(q_2)]
 \right\},\nn\\
\eea
where $\omega_\pi(q)=\sqrt{q^2+M_\pi^2}$ is the quasiparticle energy of the
pion. To simplify this expression one uses the 
$\delta(\omega-\omega_1-\omega_2)$ function to
 carry out the $\omega_2$-integration, 
\bea
\Im\,\Sigma(\omega,k)
&=&
\left(\frac{4\lambda \sigma}{N}\right)^2
\frac{N}{8\pi}\frac{1}{k}
\int_{-\infty}^{\infty} d \omega_1
[1+f(\omega_1)+f(\omega-\omega_1)]\nn\\
&\times&\int_0^{\infty}   d q_1 \, q_1 \, d q_2 \, q_2\;
\Theta\left(|q_1-q_2|\leq k\leq q_1+q_2\right)
[\omega_\pi (q_1)\omega_\pi (q_2)]^{-1}\nn\\
&\times& 
 \left\{ 
\delta [ \omega_1 - \omega_\pi(q_1)] 
\,\delta [ \omega-\omega_1 - \omega_\pi(q_2)]\right.\nn\\
&&\left.+
\delta [ \omega_1 + \omega_\pi(q_1)]
\,\delta [ \omega-\omega_1 + \omega_\pi(q_2)]\right.\nn\\
&&-\delta [ \omega_1 - \omega_\pi(q_1)] 
\,\delta [ \omega-\omega_1 + \omega_\pi(q_2)]\nn\\
&&\left.-
\delta [ \omega_1 + \omega_\pi(q_1)] 
\,\delta [ \omega-\omega_1 - \omega_\pi(q_2)] \right\}\;,
\eea
and the $\omega_1$-integration with the help of the
$\delta[\omega_1\pm \omega_\pi(q_1)]$-functions
\bea\label{cut_sunset_im_1_ln}
\Im\,\Sigma(\omega,k)
&=&\left(\frac{4\lambda \sigma}{N}\right)^2
\frac{N}{8\pi}\frac{1}{k}
\int_0^{\infty}   d q_1 \, q_1 \, d q_2 \, q_2\;\\
&\times&
\Theta\left(|q_1-q_2|\leq k\leq q_1+q_2\right)
[\omega_\pi (q_1)\omega_\pi (q_2)]^{-1}\nn\\
&\times&(\,
\{1+f[\omega_\pi(q_1)]+f[\omega-\omega_\pi(q_1)]\}
\,\delta [ \omega-\omega_\pi(q_1) - \omega_\pi(q_2)]\nn\\
&&+\{1+f[-\omega_\pi(q_1)]+f[\omega+\omega_\pi(q_1)]\}
\,\delta [ \omega+\omega_\pi(q_1) + \omega_\pi(q_2)]\nn\\
&&-\{1+f[\omega_\pi(q_1)]+f[\omega-\omega_\pi(q_1)]\}
\,\delta [ \omega-\omega_\pi(q_1) + \omega_\pi(q_2)]\nn\\
&&-\{1+f[-\omega_\pi(q_1)]+f[\omega+\omega_\pi(q_1)]\}
\,\delta [ \omega+\omega_\pi(q_1) - \omega_\pi(q_2)])\;.\nn
\eea
The $\delta[\omega+\omega_\pi(q_1)+\omega_\pi(q_2)]$ function has no support,
because $\omega>0$ and $\omega_\pi(q)>0$. Note that evaluating the 
$\delta [ \omega-\omega_\pi(q_1) + \omega_\pi(q_2)]$
and the $\delta [ \omega+\omega_\pi(q_1) - \omega_\pi(q_2)]$ function,
the latter two terms cancel each other. The remaining
delta function can be transformed to a delta function in momentum-space,
\bea
\delta [ \omega-\omega_\pi(q_1) - \omega_\pi(q_2)]=
\left|\frac{\omega_\pi(q_0)}{q_0}\right|
[\delta(q_1-q_0)+\delta(q_1+q_0)]
\eea
where
\bea
q_0\equiv\sqrt{[\omega-\omega_\pi(q_2)]^2-M_\pi^2}
\eea
is the root of the argument of the delta function. Note that 
in Eq.~(\ref{cut_sunset_im_1_ln}) $q_1>0$ and therefore there is no support 
of the $\delta(q_1+q_0)$ function. Carrying out the $q_1$-integration, 
using $\omega-\omega_\pi(q_0)=\omega_\pi(q_1)$, and relabelling
$q\equiv q_2$, one gets
\bea
\Im\,\Sigma(\omega,k)
&=&\left(\frac{4\lambda \sigma}{N}\right)^2
\frac{N}{8\pi}\frac{1}{k}
\int_0^{\infty}  \, d q \;q\,[\omega_\pi (q)]^{-1}
\Theta\left(|q_0-q|\leq k\leq q_0+q\right)
\nn\\&\times&
\{1+f[\omega_\pi(q_0)]+f[\omega_\pi(q)]\}.
\eea
To calculate the limit $k\to 0$, one starts best with
Eq.~(\ref{cut_sunset_im_1_ln}), uses the following transformation
\bea
1+f[\omega_\pi(q_1)]+f[\omega-\omega_\pi(q_1)]
&=&\left[1-\exp\left(-\frac{\omega}{T}\right)\right]\\
&\times&
\left\{1+f[\omega_\pi(q_1)]\right\}
\left\{1+f[\omega-\omega_\pi(q_1)]\right\},\nn
\eea
the identity 
\bea\label{theta_delta_ln}
\lim_{k\to 0}\frac{\Theta\left(|q_1-q_2|\leq k\leq q_1+q_2\right)}{k}
=2\,\delta(q_1-q_2)
\eea
(to perform the $q_1$-integration), and relabels $q\equiv q_2$, to get
\bea
\Im\,\Sigma(\omega,0)
&=&
\left(\frac{4\lambda \sigma}{N}\right)^2
\frac{N}{4\pi}\left[1-\exp\left(-\frac{\omega}{T}\right)\right]
\int_0^{\infty}   d q \, q^2\,\delta [\omega-2\omega_\pi(q)]
[\omega^2_\pi (q)]^{-1}\nn\\
&\times&\{1+f[\omega_\pi(q)]\}\{1+f[\omega-\omega_\pi(q)]\}.
\eea
As explained in \cite{Rischke:1998qy} this expression can be 
calculated analytically, using the Spence integral,
\be\label{im2k0_ln}
\Im\,\Sigma(\omega,0)=
\left(\frac{4\lambda \sigma}{N}\right)^2 N
\frac{1}{8\pi}\sqrt{1-\frac{4M_\pi^2}{\omega^2}}\coth\frac{\omega}{4T}\;.
\ee
Note that in \cite{Rischke:1998qy} the calculation is performed for the
special case $\omega=m_\sigma$, but this can be generalised without
further problems.

To calculate the real part of Eq.(\ref{cut_sunset_1_ln}),
one has to integrate over the principal value of 
$1/(\omega_1+\omega_2-\omega)$, denoted by $\mbox{P}\int\dots$,
\bea
[\Re\,\Sigma(\omega,k)]_2&=&
\left(\frac{4\lambda \sigma}{N}\right)^2
\frac{2N}{(2\pi)^4}\frac{1}{k}
\mbox{P}\int_{-\infty}^{\infty} d \omega_1\,d \omega_2
\frac{1+f(\omega_1)+f(\omega_2)}
{\omega_1+\omega_2-\omega}\\
&\times& 
\int_0^{\infty}   d q_1 \, q_1 \,d q_2 \, q_2\;
\Theta\left(|q_1-q_2|\leq k\leq q_1+q_2\right)
\rho_\pi(\omega_1,q_1)\rho_\pi(\omega_2,q_2).\nn
\eea
Again, one uses the fact that the spectral density 
of the pion is just a delta function to perform the
$\omega_1$- and the $\omega_2$-integration, and employs trivial
renormalisation, i.e., neglects the divergent parts. This leads to
\bea
[\Re\,\Sigma(\omega,k)]_2&=&
\left(\frac{4\lambda \sigma}{N}\right)^2
\frac{N}{8\pi^2}\frac{1}{k}
\mbox{P}\int_0^{\infty}   d q_1 \, q_1 \, d q_2 \, q_2\\
&\times&\Theta\left(|q_1-q_2|\leq k\leq q_1+q_2\right)
[\omega_\pi(q_1)\omega_\pi(q_2)]^{-1}\nn\\
&\times& \left\{ 
\frac{f[\omega_\pi(q_1)]+f[\omega_\pi(q_2)]}
{\omega_\pi(q_1)+\omega_\pi(q_2)-\omega}
+\frac{f[\omega_\pi(q_1)]+f[\omega_\pi(q_2)]}
{\omega_\pi(q_1)+\omega_\pi(q_2)+\omega}\right.\nn\\
&&\left.+\frac{-f[\omega_\pi(q_1)]+f[\omega_\pi(q_2)]}
{\omega_\pi(q_1)-\omega_\pi(q_2)-\omega}
+\frac{-f[\omega_\pi(q_1)]+f[\omega_\pi(q_2)]}
{\omega_\pi(q_1)-\omega_\pi(q_2)+\omega}\right\}\;.\nn
\eea
To evaluate the principal value numerically, in an appropriate
way, one performs the following steps. First one introduces a new variable
$x\equiv \omega_\pi(q_1)$ and transforms the $q_1$-integration to an
$x$-integration, with $dq_1=\omega_\pi(q_1)/q_1dx$, and
$q_1=\sqrt{x^2-M_\pi^2}\equiv q^*$. Relabelling $q\equiv q_2$ leads to
\bea
[\Re\,\Sigma(\omega,k)]_2
&=&\left(\frac{4\lambda \sigma}{N}\right)^2
\frac{N}{8\pi^2}\frac{1}{k}
\mbox{P}\int_0^{\infty}     \, d q\,q\,[\omega_\pi(q)]^{-1}dx
\,\Theta\;\left(|q^*-q|\leq k\leq q^*+q\right)\nn\\
&\times& \left\{ 
\frac{f(x)+f[\omega_\pi(q)]}
{x+\omega_\pi(q)-\omega}
+\frac{f(x)+f[\omega_\pi(q)]}
{x+\omega_\pi(q)+\omega}\right.\nn\\
&&\left.+\frac{-f(x)+f[\omega_\pi(q)]}
{x-\omega_\pi(q)-\omega}
+\frac{-f(x)+f[\omega_\pi(q)]}
{x-\omega_\pi(q)+\omega}\right\}.
\eea
Second one transforms the $x$-integration to a sum over $x_i$, and uses
the mean-value theorem to put the terms with the distribution functions
in front of the integrals
\bea
[\Re\,\Sigma(\omega,k)]_2
&=&\left(\frac{4\lambda \sigma}{N}\right)^2
\frac{N}{8\pi^2}\frac{1}{k}
\mbox{P}\int_0^{\infty}     \, d q\,q
\, [\omega_\pi(q)]^{-1} \sum_{i}
\Theta\;\left(|q^*-q|\leq k\leq q^*+q\right)\nn\\
&\times&\left(
\{f(\hat{x})+f[\omega_\pi(q)]\}
\int_{x_i}^{x_{i+1}}dx 
\frac{1}
{x+\omega_\pi(q)-\omega}\right.\nn\\
&&+\{f(\hat{x})+f[\omega_\pi(q)]\}\int_{x_i}^{x_{i+1}}dx 
\frac{1}
{x+\omega_\pi(q)+\omega}\nn\\
&&+\{-f(\hat{x})+f[\omega_\pi(q)]\}\int_{x_i}^{x_{i+1}}dx 
\frac{1}
{x-\omega_\pi(q)-\omega}\nn\\
&&+\left.\{-f(\hat{x})+f[\omega_\pi(q)]\}\int_{x_i}^{x_{i+1}}dx 
\frac{1}
{x-\omega_\pi(q)+\omega}\right),
\eea
where $x_i\le \hat{x}\le x_{i+1}$, and $q^*\equiv\sqrt{\hat{x}-M_\pi^2}$.
Finally, the $x$-integrals can be performed analytically, 
\bea\label{re2k_ln}
[\Re\,\Sigma(\omega,k)]_2
&=&\left(\frac{4\lambda \sigma}{N}\right)^2
\frac{N}{8\pi^2}\frac{1}{k}
\int_0^{\infty}     \, d q\,q
\, [\omega_\pi(q)]^{-1} \sum_{i}
\Theta\;\left(|q^*-q|\leq k\leq q^*+q\right)\nn\\
&\times& \left(
\{f(\hat{x})+f[\omega_\pi(q)]\}
\ln\left|\frac{[\omega_\pi(q)+x_{i+1}]^2-\omega^2}
{[\omega_\pi(q)+x_{i}]^2-\omega^2}\right|
\right.\nn\\
&&+\left.\{-f(\hat{x})+f[\omega_\pi(q)]\}
\ln\left|\frac{[\omega_\pi(q)-x_{i+1}]^2-\omega^2}
{[\omega_\pi(q)-x_{i}]^2-\omega^2}\right|\right).
\eea
In the limit $k\to 0$, 
using Eq.~(\ref{theta_delta_ln}), the $q$-integration 
can be carried out, 
\bea\label{re2k0_ln}
[\Re\,\Sigma(\omega,0)]_2
&=&\left(\frac{4\lambda \sigma}{N}\right)^2
\frac{N}{4\pi^2}\sum_{i}
\,q^*\, [\omega_\pi(q^*)]^{-1}\\
&\times& \left(
\{f(\hat{x})+f[\omega_\pi(q^*)]\}
\ln\left|\frac{[\omega_\pi(q^*)+x_{i+1}]^2-\omega^2}
{[\omega_\pi(q^*)+x_{i}]^2-\omega^2}\right|
\right.\nn\\
&&+\left.\{-f(\hat{x})+f[\omega_\pi(q^*)]\}
\ln\left|\frac{[\omega_\pi(q^*)-x_{i+1}]^2-\omega^2}
{[\omega_\pi(q^*)-x_{i}]^2-\omega^2}\right|\right).\nn
\eea
\end{appendix}

\bibliography{thesis}

\end{document}